\documentclass[showpacs, preprintnumbers, nofootinbib, aps, prd, superscriptaddress,10pt, showkeys, notitlepage, twocolumn, aastex]{revtex4-1}

\usepackage{graphicx,amssymb,amsmath,amsthm,amsfonts,epsfig, setspace}

\usepackage[linktocpage]{hyperref}
\usepackage[usenames,dvipsnames]{color}
\usepackage{epstopdf}
\usepackage{pifont}
\definecolor{darkred}{rgb}{0.5,0,0}
\definecolor{darkgreen}{rgb}{0,0.5,0}
\definecolor{darkblue}{rgb}{0,0,0.5}
\definecolor{prussian}{rgb}{0.0, 0.19, 0.33}
\definecolor{richelectricblue}{rgb}{0.03, 0.57, 0.82}
\definecolor{teal}{rgb}{0.0, 0.5, 0.5}
\definecolor{mediumseagreen}{rgb}{0.24, 0.7, 0.44}
\definecolor{lust}{rgb}{0.9, 0.13, 0.13}
\definecolor{ballblue}{rgb}{0.13, 0.67, 0.8}
\definecolor{darkcyan}{rgb}{0.0, 0.55, 0.55}
\definecolor{mountainmeadow}{rgb}{0.19, 0.73, 0.56}
\definecolor{palecarmine}{rgb}{0.69, 0.25, 0.21}
\definecolor{richcarmine}{rgb}{0.84, 0.0, 0.25}
\definecolor{tangelo}{rgb}{0.98, 0.3, 0.0}
\definecolor{venetian}{rgb}{0.784,0.031,0.082}
\definecolor{bdfrance}{rgb}{0.192,0.549,0.906}

\hypersetup{colorlinks=true, citecolor=venetian,
linkcolor=bdfrance, urlcolor=lust}
\usepackage{amsmath,amssymb}
\usepackage{tensor}
\usepackage{mathtools}
\usepackage{amsbsy}
\usepackage{bm}
\usepackage{float}
\usepackage{cleveref}

\usepackage{appendix}

\newcommand{\be}{\begin{equation}}
\newcommand{\ee}{\end{equation}}
\newcommand{\bear}{\begin{eqnarray}}
\newcommand{\eear}{\end{eqnarray}}

\newcommand\tab[1][0.8cm]{\hspace*{#1}}

\begin{document}

\title{Fractal signatures of non-Kerr spacetimes in the shadow of light-ring bifurcations}

\author{Konstantinos Kostaros}
\email{kkostaro@auth.gr}
\affiliation{Department of Physics, Aristotle University of Thessaloniki, Thessaloniki 54124, Greece}

\author{Padelis Papadopoulos}
\email{padelis@auth.gr}
\affiliation{Department of Physics, Aristotle University of Thessaloniki, Thessaloniki 54124, Greece}
\affiliation{Research Center for Astronomy, Academy of Athens, Soranou Efesiou 4, GR-11527 Athens, Greece}
\affiliation{School of Physics and Astronomy, Cardiff University, Queens Buildings, The Parade, Cardiff CF24 3AA, UK}

\author{George Pappas}
\email{gpappas@auth.gr}
\affiliation{Department of Physics, Aristotle University of Thessaloniki, Thessaloniki 54124, Greece}

\begin{abstract}
Light-ring bifurcations that can occur for prolate non-Kerr compact objects can leave an indelible signature on SMBH shadows as a fractal sequence of eyebrow-like formations. These fractal features are the result of two properties of these spacetimes. The first is that they allow for multiple escapes for the photons (throats in the effective potential of photon geodesic motion). The second is that photon geodesics can resonate between different generalized light-rings related to the escapes, called fundamental photon orbits, that lead photons to alternate between the different exits either towards the compact object or infinity. The resulting fractal structures of the shadow seem to be a generic feature of prolate non-Kerr objects that may be observable in (accretion-disk)-illuminated compact objects, especially along equatorial lines of sight, but the best orientation depends on the specific parameters. Such fractal features if observed in the shadows of singular supermassive black holes at the centers of galaxies, would be smoking gun signals of non-Kerr compact objects. 
\end{abstract}

\maketitle

%%%%%%%%%%%%%%%%%%%%%%%%%%%%%%%%%%%%%%%%%%%%%%%%%%%%%%%%%%%%%%%%%%%
\section{Introduction}
\label{sec:intro}
%%%%%%%%%%%%%%%%%%%%%%%%%%%

The recently published images of the supermassive black holes (SMBH) in M87 and Sgr A* using Very Long Baseline Interferometry (VLBI) \cite{Akiyama_2019,EHT_M87_2,EHT_M87_3,EHT_M87_4,EHT_M87_5,EHT_M87_6,EHT_Sgr1,EHT_Sgr2,EHT_Sgr3,EHT_Sgr4,EHT_Sgr5,EHT_Sgr6,EHT_M87_7}, have opened up new paths in probing the spacetimes of black holes (BHs) and testing the theory of General Relativity (GR). With the shadow of a BH and the light-rings of various orders containing a wealth of information, relevant not only for BH imaging but also for gravitational waves and BH perturbations, much work has been done on light-ring computations within GR as well as in settings beyond GR and for non-BH compact objects \cite{Glampedakis:2017dvb,Glampedakis_2018,Glampedakis:2019dqh,Cardoso:2019mqo,McManus:2019ulj,Silva:2019scu,Volkel:2020xlc,Glampedakis:2021oie,Lima:2021las,Bryant:2021xdh,konoplya2020shadow,Guo_2020,Contreras_2020,Wang_2019,Wang_2020,Ayzenberg_2018,FPOs,Cunha2016IJMPD,Cunha2018GReGr,Kostaros_2022,Glampedakis:2023eek,Gralla_2020,Younsi:2021dxe,Bauer:2021atk,Gralla:2020nwp,Lima:2021las,Medeiros:2019cde,Gralla:2019xty,Cunha:2017qtt,Cunha2017PhysRevD,Johannsen_2010,Patil_2017,Kocherlakota:2023qgo,Kocherlakota:2024hyq,Kocherlakota:2022jnz}. 

The shadows are characterized by a central brightness depression surrounded by a bright ring created by the photons emitted by the accreting matter. The radiation emitted in the vicinity of the compact object illuminates the space, gets scattered, and propagates from the deep gravitational field of the compact object to an observer at infinity. This leaves an imprint of the strong field regime on the resulting images. 
Even though the bright ring that surrounds the shadow provides a way to test the metric, used as a proxy to measure the shadow's size, the image itself also folds in the  distribution and emission characteristics of the accreting matter and its geometric configuration. Therefore, one has to go beyond the shadow's pure features and calculate shadow images using different accretion disk models and their resulting illumination backgrounds in order to provide a more realistic description.

The study of the properties of the null geodesics threading spacetime is important in analyzing the gravitational lensing effects that take place and shape the observed shadow. 
The no-hair theorem states that all black holes are described by the Kerr metric and are uniquely defined by their mass and spin (for astrophysical black holes charge is not deemed important), but evidence that astrophysical black hole candidates are actually Kerr black holes is still not conclusive. 
In order to test the nature of these compact objects, a framework for quantifying the deviations from the Kerr metric is required.  Alternative models to Kerr black holes exist both within GR and beyond it \cite{Manko_1992,friedberg1987,Mazur:2001fv,barcelo2008,Glampedakis_2018,Yagi:2016ejg,psaltis2008,Herdeiro:2015waa,Eichhorn_2021}. It is therefore important to test for the ``Kerrness'' of such astrophysical objects, in order to both identify the nature of these dark objects and test the predictions of GR in the strong-field regime \cite{Barack:2018yly,Berti:2015itd,LIGOScientific:2016lio}.

An important aspect of non-Kerr spacetimes is that they can generally describe compact objects that even when they lack an event horizon, they possess characteristics that make them appear similar to BHs. These objects are termed \textit{black hole mimickers}. Such compact objects alternative to Kerr BHs can be ultra-compact objects (UCOs) like gravastars, boson stars, and other matter configurations with exotic equations of state, such as anisotropic pressure or non-GR objects like BHs and compact objects in scalar-tensor theories and f(R) gravity \cite{Yagi:2016ejg,Mazur:2001fv,Visser:2003ge,Eby_2016,Schunck:2003kk,darkobjcardoso}. In these cases one would have to construct specific models and study their properties on a case-by-case basis. 

An alternative avenue for describing non-Kerr compact objects has also been developed in the literature. There has been a lot of work on constructing spacetimes that are not particular solutions of a particular theory and do not correspond to a particular mass/energy distribution. They are instead agnostic parametric deviations from the Kerr spacetime, with a number of free parameters selected to quantify (Kerr-spacetime)-deviations in a manner the user prefers  \cite{Glampedakis:2005cf,Collins:2004ex,Vigeland:2011ji,Johannsen:2011dh,Johannsen:2013szh,Rezzolla:2014mua,Cardoso:2014rha,Konoplya:2016jvv,Papadopoulos:2018nvd,Carson:2020dez,Yagi:2023eap}. 
Such spacetimes can have their parameters adjusted to approximate known solutions in alternative theories of gravity or simply behave as stand-alone generic spacetimes with such versatility that are a useful tool for testing the Kerr hypothesis.

The spacetimes that we are investigating here are non-Kerr, i.e. they can be considered as deviations from a Kerr spacetime. 
They are also non-integrable, namely the equations of motion do not admit a full set of constants of motion and cannot be cast in a separable form. As such, particle motion in these spacetimes may exhibit chaotic behavior that can lead to interesting phenomenology such as fractal structures in the shadow that the object casts \cite{Kostaros_2022,Shipley:2019kfq,FPOs,WangPhysRevD2018,Wang_2020,Sengo_2023,Wang_long2020shadow,Wang:2022kvg} or other chaotic behavior in both their null or timelike geodesics \cite{Dettmann:1994dj,Dettmann:1995ex,Sota:1995ms,Gueron:2001ey,Han:2008zzd,Zelenka:2017aqn,Destounis:2023cim}. 

In previous work, these fractal structures have been mainly found in cases where the Hamiltonian system can take the form of a closed system, i.e., the separatrix that marks the allowed region of motion for light rays can form a pocket where light rays get trapped for long periods of time. These pocket formations while providing a suitable lab to study the chaotic motion of geodesics are usually not accessible from spatial infinity, meaning that the quest to find any observational signatures is rather challenging \cite{kopacek2010,CONTOPOULOS_2011,Zelenka:2017aqn,Dubeibe_2007,Igata_2011,Gueron:2001ey,Sota:1995ms}. In cases where the connection of the pocket with infinity is possible, i.e., a light ray can reach an observer after its chaotic motion, interesting features emerge on the shadow of the object in the form of cusps, eyebrows and eyelashes \cite{Sengo_2023,FPOs,Shipley:2019kfq,Kostaros_2022}. It is worth mentioning that similar features have been observed in rotating non-Kerr BHs regularized by quantum gravity effects \cite{Eichhorn_2021,Eichhorn:2021etc}.

In this work, we will further explore the conditions under which one can have such fractal features in the shadows of compact objects against an illumination background. We do so by focusing on spacetimes and parameters that leave the systems open (i.e., without forming pockets) and the photons can therefore escape from the scattering region to infinity or onto the compact object via one or more \textit{escapes}. These systems are \textit{Open Hamiltonian systems} and we will show that they can still form fractal features in their shadows that are essentially caused by the presence of multiple escapes caused by the bifurcation of the equatorial photon orbit to multiple off-equatorial photon orbits. Such fractal structures are smoking gun signals of non-Kerr compact objects and a solid indicator of their non-Kerr nature.

In Section \ref{sec:hamiltonform} we present the formulation used to solve the photon geodesics and also present the spacetimes that we use, which are typical and general examples of the spacetimes that possess the properties we are interested in. In Section \ref{sec:photonorbits} we present the properties that photon orbits have in the spacetimes we investigate and give the results of our analysis of the corresponding mathematical shadows and their fractal structure. Finally, in Section \ref{sec:accretion} we apply our results in an astrophysical situation, where an accretion disk illuminates a compact object and we observe its shadow. We finish with our conclusions and aims of future work in Section \ref{sec:conclusions}.

In what follows we use units with $G=c=1$.

%%%%%%%%%%%%%%%%%%%%%%%%%%%
\section{Hamiltonian Formalism and Non-Kerr Spacetimes}
\label{sec:hamiltonform}
%%%%%%%%%%%%%%%%%%%%%%%%%%%

For a stationary and axisymmetric spacetime, the metric takes the general form \cite{Wald:1984rg}:
\begin{equation}
ds^2=g_{tt} dt^2 +g_{rr}dr^2 + g_{\theta\theta}d\theta^2+ g_{t\phi}dt d\phi +g_{\phi\phi}d\phi^2.
\label{stataxismetric}
\end{equation}
The two Killing vector fields admitted by this spacetime are associated to two conserved quantities, the energy $E$ and the angular momentum $L$, both normalized per unit mass,
\begin{equation}
E=-\xi^\alpha u_{\alpha}= -\left(g_{tt} \frac{dt}{d\lambda} + g_{t\phi}\frac{d\phi}{d\lambda}\right),
\label{eq:9}
\end{equation}
\begin{equation}
L=\eta^\alpha u_{\alpha}=g_{t\phi}\frac{dt}{d\lambda}+ g_{\phi\phi}\frac{d\phi}{d\lambda},
\label{eq:10}
\end{equation}
where $\lambda$ is an affine parameter that parameterizes the geodesic. The equations of motion are derived from the Lagrangian:
\begin{equation}
\mathcal{L}=\frac{1}{2}g_{\alpha b} \dot{x}^\alpha \dot{x}^b,
\end{equation}
and correspond to the equations that govern geodesic motion in this spacetime. One can then define the associated generalized momenta as,
\begin{equation}
p_{\alpha}=\frac{\partial \mathcal{L}}{\partial \dot{x}^\alpha},
\end{equation}
and express the Hamiltonian as
\begin{equation}
\mathcal{H}=\frac{1}{m} \sum p_{\alpha} \dot{x}^\alpha - \mathcal{L} = -E \dot{t}+L \dot{\phi} + g_{rr}\dot{r}^2 +g_{\theta\theta} \dot{\theta}^2 -\mathcal{L}.
\end{equation}
In this context, $p_r=g_{rr}\dot{r}$ and $p_{\theta}=g_{\theta\theta} \dot{\theta}$ represent the momenta in the radial and poloidal directions, respectively. Solving eqs.(\ref{eq:9}-\ref{eq:10}) for $\dot{t}$ and $\dot{\phi}$ and replacing in the previous expression, we get the Hamiltonian in the form 
\begin{equation}
\begin{split}
\mathcal{H}&= \frac{1}{2}\left(g_{rr}\dot{r}^2 +g_{\theta\theta}\dot{\theta}^2 - \frac{L^2 g_{tt} + 2EL g_{t\phi} + E^2 g_{\phi\phi}}{\mathcal{D}}\right) \\
&=\frac{1}{2}\left(\frac{p_{r}^2}{g_{rr}}+\frac{p_{\theta}^2}{g_{\theta\theta}}-\frac{L^2 g_{tt} + 2EL g_{t\phi} + E^2 g_{\phi\phi}}{\mathcal{D}}\right),
\end{split}
\label{eq:hamiltonian}
\end{equation}
where we have defined $\mathcal{D}=g_{t\phi}^2 - g_{tt}g_{\phi\phi}$, which also defines the Killing horizon. The above Hamiltonian is expressed in terms of the effective potential
\begin{equation}
V_{\text{eff}}=-\frac{L^2 g_{tt} + 2EL g_{t\phi} + E^2 g_{\phi\phi}}{\mathcal{D}}.
\end{equation}
The contour defined by $V_{\text{eff}}=0$ marks the region of allowed motion, signifying a boundary of zero velocity. The respective conditions for massive and massless particles are $\mathcal{H}=-1/2$ and $\mathcal{H}=0$. Finally, Hamilton's canonical equations yield the equations of motion
\begin{equation}
\dot{x}^\alpha=\frac{\partial \mathcal{H}}{\partial p_\alpha},
\quad 
\dot{p}_\alpha=-\frac{\partial \mathcal{H}}{\partial x^\alpha},
\end{equation}
where the dot refers to differentiation with respect to the affine parameter. These equations written explicitly take the form
\begin{equation}
\begin{aligned}
\dot{r}&=\frac{p_r}{g_{rr}} , &\dot{p_r}&=-\frac{\partial \mathcal{H}}{\partial r}, \\
\dot{\theta}&=\frac{p_\theta}{g_{\theta\theta}} , &\dot{p_\theta}&=-\frac{\partial \mathcal{H}}{\partial \theta}, \\
\dot{t}&=\frac{E g_{\phi\phi}+L g_{t\phi}}{\mathcal{D}} , &\dot{p_t}&=0, \\
\dot{\phi}&=-\frac{L g_{tt} + E g_{t\phi}}{\mathcal{D}} , &\dot{p_\phi}&=0.
\end{aligned}
\end{equation}
In our analysis, we will either use the momenta $p_r,p_\theta$ or alternatively the velocities $u^r=\dot{r},$ $ u^\theta=\dot{\theta}$. 
For photon geodesics, one defines the two impact parameters
\begin{equation}
  b\equiv- \frac{p_{\phi}}{p_t}=\frac{L}{E}, \quad \textrm{and} \quad \alpha=\frac{p_{\theta}}{p_t}.
\end{equation} 
It is important to emphasize that this is the conventional definition of the orbital angular momentum impact parameter, where positive values of $b$ refer to co-rotating orbits. In scenarios where one wants to visualize the shadow of a compact object, a variant of this definition is utilized, i.e., $b=p_{\phi}/p_t$, which is the opposite of the conventional definition. This adjustment accounts for photons being emitted near the compact object and reaching the image plane of a distant observer, which is the reversed situation to the usual case where one emits photons towards the compact object. We will use the appropriate definition depending on the context.

Having put forward the setup for calculating null trajectories, we proceed to briefly present the spacetimes that we will use in our analysis.

%%%%------------------------
\subsection{HT}
%%%%------------------------

The Hartle-Thorne (HT) spacetime in our analysis serves as a model for the external region of a rotating compact object. Proposed by Hartle and Thorne \cite{1967ApJ...150.1005H,hartle2}, this framework was designed to provide both the internal and external structure of compact slowly rotating fluid configurations. One starts from a static configuration %
and introduces rotational effects as perturbations, using the rate of rotation as the expansion parameter. 
The spacetime is then described by the line element \cite{hartle2},
\begin{equation}
    \begin{split}
        ds^2= &  -e^\nu \left(1+2h\right) dt^2 + e^\lambda \left(1+\frac{2\mu}{r-2m}\right)dr^2  \\
    &+ r^2\left(1+2k\right)\{d\theta^2 +  \sin^2\theta \left[ d\phi -\left(\Omega - \omega\right)dt\right]^2 \}  \\
    &+ \mathcal{O}\left(\Omega^3 \right),
    \end{split}
    \label{eq:htgen}
\end{equation}
where $\Omega$ is the stellar angular velocity. The metric potentials $h\left(r,\theta\right)$, $\mu\left(r,\theta\right)$, $k\left(r,\theta\right)$ and $\omega \left( r, \theta\right)$ are expanded in terms of Legendre polynomials,
\begin{equation}
    \begin{split}
     &h\left(r,\theta\right)=h_0\left(r\right) +h_2\left(r\right) P_2, \\
    &\mu\left(r,\theta\right)=\mu_0\left(r\right) + \mu_2\left(r\right) P_2\\ 
    &k\left(r,				\theta\right)=k_2\left(r\right) P_2, \tab \omega \left(r,\theta\right)=			\omega_1 \left(r\right) P_1 ' .
    \end{split}
\end{equation}
For the exterior of the compact object the HT metric is parameterized in terms of the non-rotating mass $M$, the $\mathcal{O}\left(\Omega\right)$ angular momentum $J$ and the spin parameter $\chi= J/M^2$, the quadrupole moment $Q=\chi^2 M^3 \left(1-\delta q\right)$ in terms of $\delta q$, which is the deviation from the Kerr quadrupole, $\delta m$, that is the $\mathcal{O}\left(\Omega^2\right)$ correction in the mass, and $x=r/M$, the reduced radial coordinate \cite{Glampedakis:2018blj}
\begin{equation}
\begin{split}
&m=M, \tab  e^\nu = e^{-\lambda} = 1- \frac{2}{x} , \tab \omega_1= \Omega - \frac{2\chi}{Mx^3},\\
&\frac{\mu_0}{M}= \chi^2 \left(\delta m - \frac{1}{x^3}\right),\tab h_0=\frac{\chi^2}{x-2}\left(\frac{1}{x^3}-\delta m \right),
\end{split}
\end{equation}
\begin{equation}
\begin{split}
&h_2= \frac{5}{16}\chi^2 \delta q \left(1- \frac{2}{x}\right) \left[ 3x^2 \log\left(1-\frac{2}{x}\right) + \frac{2}{x}\frac{\left(1-1/x\right)}{\left(1-2/x\right)^2}\right. \\
&\Biggl.\left(3x^2-6x-2\right) \Biggr]+\frac{\chi^2}{x^3}\left(1+\frac{1}{x}\right),
\end{split}
\end{equation}
\begin{equation}
\begin{split}
&k_2 = - \frac{\chi^2}{x^3}\left(1+\frac{2}{x}\right)- \frac{5}{8} \chi^2 \delta q\left[3\left(1+x\right)-\frac{2}{x}-\right.\\
&\left.3\left(1-\frac{x^2}{2}\right)\log\left(1-\frac{2}{x}\right)\right],
\end{split}
\end{equation}
\begin{equation}
\begin{split}
&\frac{\mu_2}{M} =  - \frac{5}{16}\chi^2 \delta q x \left(1-\frac{2}{x}\right)^2 \left[3x^2 \log\left(1-\frac{2}{x}\right) + \right.\\ &\left.+\frac{2}{x}\frac{\left(1-1/x\right)}{\left(1-2/x\right)^2}\left(3x^2-6x-2\right)\right]-\frac{\chi^2}{x^2}\left(1-\frac{7}{x}+\frac{10}{x^2}\right).
\end{split}
\end{equation}
To get the final form of the metric, we redefine $M$ as the spin-modified stellar mass, i.e., the corrected total mass, which amounts to setting $\delta m=0$ in the above equations. 
One may use the HT metric as it is given here at the specified $\Omega$-order, to describe the exterior of a compact object where both the spin parameter $\chi$ and the quadrupole deviation $\delta q$ can be considered as free parameters of the spacetime together with the mass $M$. When calculating shadows, we will assume that the HT compact objects have a surface at a radius that encloses all possible pathologies of the spacetime and doesn't interfere with the region where photons are allowed to move.  

%%%%------------------------
\subsection{JP}
%%%%------------------------

The Johannsen-Psaltis (JP) metric \cite{Johannsen:2011dh}, is a spacetime designed to deviate from the Kerr metric in a nonlinear manner through a set of free parameters. 
For our analysis, we will use a special case of this deformed Kerr spacetime with the deformation given specifically by the function
\begin{equation}
    h(r,\theta)=\epsilon_3 \frac{M^3 r}{\Sigma^2}
\end{equation}
where $\epsilon_3$, the only available deformation parameter in this case, is a constant. 
It's important to note that different parameterizations of the deformation function can lead to different physical predictions. It is convenient to express the JP metric in terms of the Kerr one as \cite{Glampedakis:2018blj},
\begin{equation}
\begin{split}
    &g^{JP}_{tt}=(1+h)g_{tt}^K,\tab g_{t\phi}^{JP}=(1+h)g_{t\phi}^K\\   &g_{rr}^{JP}=g_{rr}^K(1+h)\left(1+h\frac{a^2\sin^2\theta}{\Delta}\right)^{-1},\\ &g_{\theta\theta}^{JP}=g_{\theta\theta}^K,\tab g_{\phi\phi}^{JP}=g_{\phi\phi}^K+ha^2\left(1+\frac{2Mr}{\Sigma}\right)\sin^4\theta,
\end{split}
\end{equation}
where
\begin{equation}
\Sigma=r^2+a^2\cos\theta^2,\quad \Delta=r^2-2mr+a^2, \quad a=\chi M,
\end{equation}
and for $\epsilon_3 \longrightarrow 0$ we get the Kerr metric. The $\epsilon_3$ parameter can be considered as a parameter that drives the deformation of the mass quadrupole of the spacetime at the leading order. 

%%%%------------------------
\subsection{MP di-hole}
%%%%------------------------

The Majumdar-Papapetrou spacetime is a static solution discovered in 1947 by Majumdar and Papapetrou, independently \cite{Papapetrou:1947,Majumdar:1947eu}. In cylindrical coordinates, $\{t,\rho,\phi,z\}$, its geometry is described by the line element \cite{Patil_2017}
\begin{equation}
    ds^2=-\frac{dt^2}{U^2}+U^2(d\rho^2+\rho^2d\phi+dz^2),
\end{equation}
where $U(\rho,z)$ is a function that satisfies Laplace's equation $\nabla^2 U=0$ on a three-dimensional auxiliary Euclidean space. We will focus on the case where this spacetime describes two equal mass extremal Reissner-Nordstr\"om (RN) black holes in static equilibrium. In this case $U(\rho,z)$ has the form
\begin{equation}
    U(\rho,z)=1+\frac{M}{\sqrt{\rho^2+(z-z_+)^2}}+\frac{M}{\sqrt{\rho^2+(z-z_-)^2}}.
\end{equation}
The two equal mass RN black holes are located at $z_{\pm}=\pm d/2$ where $d$ is their coordinate distance. Their center of mass is at the origin of the coordinate system and their event horizons are around their positions at $(0,0,z_{\pm})$ \cite{Shipley:2019kfq}. This system clearly describes a prolate configuration which is though non-rotating. It is therefore qualitatively different from the other two with respect to aspects other than its prolateness. We thus include the di-hole in our analysis as a model that serves to broaden our investigation of the feature of fractal shadows.

%%%%%%%%%%%%%%%%%%%%%%%%%%%
\section{FPOs, Exit Basins and Shadows }
\label{sec:photonorbits}
%%%%%%%%%%%%%%%%%%%%%%%%%%%

We will now proceed by first providing some useful definitions regarding the properties of photon orbits and the tools we will be using to analyze them, and we will then continue with the analysis for each of the spacetimes we presented.

%%%%------------------------
\subsection{Properties and description of photon orbits}
%%%%------------------------

\textit{Fundamental Photon Orbits}. In stationary and axisymmetric spacetimes, there can exist photon orbits that neither fall into the BH or the surface of an ultra-compact object (UCO) nor escape to infinity. These bound photon orbits, which have been named \textit{Fundamental Photon Orbits} (FPOs), are a generalization of light-rings (LRs) and they are important to the understanding of shadows of more general spacetimes than Kerr. These orbits are formally defined following \cite{Cunha2017PhysRevD} as:

\noindent\textit{Definition:} Let $s(\lambda):\mathbb{R}\longrightarrow \mathcal{M}$ be an affinely parameterized null geodesic. $s(\lambda)$ is a FPO if it is restricted to a compact spatial region and there is a value $T>0$ for which $s(\lambda)=s(\lambda+T),\forall \lambda \in \mathbb{R}$, up to isometries, and they are categorized as $X_{n_s}^{n^r\pm}$, where $X=\{O,C\}$, and $n_r,n_s\in\mathbb{N}_0$:

\begin{itemize}
    \item \textit{class $O$}(open) if they reach the boundary or \textit{class $C$}(closed) if they do not (form a loop onto themselves);
    \item \textit{subclass}$^+$ : even under $\mathbb{Z}_2$ or \textit{subclass}$^-$ : odd under $\mathbb{Z}_2$ ;
    \item they intersect the equatorial plane $(\theta=\pi/2)$ at $n_r$ distinct $r$ values (subclass$^{n_r}$);
    \item orbits \textit{on} the equatorial plane (LRs), have $n_r=0$;
    \item subclass$_{n_s}$ for $n_s$ self-intersection points.
\end{itemize}
In simple terms, this definition essentially identifies the orbits that trap photons in bound periodic trajectories as FPOs and then characterizes their different possible morphologies. In what follows, we will identify such FPOs in the cases that we will work on, but we will not go into more details about these orbits. 

Equatorial light-rings (which in this scheme are designated as $O_0^{0+}$) can be simpler to analyze since the problem becomes effectively 1-dimensional and the effective potential is $V_{\rm eff} = V_{\rm eff}(r,b)$. This simplification reduces the problem of finding FPOs to the standard conditions for circular geodesics, i.e., \(V_{\text{eff}}(r_0, b_0) = 0\) and \(V_{\text{eff}, r}(r_0, b_0) = 0\), from which one can derive equations for both the radius of a photon ring and the impact parameter \(b\) \cite{Glampedakis:2018blj}. 
More general LRs can be identified using the $h_{\pm}$ potentials \cite{Cunha2016PhRvD} defined as 
\begin{equation}
    h_{\pm}\equiv \frac{-g_{t\phi}\pm \sqrt{D}}{g_{tt}}.
\end{equation}
A LR is either an extremum or a saddle point of the potentials $h_{\pm}$ at fixed $(r,\theta)$.
These points correspond to specific values of the $b$-impact parameter which when perturbed, give rise to an escape channel in the phase space. Such an escape channel is defined by the neighboring contours of $h_{\pm}$ to the contour of the saddle point, as we will see in specific examples in the next subsections. These contours are almost parallel to each-other and on either side of the saddle, forming a throat. In these throats, periodic unstable orbits form that bounce between the contours that define the throat forming FPOs. 
In Non-Kerr spacetimes it is furthermore possible for generic periodic unstable orbits to form that bounce around inside the region of allowed motion moving between FPOs and in this sense, these FPOs are \textit{dynamically connected} \cite{Shipley:2019kfq}.
Additional FPOs that are not saddle points or extrema of the $h_{\pm}$ functions can also exist 
in non-separable spacetimes \cite{Cunha2017PhysRevD}. 
These are found numerically through trial and error, by searching for orbits that spend a lot of coordinate time at specific locations on the $(r,\theta)$ plane.

\textit{Exit Basins.} A set of light rays that enter the scattering region, may remain confined in it through different 
resonances with the dynamically connected FPOs \cite{Cunha2016PhRvD}. 
When the system is an \textit{open Hamiltonian system}, there exist more than one escapes from the scattering region and in order to better understand the system's dynamics we construct \textit{exit basins} in the phase space. An exit basin is a subset of the state space such that all the initial conditions that lie in it escape through the same exit. To draw an exit basin diagram, we integrate the equations of motion for a fine grid of initial conditions and color code our data based on the escape through which the light rays either plunge into the compact object or reach infinity. The exit basins can be either wide and well-defined or elongated with a more complicated structure that may also be self-similar \cite{Zotos_2017}. 

\textit{Shadows and self-similarity.} The shadows that compact objects cast can also be viewed as exit basins since they are the set of all initial conditions on the observer's image plane that when traced backwards in time, lead to the surface of the object through the escapes of the open Hamiltonian system \cite{Shipley:2019kfq}. The bright boundary that defines the shadow of a BH (or a BH mimicker compact object) as seen by a distant observer is essentially an image of the unstable photon orbits, i.e., the BH's photon spheres or FPOs, from which photons marginally escape to infinity. For the spacetimes that we will investigate that are non-Kerr and non-separable, we have to use numerical ray-tracing methods. The simplest setup that will give the shadow is one where we assume a uniformly lit BH or compact object where the light source is essentially a spherical screen surrounding the object and the observer that emits isotropically with uniform brightness. Each light ray that connects the observer to a point of the light source will correspond to a bright spot on the observer's field of view, while those that do not, will correspond to dark spots. The boundary between the bright and dark spots will thus define the photon spheres  \cite{Perlick_2022} in the observer's image plane, which is essentially the shadow. This setup is of little astrophysical interest since it only provides the \textit{mathematical} shape of the shadow but it is, however, the textbook way of understanding key features of gravitational lensing \cite{Bardeen:1972fi, Cunha2016PhRvD, Cunha2016IJMPD, Cunha2016PhRvD, Cunha2017PhysRevD, Cunha2018GReGr, Lima:2021las, Medeiros:2019cde, Perlick_2022, WangPhysRevD2018, Shipley:2019kfq}.

A structure can be characterized as \textit{self-similar} when it can be broken down into arbitrarily smaller pieces that replicate the entire structure. One way to measure the degree of complexity in such fractal structures is through their \textit{dimension}. For our work, we are particularly interested in the \textit{box-counting dimension} which is related to the self-similarity dimension. The structure is initially put onto a grid of mesh size $s$ and the number of grid boxes that contain a part of the structure are counted to be $N(s)$. We then continuously decrease the size of $s$ and keep counting the number $N(s)$ to make a log/log diagram of the $N$ versus $s$. The slope $D$ of the resulting line is the \textit{box-counting dimension} $D_s$ \cite{peitgen2004chaos}.  
The box-counting method can be applied considering any part of the fractal structure or it can be applied considering just its boundary. In the first case, we are interested in measuring how much space the structure takes up at different scales and we will use this implementation to calculate the dimension of the exit basin diagrams. In the second case, we are interested in understanding the complexity of a specific boundary and we will use it to find the dimension of a shadow's boundary. The most well-known example of this approach with respect to fractals is the Coastline paradox, i.e., the observation that a coastline does not have a well-defined length due to having additional structure as one goes to smaller scales \cite{Mandelbrotbritain}.

Now that we have described the subjects of our investigation and the tools that we use to explore their properties, we proceed to the application to the specific spacetimes introduced in the previous section.

%%%%------------------------
\subsection{HT}
%%%%------------------------

Glampedakis and Pappas, \cite{Glampedakis:2018blj}, showed that the HT spacetime possesses photon orbits and light-rings with unique properties which can diverge distinctly from those seen in a Kerr BH. When the quadrupole deviation parameter $\delta q$ assumes specific positive values, which correspond to having a prolate compact object, and the spin parameter exceeds a critical value $\chi_c(\delta q)$ that is different for different $\delta q$s, the Kerr-like co-rotating equatorial light-ring bifurcates into two non-equatorial rings.

\begin{figure}[h]
\begin{center}
\includegraphics[width=0.2\textwidth]{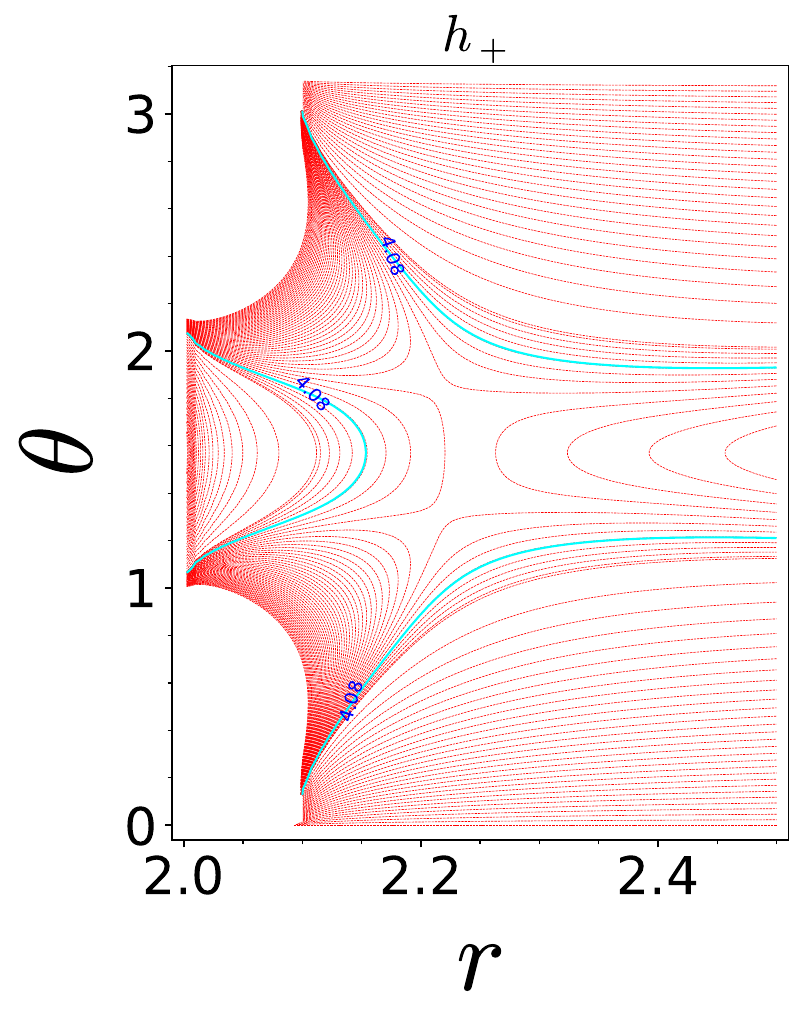}
\includegraphics[width=0.2\textwidth]{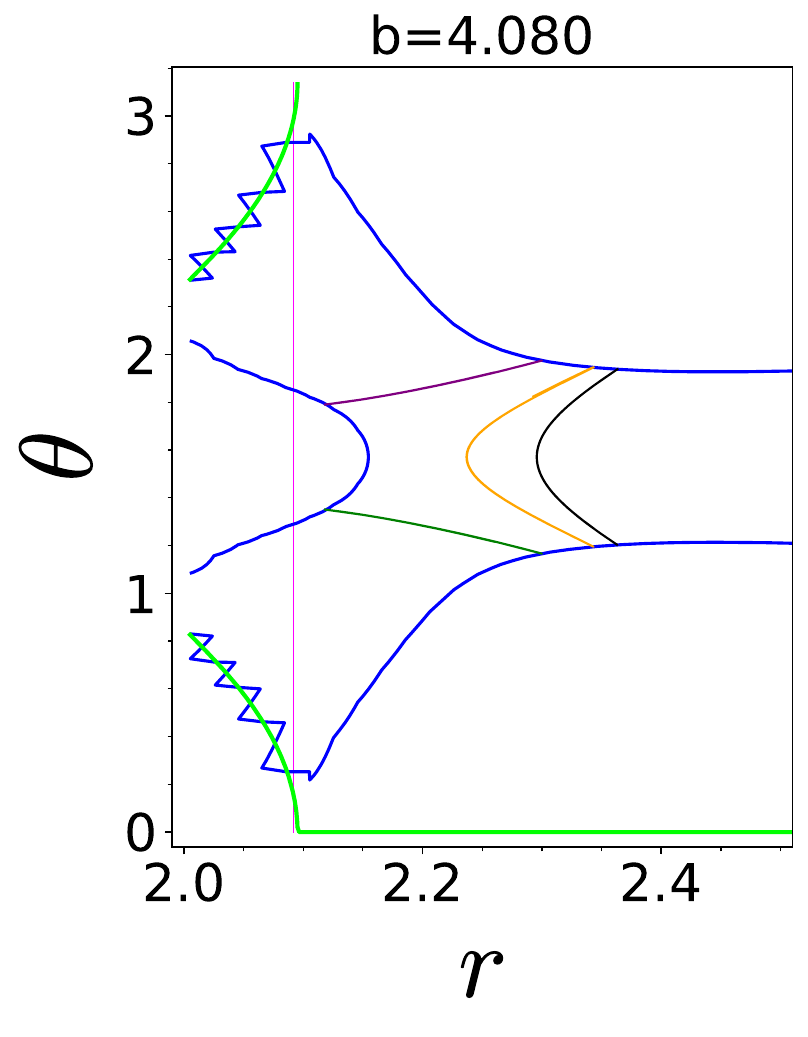}
\end{center}
\caption{Left: The potential function $h_+$ for the HT spacetime. The two non-equatorial LRs appear as saddle points that are in low latitudes due to the small value of the spin parameter, $\chi=0.35$. Right: The separatrix for an impact parameter $b=4.08M$ and the FPOs admitted. The blue line marks the forbidden region for photon motion, the lime line marks the horizon and the magenta line marks the surface that we set for the compact object at $r=2.092M$.}
\label{fig:htfpos}
\end{figure}

For a relatively narrow range of the spin parameter $\chi$, these two off-equatorial light-rings coexist with the equatorial one and can trap photon orbits through the formation of a pocket. This phenomenon is consistently observed in the HT metric and extends to higher order expansions, at least to $\mathcal{O}(\Omega^3)$. Previous work, \cite{Kostaros_2022}, investigated how the photon orbits' properties change by the formation of this pocket and by the transition from an open Hamiltonian system with three escapes, to a system disconnected from the compact object with only one escape to infinity (a pocket with a narrow throat). 

As the parameters change further moving the system away from the light-ring bifurcation, 
the equatorial co-rotating light-ring vanishes and the off-equatorial rings move to higher latitudes, and the system becomes an open Hamiltonian one. 

In this paper, we focus on this open HT setup that can exist for a wider range of parameters and is a more general case of open Hamiltonian systems in GR. For the results presented here, we set the spin parameter to $\chi=0.35$, the quadrupole deviation parameter to $\delta q =1$ and the mass of the compact object to $M=1$ (which sets the length scale of the system to 1 unit of length). For these parameters, the spacetime does not admit an equatorial Kerr-like LR and we now only have two non-equatorial LRs. The potential $h_+$, where the two unstable LRs appear as saddle points, and the separatrix for an impact parameter $b=4.08M$ along with the FPOs admitted in this setup are shown in \cref{fig:htfpos}.
The system is open with three escapes; two throats that connect it to the surface of the compact object and one connecting it to infinity. 
In \cref{fig:htfpos} on the right panel we can see three FPOs. The two curves on the two escape throats, in purple and green, constitute one FPO of type $O^{0_{-}}_0$, i.e., $O$: indicates that the orbit reaches the boundary of the separatrix, $O^{0}$: indicates that there is no intersection with the equatorial plane, $O^{0-}$: indicates that it is odd under $\mathbb{Z}_2$, $O^{0_{-}}_0$: indicates there are no self-intersection points. The other two curves, the orange and the black that form on the side of the escape to infinity, each constitute one FPO of type $O^{1_{+}}_0$,i.e., $O$: the orbit reaches the boundary of the separatrix, $O^{1}$: there is one intersection with the equatorial plane, $O^{1_{+}}$: it is even under $\mathbb{Z}_2$, $O^{1_{+}}_0$: it has no self-intersection points. We will not continue with the detailed description of the FPOs beyond this point.

\begin{figure}[h]
\centering
    \includegraphics[width=0.15\textwidth]{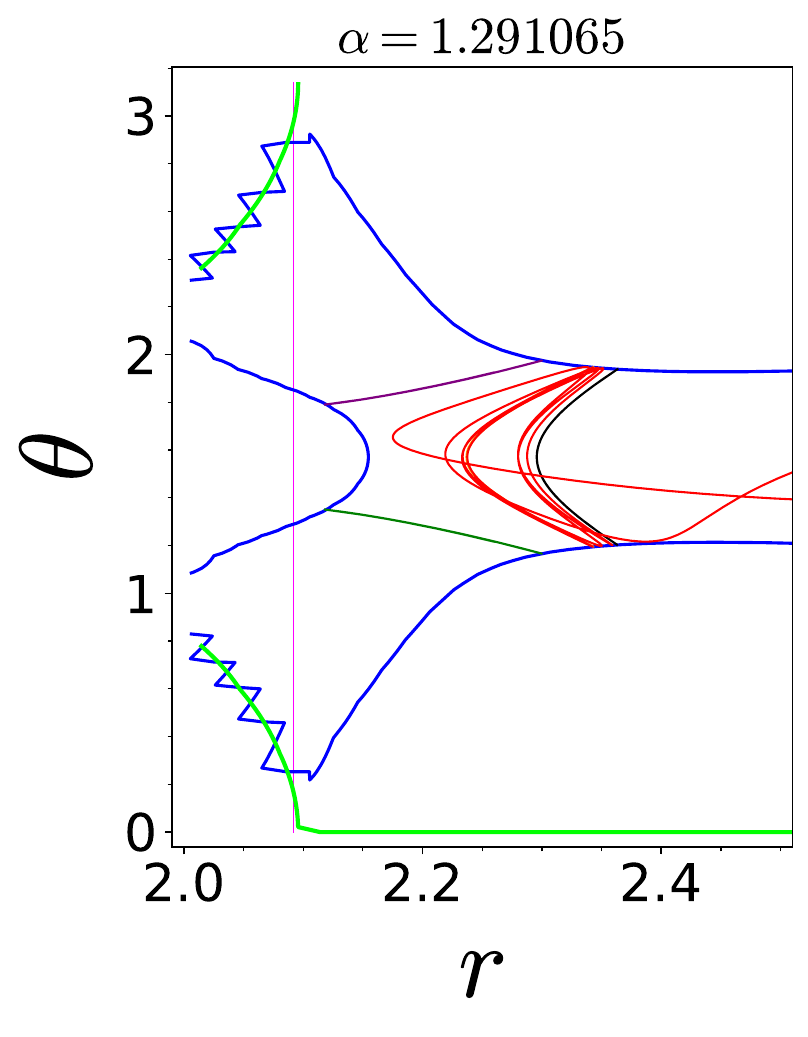} 
    \includegraphics[width=0.15\textwidth]{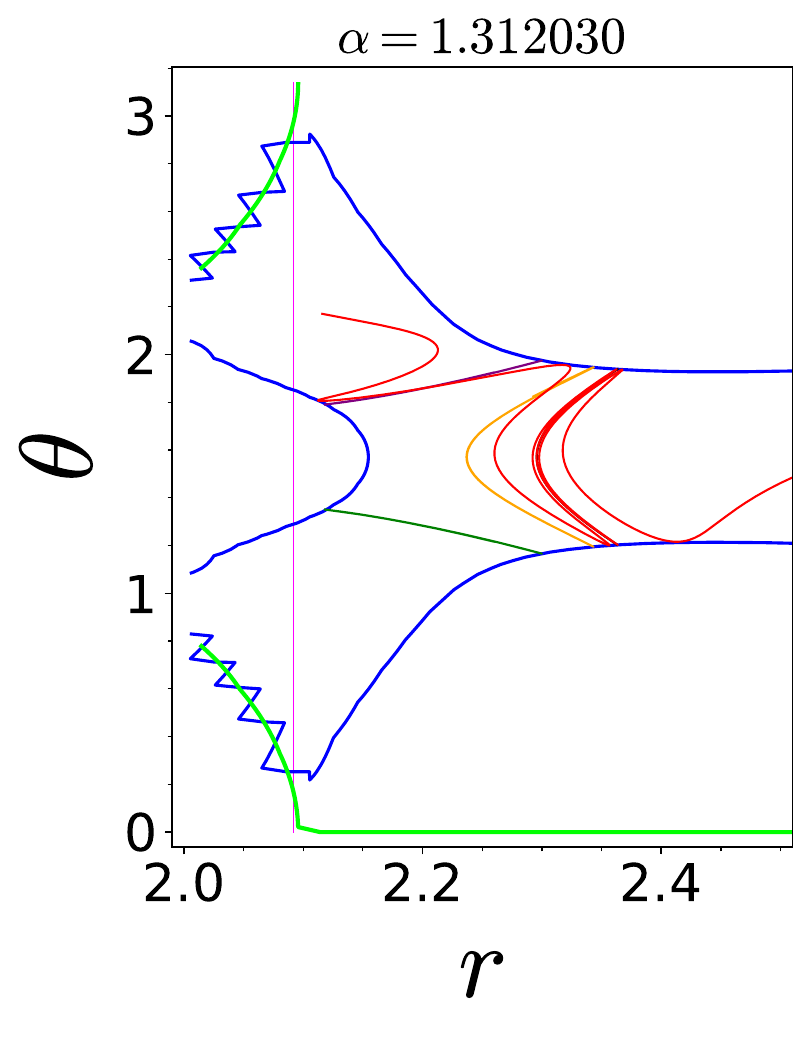}
    
    \includegraphics[width=0.15\textwidth]{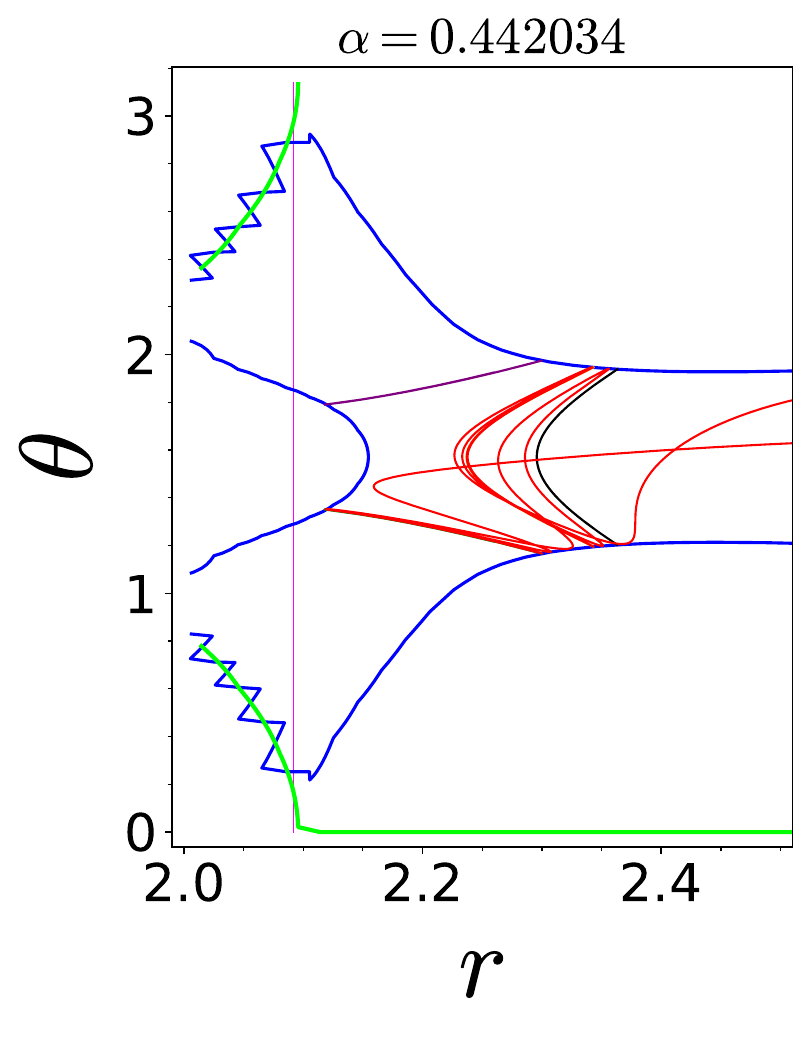} 
    \includegraphics[width=0.15\textwidth]{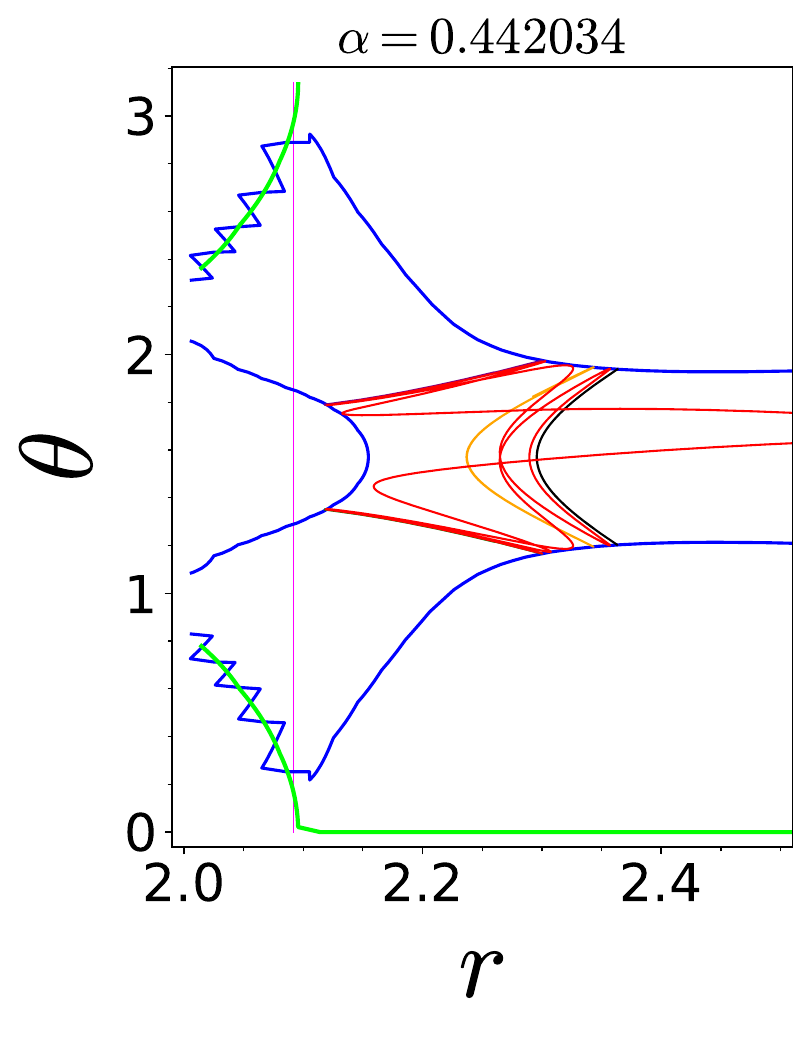} 
  \caption{Top: Two photons launched with impact parameters $\alpha\approx 1.2910M$ and $\alpha\approx 1.3120M$ respectively. They both resonate with one of the $O^{1_{+}}_0$ FPOs. Bottom: Two photons launched with impact parameters $\alpha\approx 0.442034M$ but with a difference in their 7th decimal. They both resonate with two FPOs, but the first one fails to resonate with both distinct orbits of $O^{0_{-}}_0$.}
  \label{fig:diffcombsHT}
\end{figure}

 For spacetimes with potential configurations like the ones we investigate, i.e., where multiple FPOs may exist, it is possible for photons to resonate with each FPO individually or in principle any combination thereof. This practically means that a photon orbit can get trapped near an FPO for a long period of time or it may alternate between more than one FPOs, being trapped initially near one and then moving to being trapped near another and so on. There can exist a hierarchy of resonances where a photon that is more ``excited'' in the context of the resonance can resonate more times with different FPOs \cite{Cunha2016PhRvD,Kostaros_2022}. 
This is the case for photons (red) with different $\alpha$-impact parameters that we follow in \cref{fig:diffcombsHT}. The FPOs are dynamically connected i.e., the light rays can transition between the vicinity of one FPO to the vicinity of another FPO in a sensitive to the initial conditions manner.

\begin{figure}[h] 
\centering
    \includegraphics[width=0.2\textwidth]{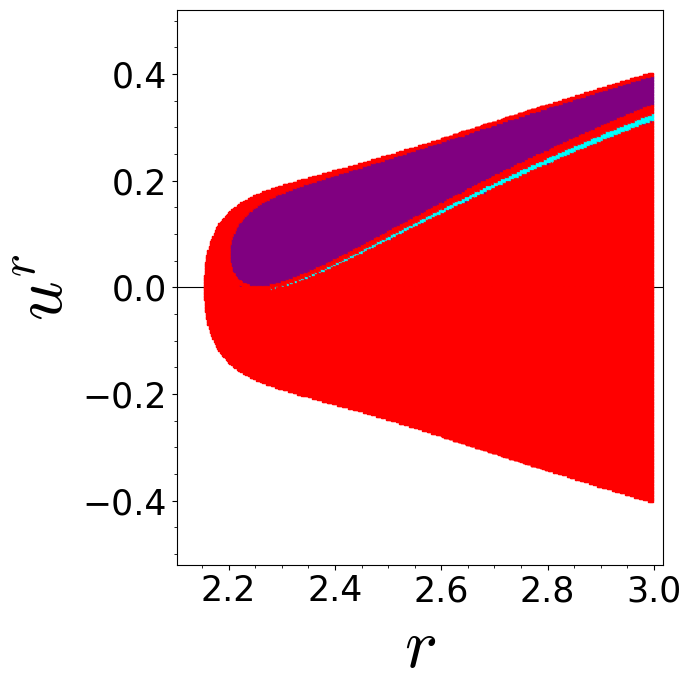} 
    \includegraphics[width=0.2\textwidth]{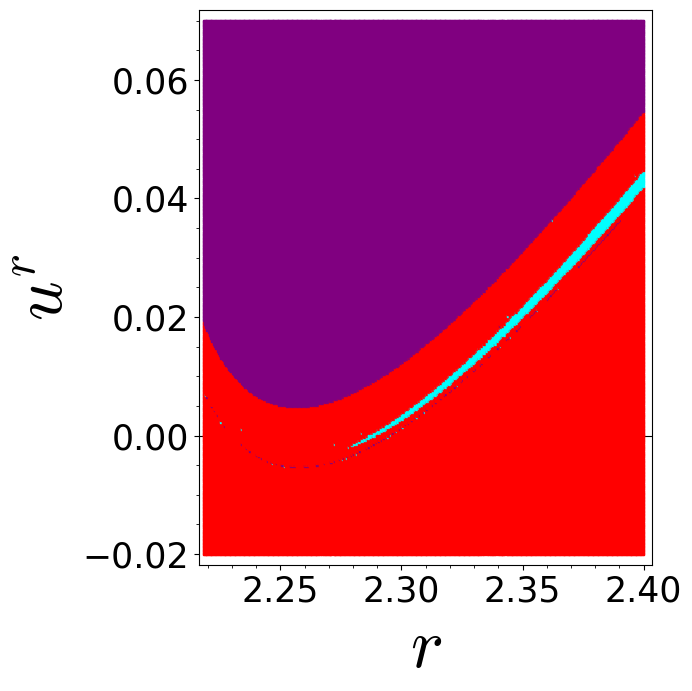}
    
    \includegraphics[width=0.2\textwidth]{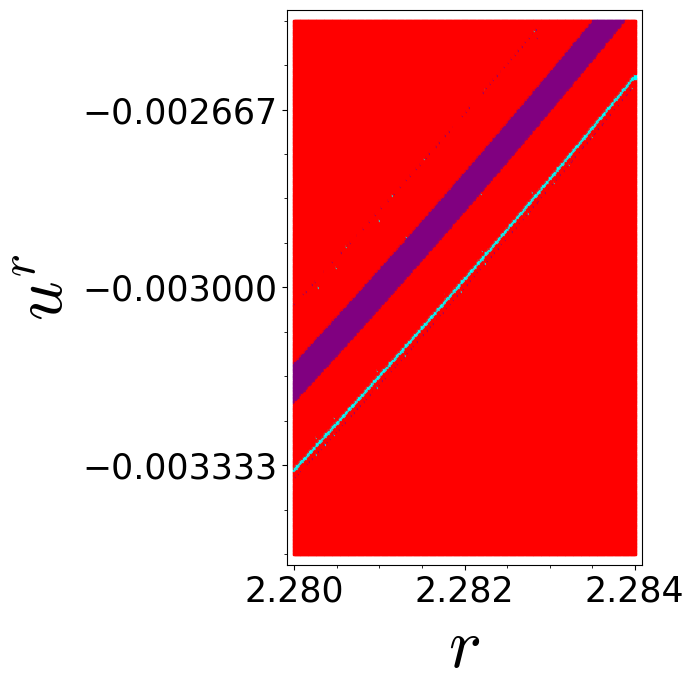} 
    \includegraphics[width=0.2\textwidth]{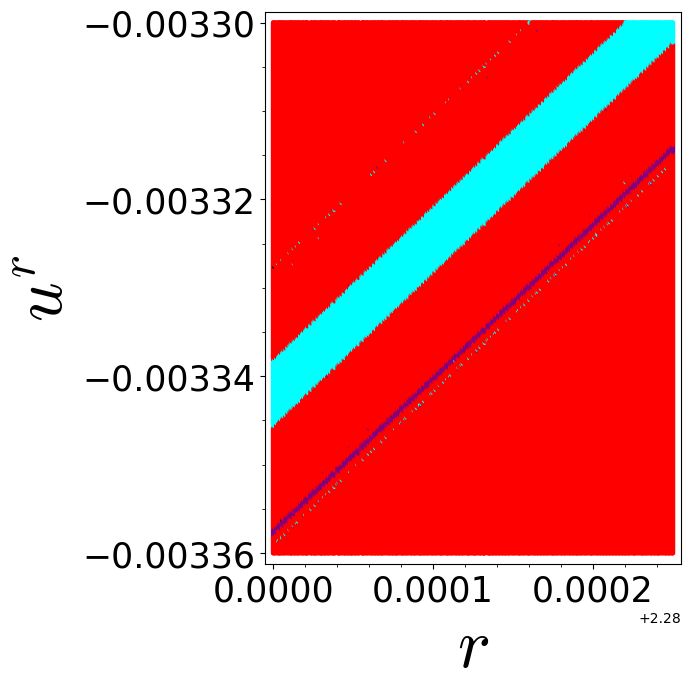}
  \caption{Exit basin diagrams for the HT spacetime discussed. 
  Purple (cyan) basins correspond to light rays that fall into the object through the upper (lower) escape and red basins to ones that escape to infinity. The diagram is self-similar with a fractal dimension $D_s=1.680514$.} 
  \label{fig:exbsHT}
\end{figure}
The exit basin diagrams for the HT spacetime with the aforementioned parameters, i.e., $\chi=0.35$, $\delta q=1$ and $b=4.08M$, are presented in \cref{fig:exbsHT}. We color code the initial conditions as red when the light rays escape to infinity, purple when they fall into the compact object through the upper throat and cyan through the
lower one. The exit basins are self-similar with formed eyebrows on top of eyebrows, each corresponding to a different kind of resonance. The light rays essentially get temporarily trapped in the vicinity of different combinations of FPOs or they resonate for larger time intervals with one of the FPOs in a particular combination. The boundaries of the basins are well defined and lack the highly fractalized structure found when a "pocket" feature is present \cite{Shipley:2019kfq,Kostaros_2022}. To quantify the \textit{degree of self-similarity} of the exit basin diagram, we use the box-counting method and calculate its fractal dimension $D_s$. We find the dimension to be $D_s=1.680514$ with a goodness of fit value $R^2=0.99811$ suggesting that as we zoom in, details will consistently emerge, keeping the complexity high. This behavior is a result of the observed behavior of photon orbits resonating with FPOs. As the parameters change the photon orbits shift between different combinations of resonances in a way sensitive to the initial conditions, thus changing exits which in turn shows in the exit basin diagram as a fractal structure. 

 We can expect the shadow to also have regions with self-similar structures for the same reason we observed self-similar structures in the exit basins. To calculate the shadow we start by setting up a grid of initial conditions $N\times M$ where $N$ and $M$ correspond to the number of different values for our two impact parameters $b$ and $\alpha$ respectively. We assume the observer to be on the equatorial plane $\theta_0=\pi/2$ and at a distance $r_0=100M$, and numerically integrate backwards the geodesic equations for every $\{b,\alpha\}$. If the light ray falls into the object we assign a dark pixel to it. If the light ray reaches our illuminating screen at infinity we assign a bright pixel to it with the following color code for its origin: green when $\sin\phi>0$, $\theta>\pi/2$; pink when $\sin\phi<0$, $\theta>\pi/2$; blue when $\sin\phi>0$, $\theta<\pi/2$; yellow when $\sin\phi<0$, $\theta<\pi/2$.

\begin{figure}[h]
\begin{center}
    \includegraphics[width=0.2\textwidth]{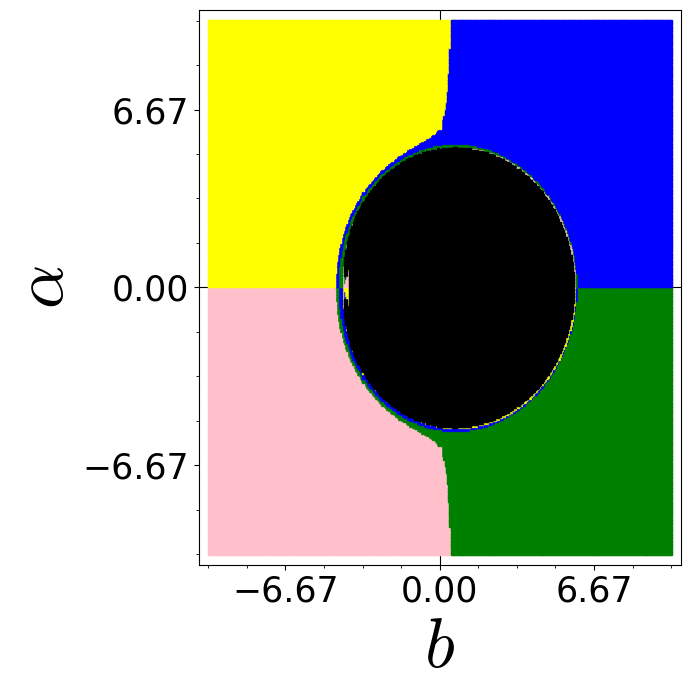}\\ 
    \includegraphics[width=0.2\textwidth]{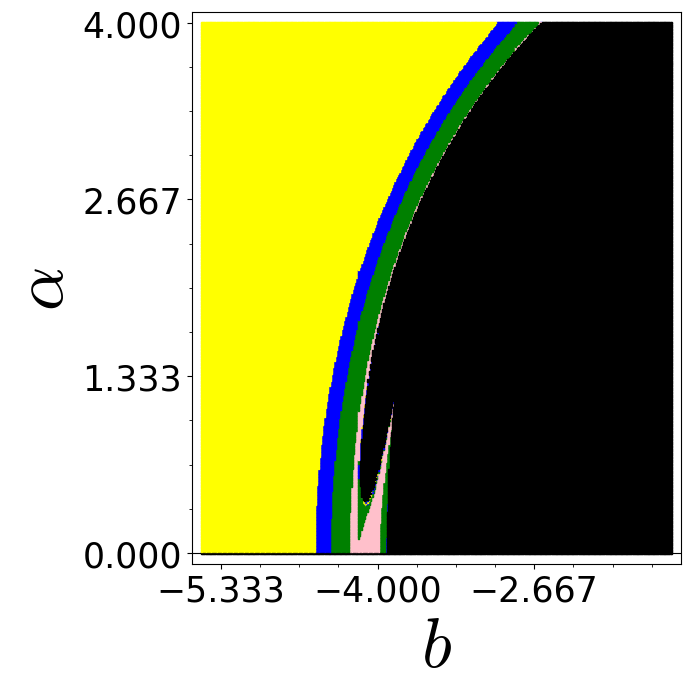} 
    \includegraphics[width=0.2\textwidth]{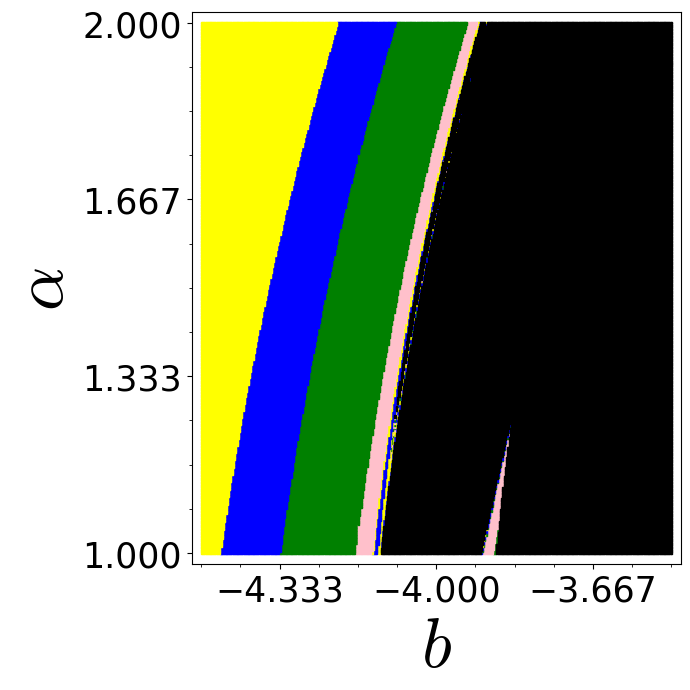} 
    \includegraphics[width=0.2\textwidth]{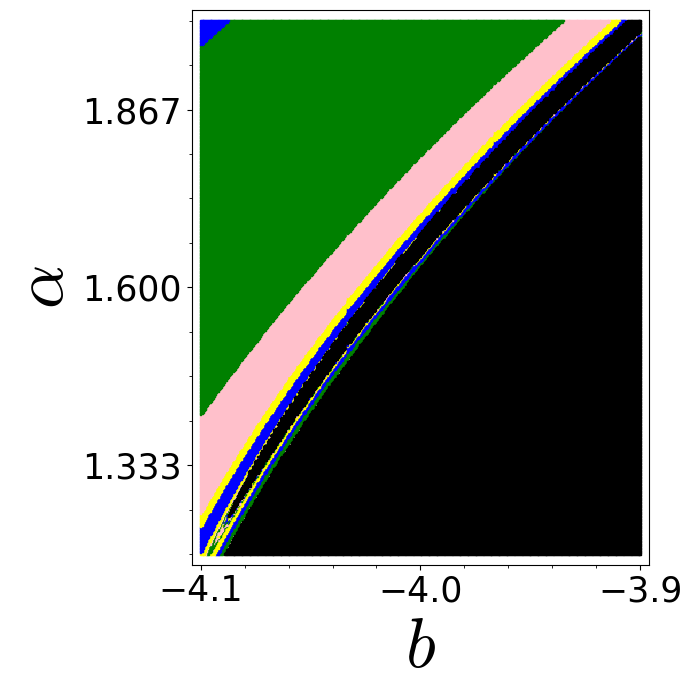} 
    \includegraphics[width=0.2\textwidth]{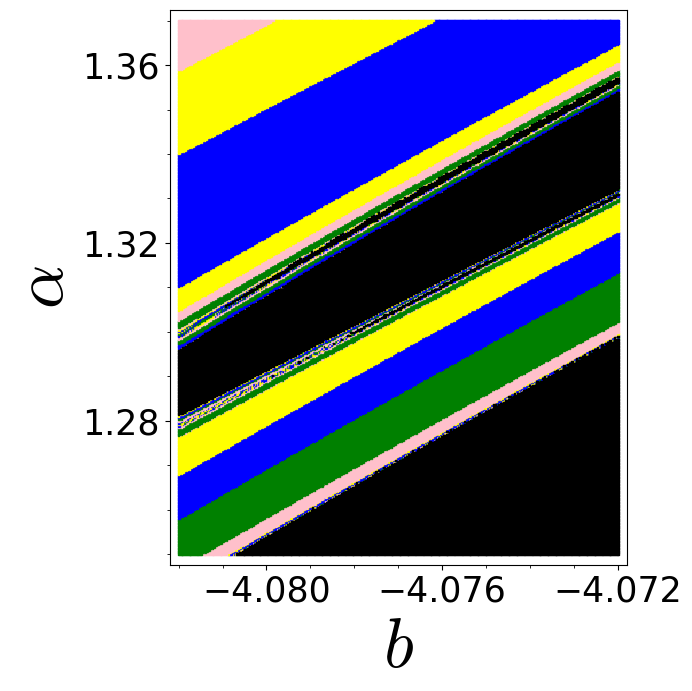}
  \caption{Shadow of the HT compact object discussed.
  The shadow has self-similar fractal eyebrows that correspond to different resonances of the light rays with the FPOs.} 
  \label{fig:HT_shadows_black}
\end{center}
\end{figure}
 
 The shadow is presented in \cref{fig:HT_shadows_black} with the  resolution of each image being $384\times384$. The overall circular shadow exhibits ``eyebrow'' features, i.e., arc-shaped shadows that their formation can be understood if we consider the shape of the separatrix (\cref{fig:htfpos}) and the behavior of the exit basins (\cref{fig:exbsHT}). 
 
\begin{figure}[h]
    \center
    \includegraphics[width=0.23\textwidth]{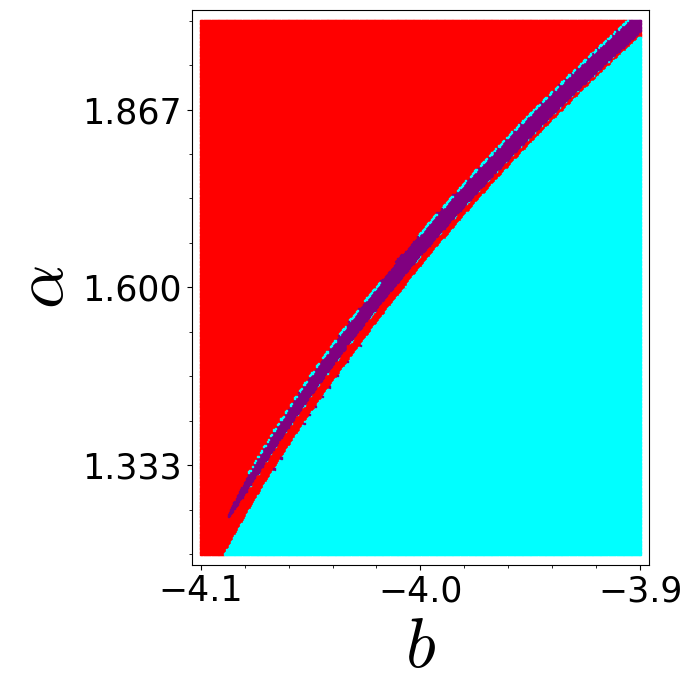} 
    \includegraphics[width=0.23\textwidth]{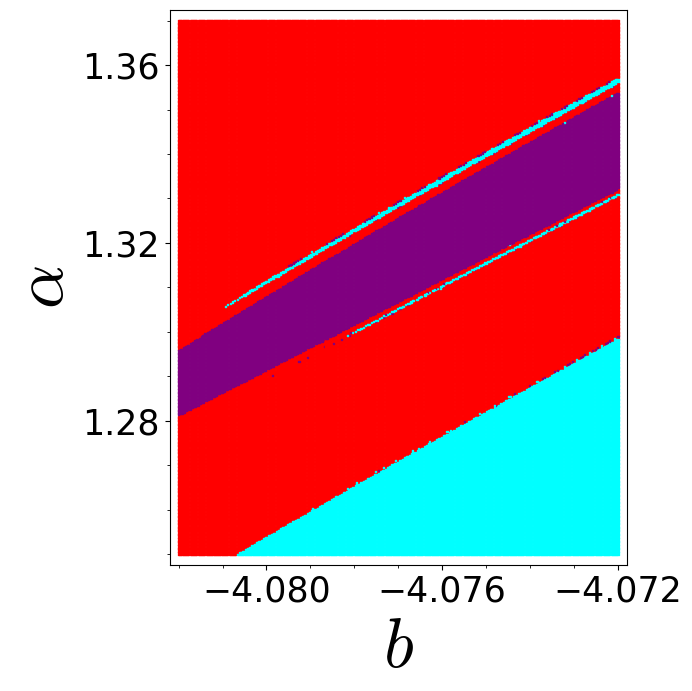}
  \caption{Shadow of the HT compact object discussed. 
  The light rays are color-coded as purple(cyan) if they fall into the surface of the object through the upper(lower) exit and red if they reach infinity.} 
  \label{fig:HT_shadows_clrd}
\end{figure}

 To make this clearer, we can use the same color code as we did for the exit basin diagram to paint the shadow, where purple (cyan) pixels correspond to light rays that reach the objects surface through the upper (lower) throat and red pixels to ones that escape to infinity. This is shown in \cref{fig:HT_shadows_clrd}. One can see that the "eyebrows" are separated by red regions of photons that escape to infinity, while the "eyebrows" themselves are of consecutive exit color, as the photons first resonate with one combination of FPOs that eventually leads to one of the escapes and then resonate to a different combination of FPOs that leads to a different escape and so on cycling between the escapes.   

 We can now find the fractal dimension of the shadow by using the box-counting method. We work on the sets of \cref{fig:HT_shadows_black} but abandon the background color map so we consider all light as white. For every grid size we count the number of boxes that contain neighboring black and white pixels i.e. the boundary of the shadow. We find the dimension to be $D_s=1.78308$ with a goodness of fit $R^2=0.99974$ suggesting that the boundary of the shadow is also self-similar across all scales \cite{Mandelbrotbritain,peitgen2004chaos}.

%%%%------------------------
\subsection{JP}
%%%%------------------------

 Similarly to the HT metric, for positive values of the deformation parameter $\epsilon_3>0$, i.e., prolate deformation of the compact object, and above a certain spin threshold $a_*$, the Kerr-like prograde LR of the JP spacetime \textit{bifurcates} into a pair of non-equatorial and symmetric LRs ($O_0^{0+}\longrightarrow O^{0_{-}}_0$). For $\epsilon_3$ in the range $0.1\lesssim \epsilon_3\lesssim 10$, the spin parameter for which the bifurcation takes place lies in the range $0.95\gtrsim a/M\gtrsim 0.4$ and is a function of the deformation parameter, $a_*=a_*(\epsilon_3)$ \cite{Glampedakis:2018blj}. 

\begin{figure}[h]
\begin{center}
\includegraphics[width=0.2\textwidth]{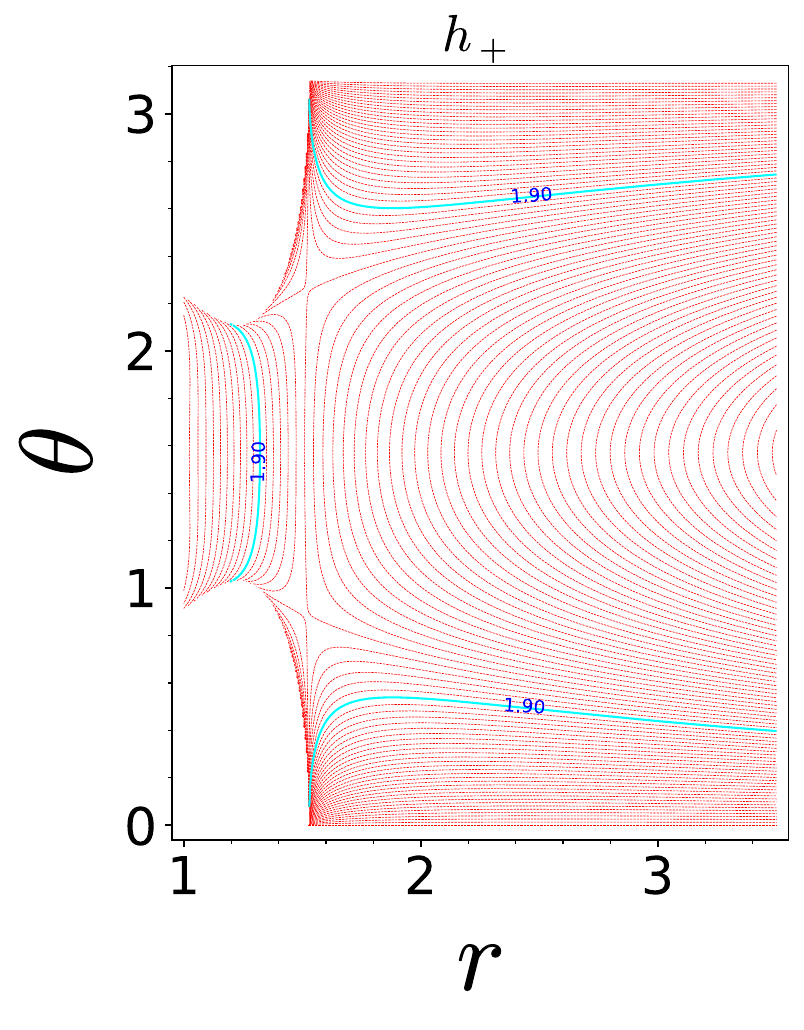}
\includegraphics[width=0.2\textwidth]{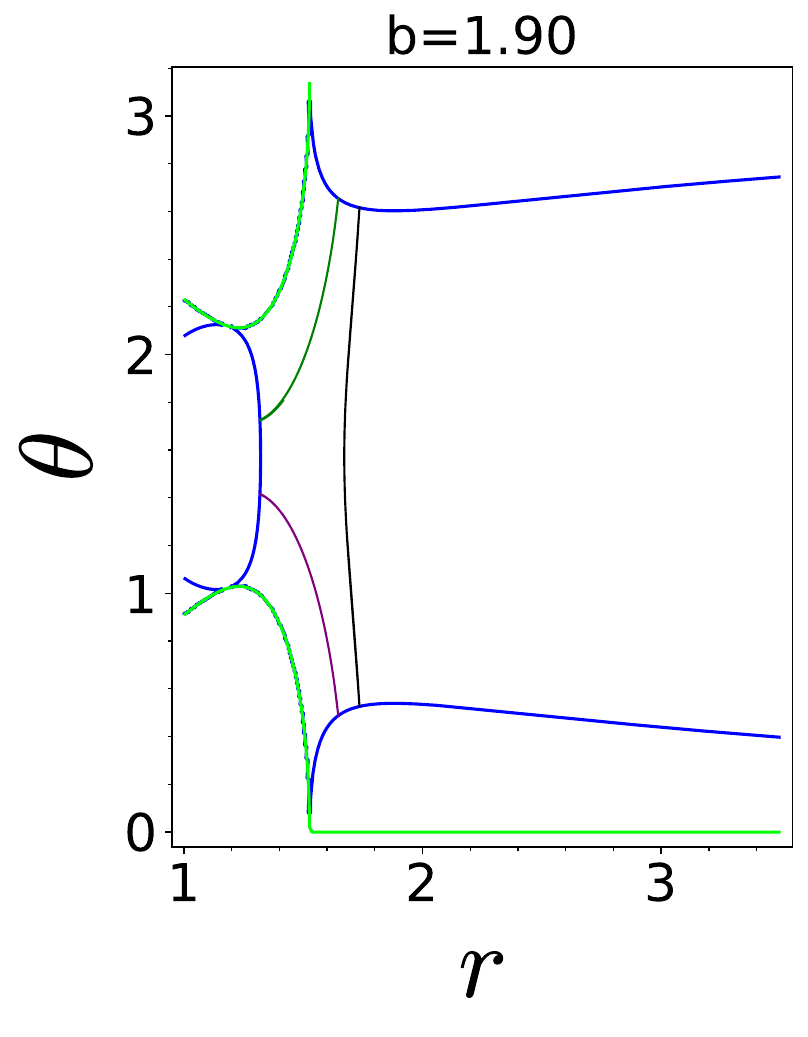}
\end{center}
\caption{Left: The potential function $h_+$ for the JP spacetime with parameters $a=0.85M$, $\epsilon_3=1$ and $M=1$. Right: The separatrix for an impact parameter $b=1.90M$ and the FPOs admitted. The lime line marks the horizon.}
\label{fig:jpfpos}
\end{figure}

\begin{figure}[H] 
\begin{center}
    \includegraphics[width=0.15\textwidth]{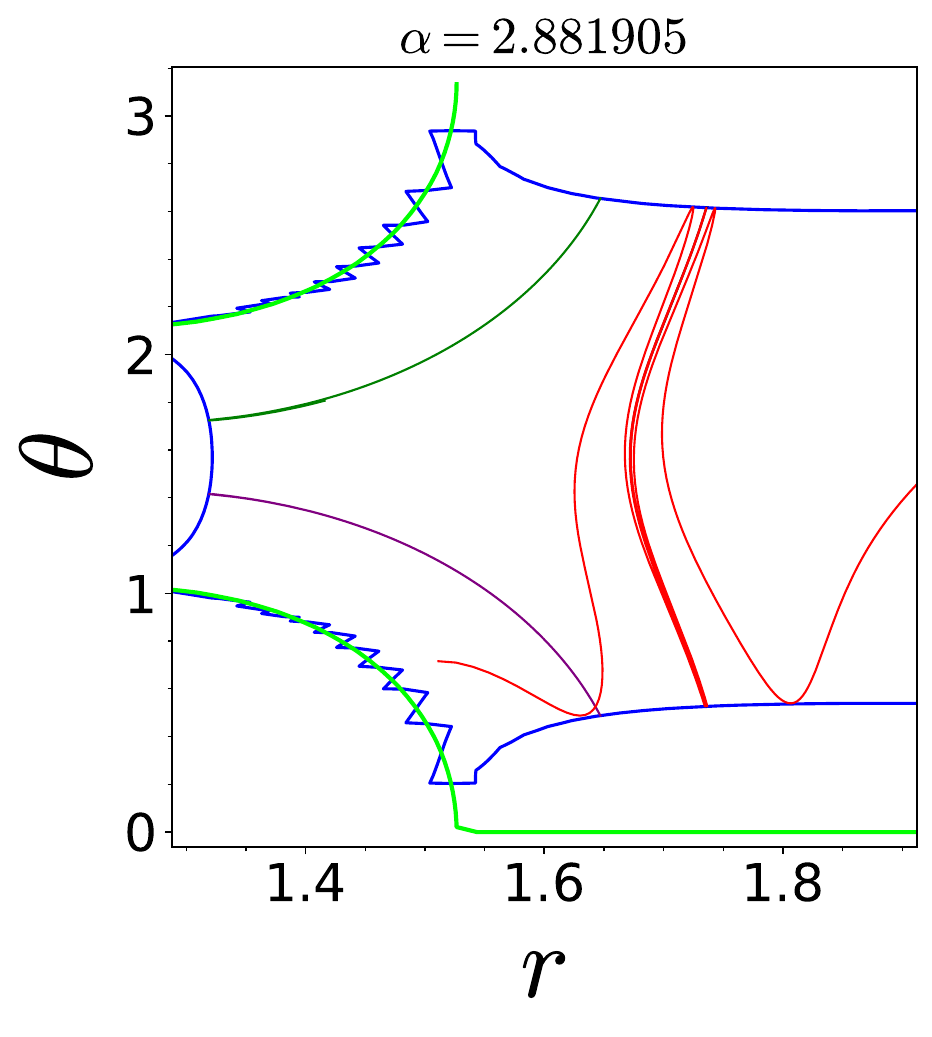} 
    \includegraphics[width=0.15\textwidth]{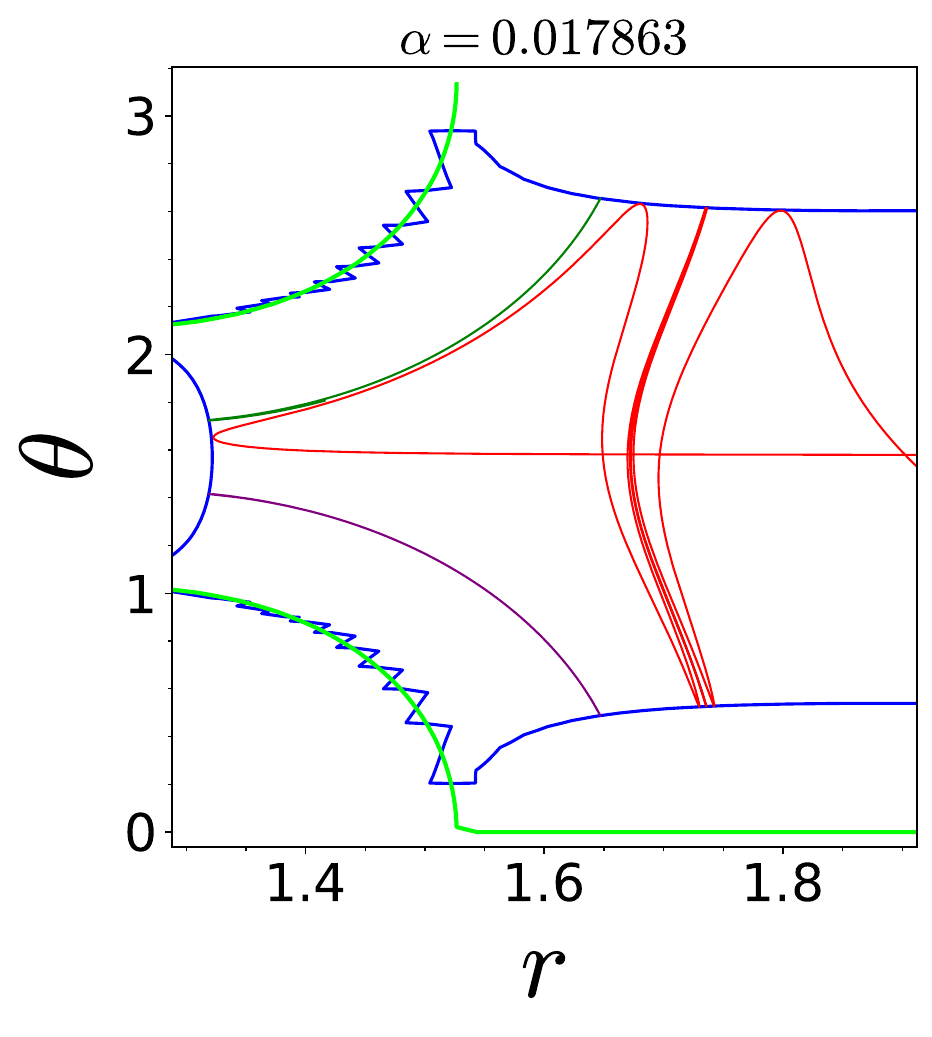}  
    
    \includegraphics[width=0.15\textwidth]{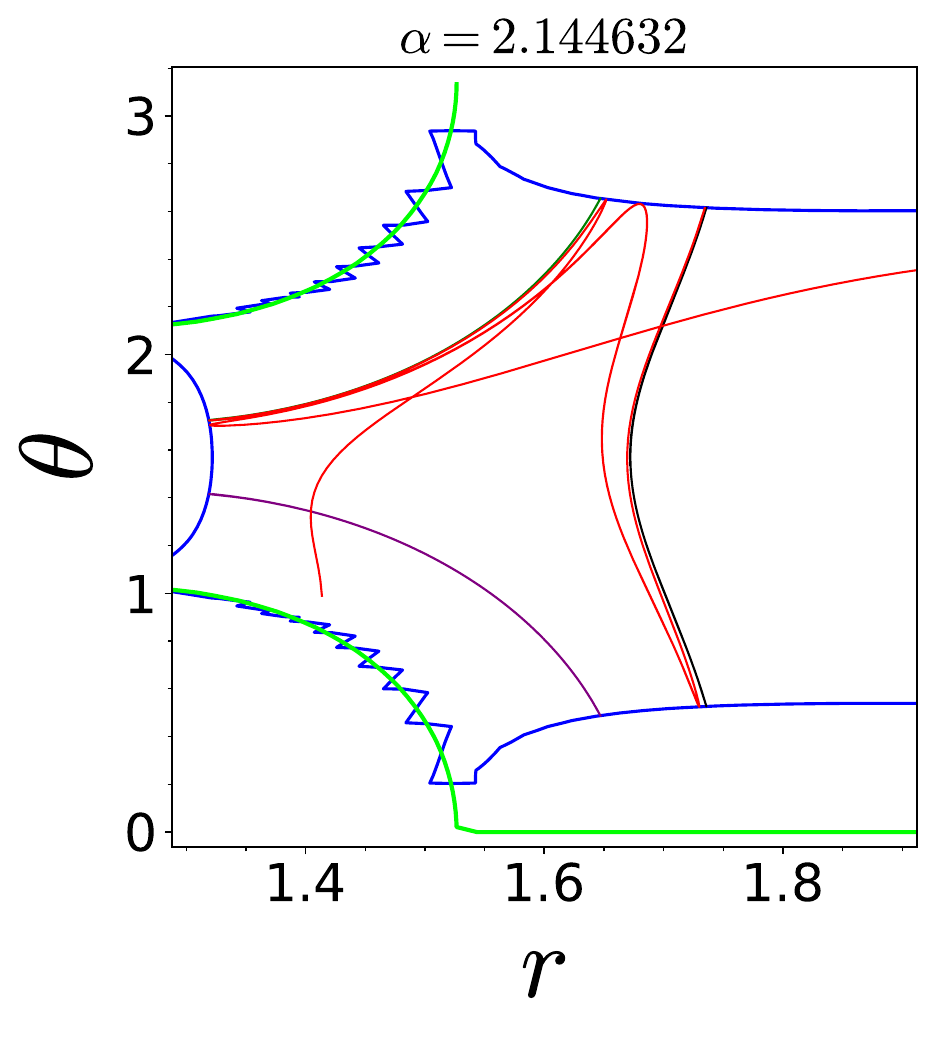} 
    \includegraphics[width=0.15\textwidth]{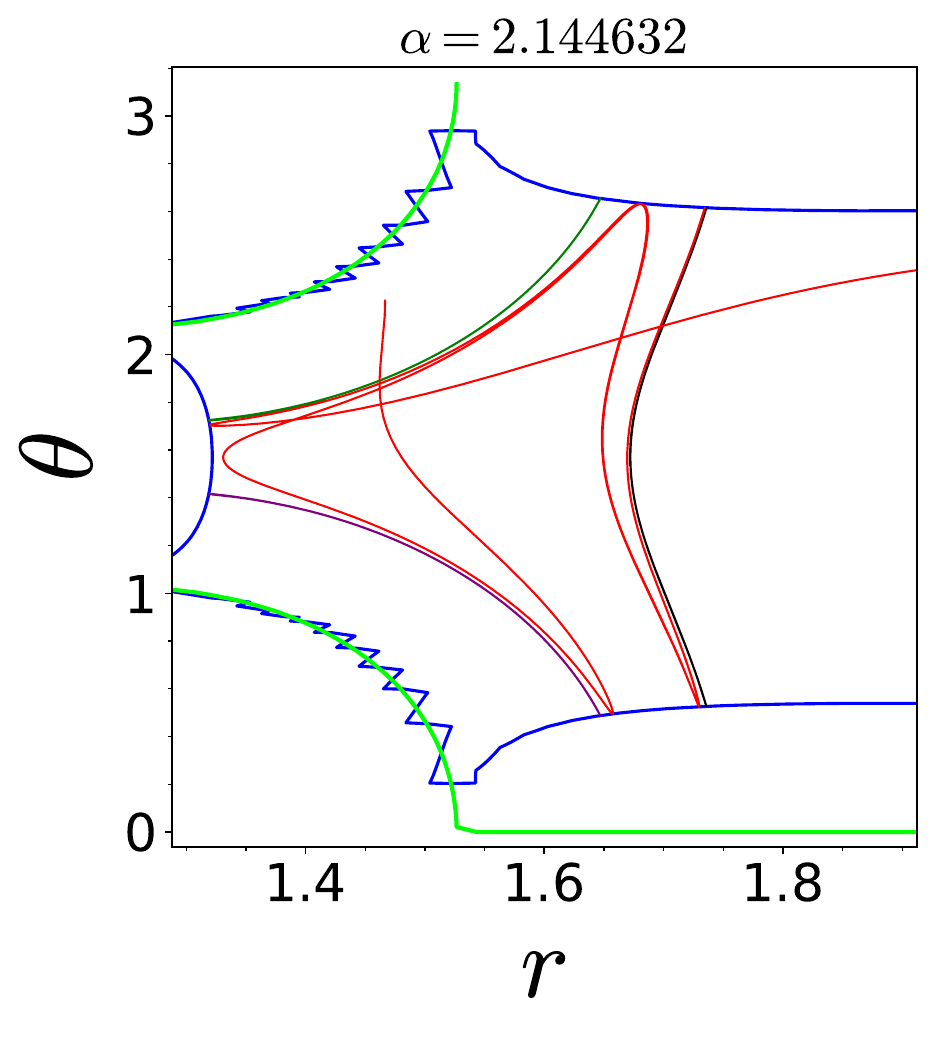} 
  \caption{Top: Two photons launched with impact parameters $\alpha\approx 2.8819M$ and $\alpha\approx 0.01786M$ respectively. They both resonate with the $O^{1_{+}}_0$ FPO. Bottom: Two photons launched with impact parameters $\alpha\approx 2.144632M$ but with a difference in their 7th decimal. The first one resonates with two FPOs while the second one with all three.} 
  \label{fig:diffcombsJP}
\end{center}
\end{figure}

 The key difference with the HT spacetime is that in the JP case, there is no simultaneous presence of all three LRs, therefore the formation of a pocket is not possible for JP. The potential function $h_+$ for the JP spacetime, with parameters $M=1$ unit of length, $a=0.85M$ and $\epsilon_3=1$, and the separatrix for an impact parameter $b=1.90M$ along with the FPOs are shown in \cref{fig:jpfpos}.
 The two unstable LRs appear as saddle points which in this case are in higher latitudes compared to the HT case.

\begin{figure}[H] 
\begin{center}
    \includegraphics[width=0.18\textwidth]{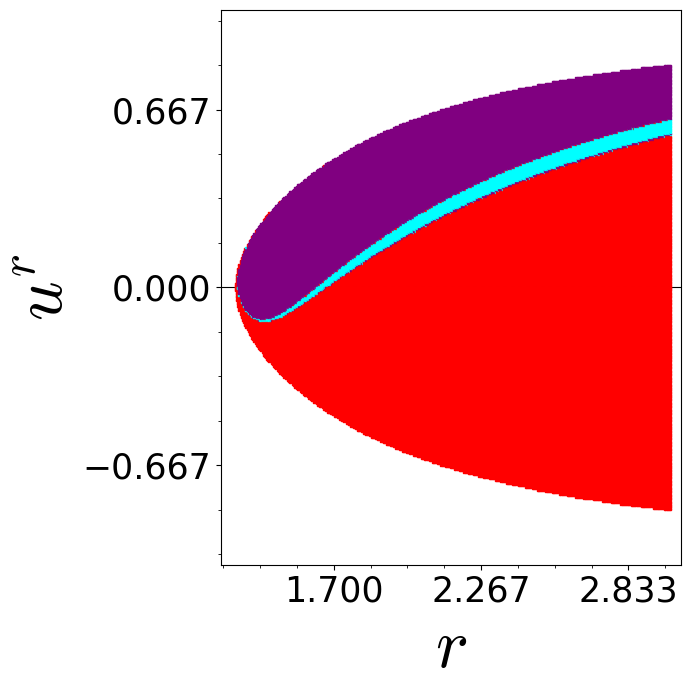} 
    \includegraphics[width=0.18\textwidth]{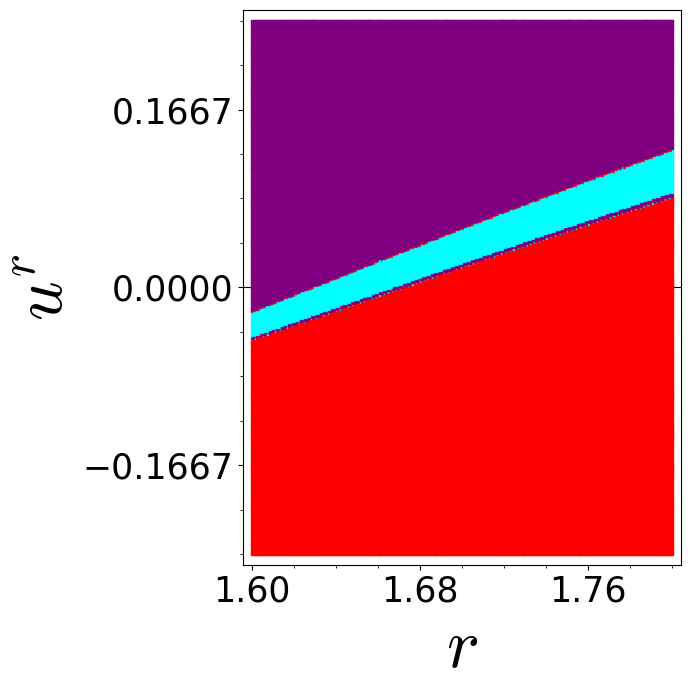} 
    
    \includegraphics[width=0.18\textwidth]{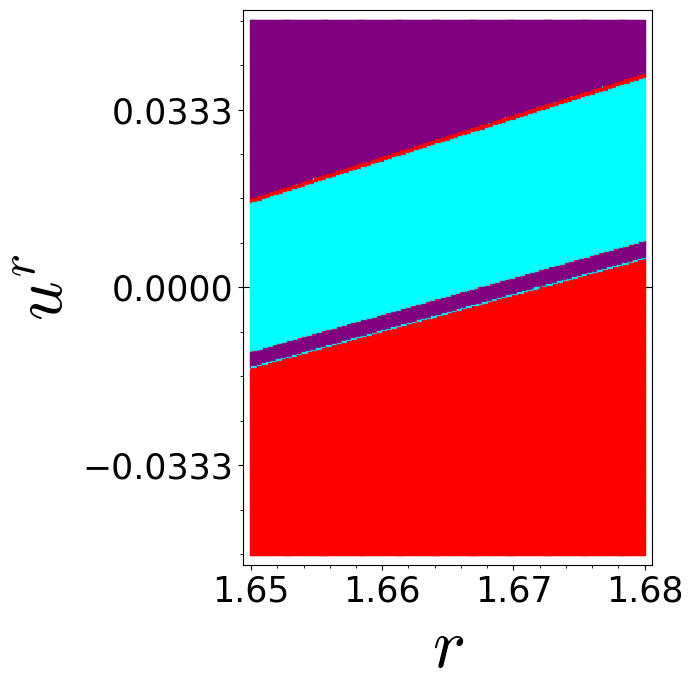} 
    \includegraphics[width=0.18\textwidth]{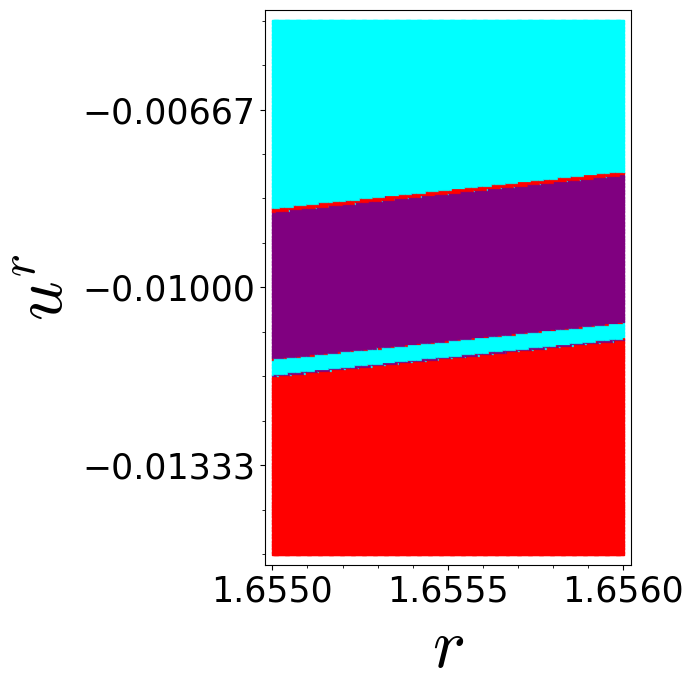} 
  \caption{Exit basin diagrams for the JP spacetime discussed. 
  The diagram is self-similar with a fractal dimension $D_s=1.80807$.} 
  \label{fig:exbsJP}
\end{center}
\end{figure}

 The various light rays that resonate with the dynamically connected FPOs are shown in \cref{fig:diffcombsJP}. The exit basin diagrams for the JP spacetime with the same parameters are presented in \cref{fig:exbsJP}. Similarly to the HT spacetime (\cref{fig:exbsHT}) the basins are self-similar with a fractal dimension $D_s=1.80807$ and a goodness of fit $R^2= 0.99937$. 

\begin{figure}[H]
\begin{center}
    \includegraphics[width=0.2\textwidth]{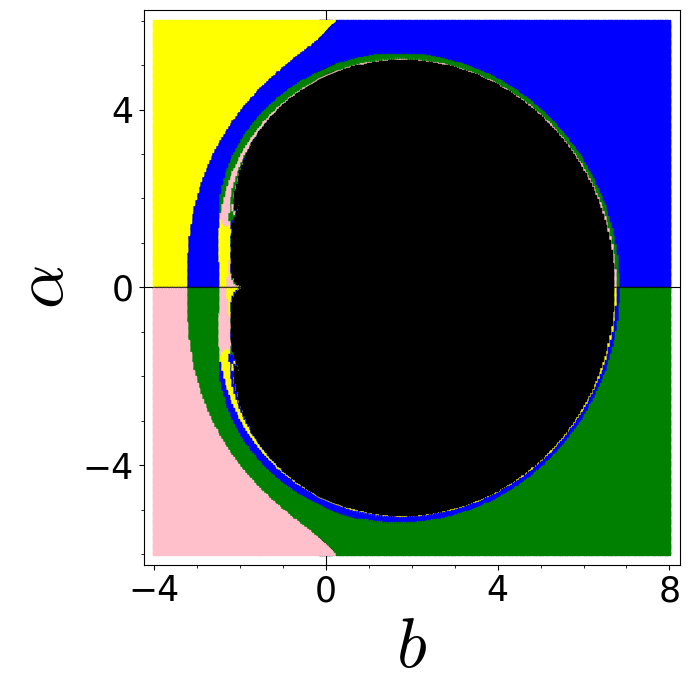} \\
    \includegraphics[width=0.2\textwidth]{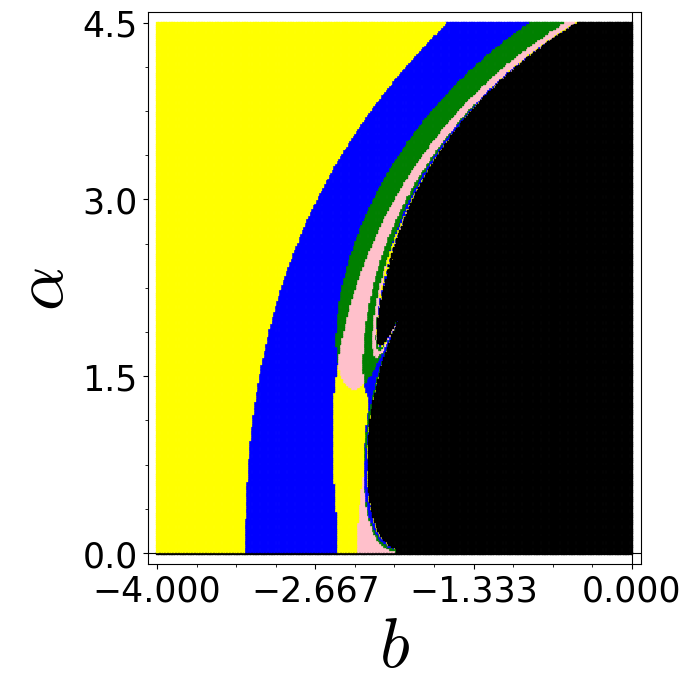} 
    \includegraphics[width=0.2\textwidth]{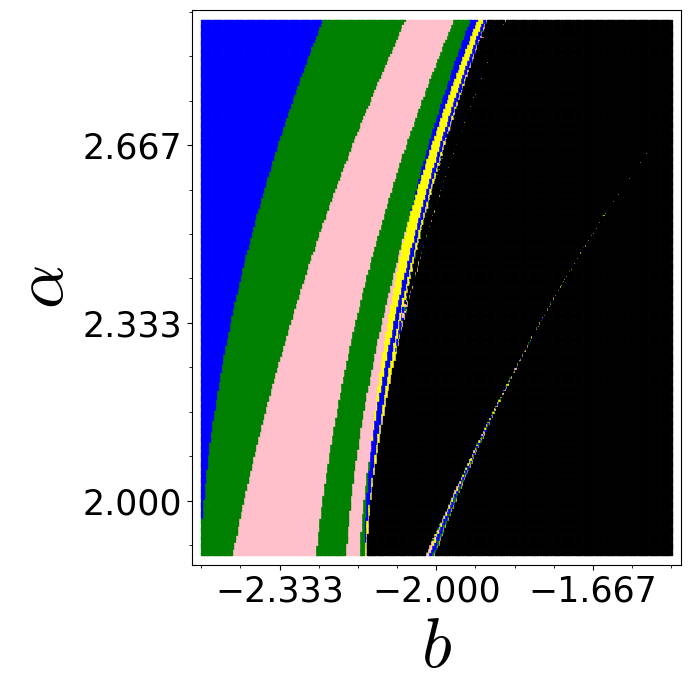}\\
    \includegraphics[width=0.2\textwidth]{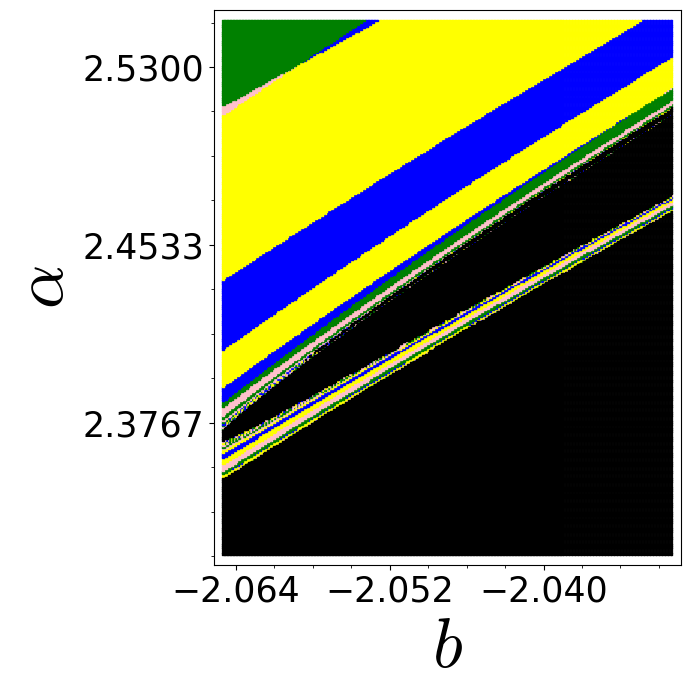} 
    \includegraphics[width=0.2\textwidth]{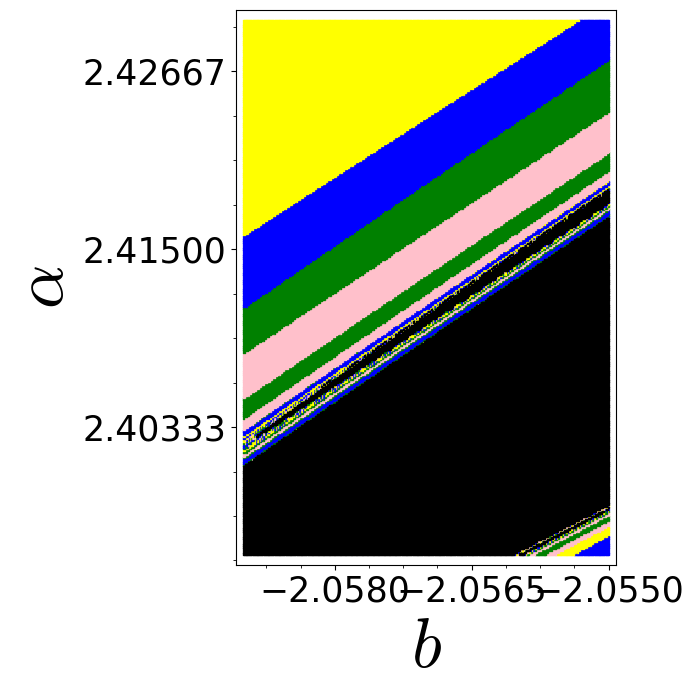}
    \caption{Shadow of the JP compact object discussed. 
    The shadow has self-similar fractal eyebrows that correspond to different resonances of the light rays with the FPOs.}
    \label{fig:JP_shadows_black}
\end{center}
\end{figure}

The shadow of the JP compact object and for an observer located at $\theta_0=\pi/2$ is presented in \cref{fig:JP_shadows_black}. Self-similar fractal eyebrows are once again formed due to the type of resonances we discussed in the HT case as well. Using the box counting method on \cref{fig:JP_shadows_black}, we find the fractal dimension of the shadow's boundary to be $D_s=1.76592$ with a goodness of fit $R^2=0.99955$. The JP shadow also maintains its self-similar structure on all scales. 

\begin{figure}[H]
\begin{center} 
    \includegraphics[width=0.23\textwidth]{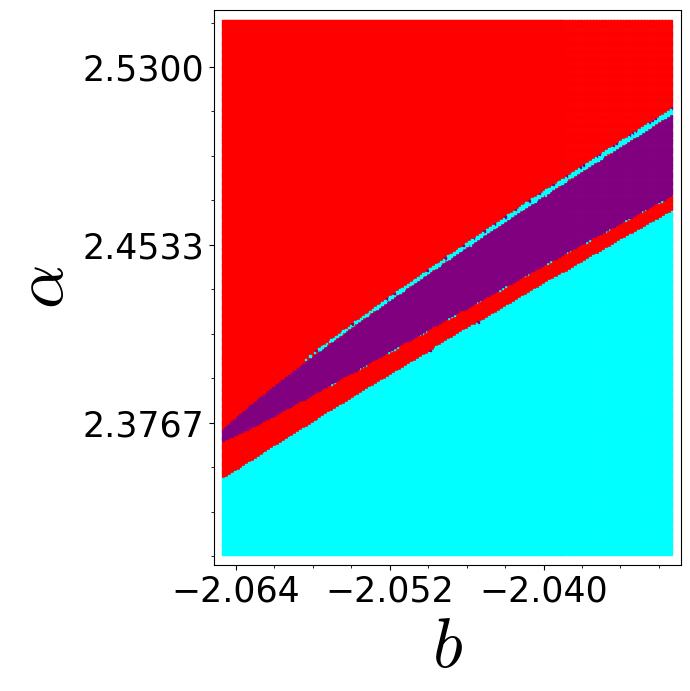} 
    \includegraphics[width=0.23\textwidth]{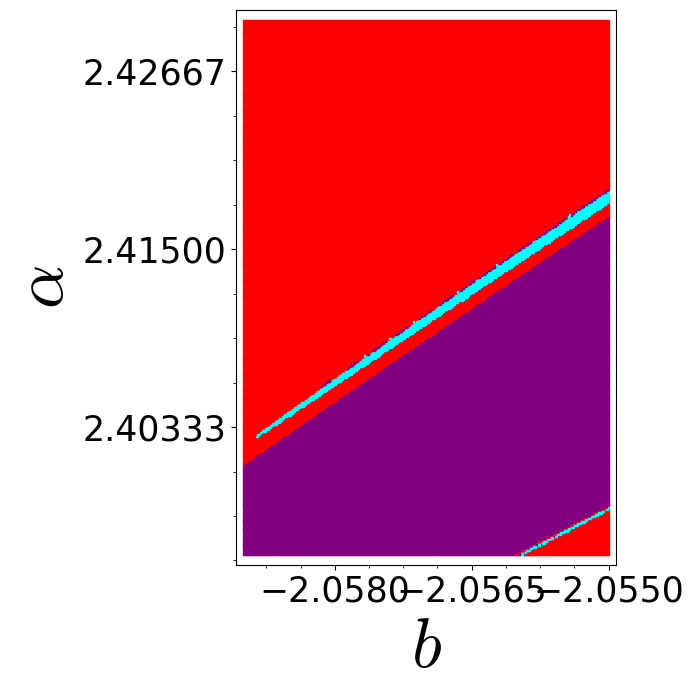}
    \caption{Shadow of the JP compact object, 
    The light rays are color-coded as purple(cyan) if they fall into the surface of the object through the upper(lower) exit and red if they reach infinity.}
    \label{fig:JP_shadows_clrd}
\end{center}
\end{figure}

Lastly, in \cref{fig:JP_shadows_clrd} we color code the parts of the shadow corresponding to the various exits and demonstrate how the fractal eyebrow forms by the resonant orbits shifting from one exit towards the central object to the other exit towards the central object, and then to the exit towards infinity, and repeating the cycle, as in the HT case.

%%%%------------------------
\subsection{Majumdar-Papapetrou di-hole}
%%%%------------------------

\begin{figure}[h]
\begin{center}
\includegraphics[width=0.2\textwidth]{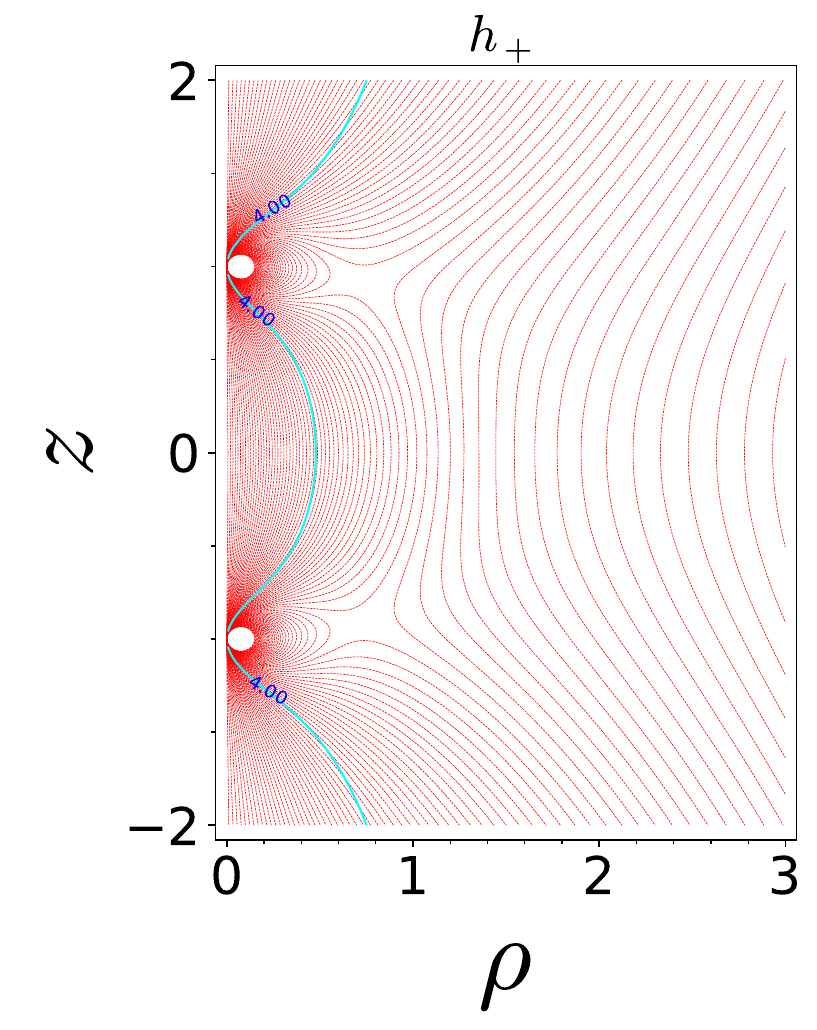}
\includegraphics[width=0.2\textwidth]{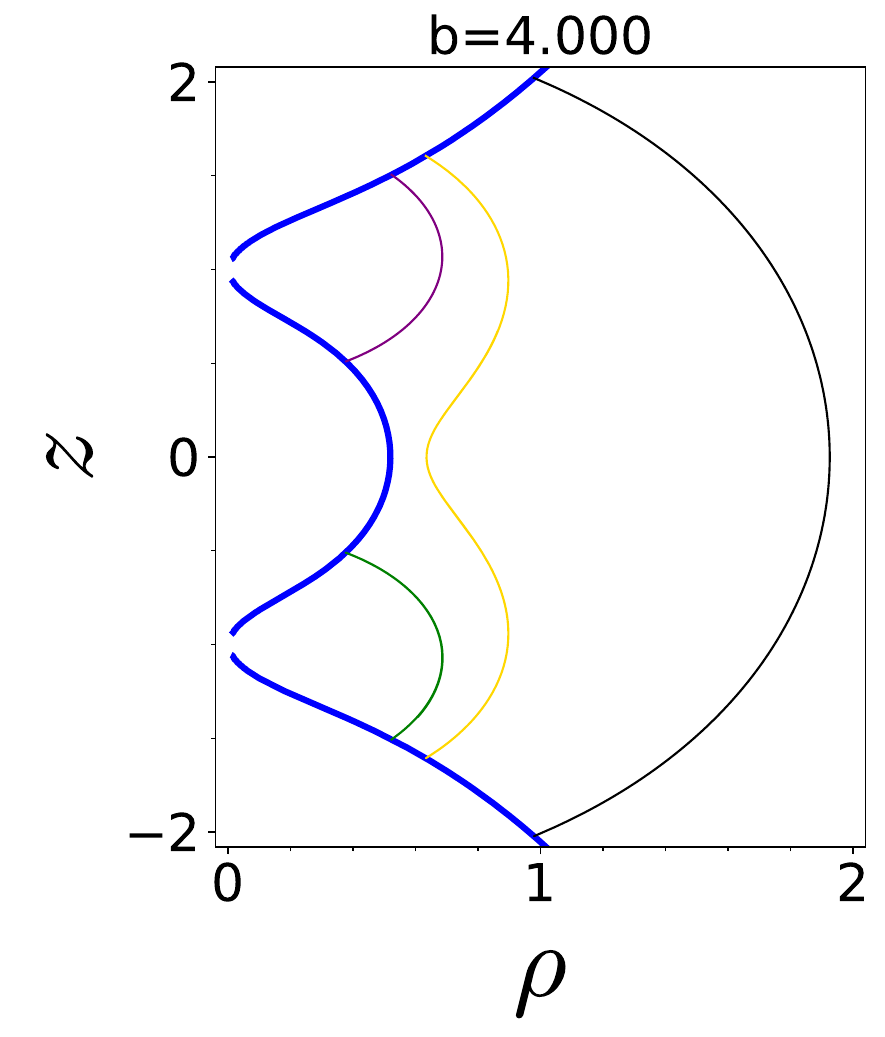}
\end{center}
\caption{Left: The potential function $h_+$ for the MP spacetime with parameters $M=1$. Right: The separatrix for an impact parameter $b=4M$ and the FPOs admitted. The two BHs are located at $z_+=M$ and $z_-=-M$.}
\label{fig:mpfpos}
\end{figure}

\begin{figure}[h] 
\begin{center}
    \includegraphics[width=0.15\textwidth]{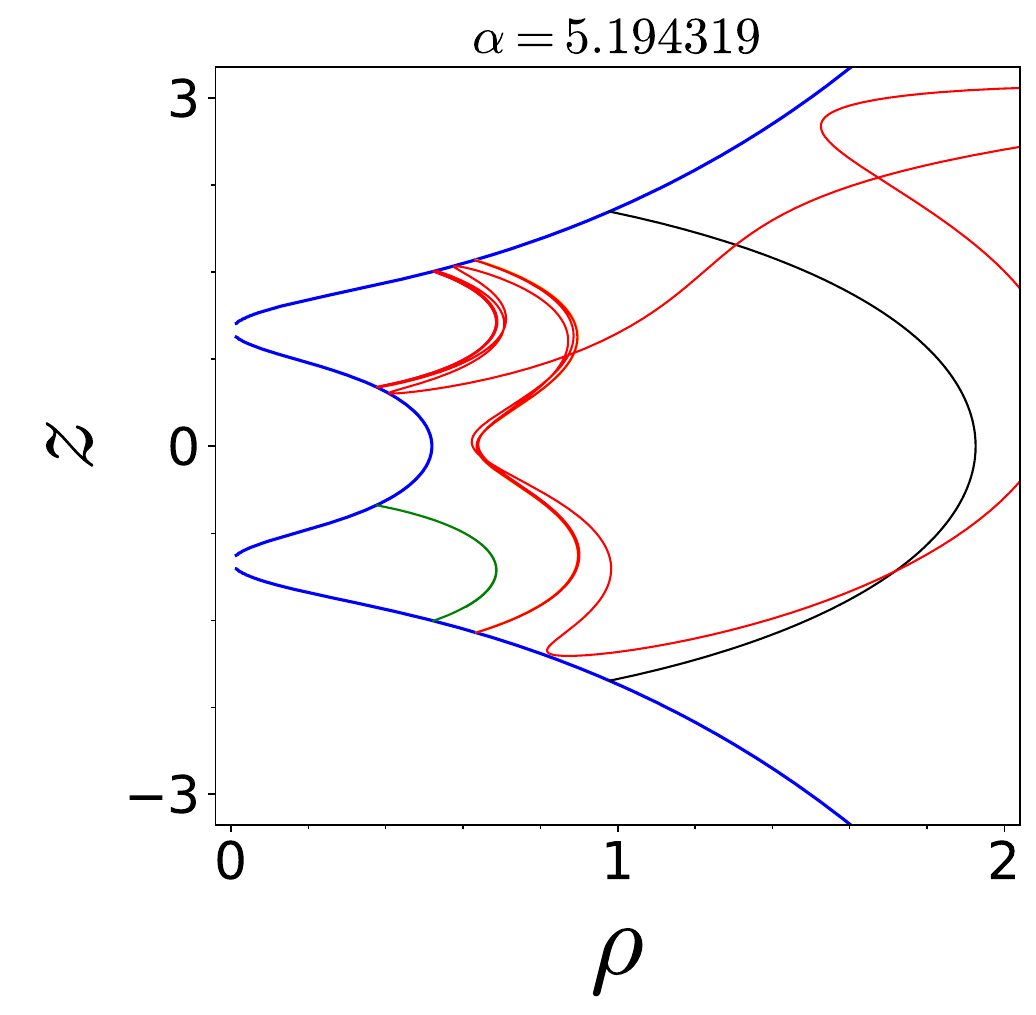} 
    \includegraphics[width=0.15\textwidth]{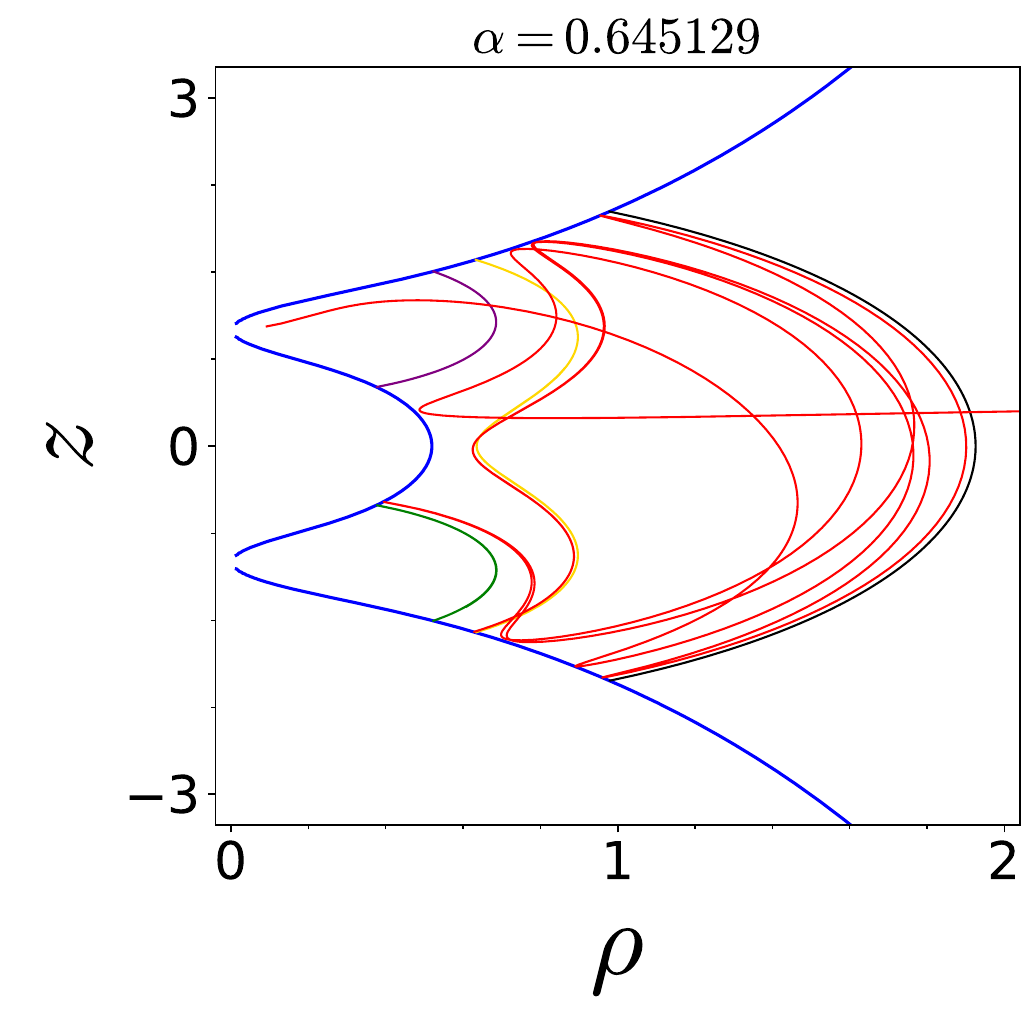}
    
    \includegraphics[width=0.15\textwidth]{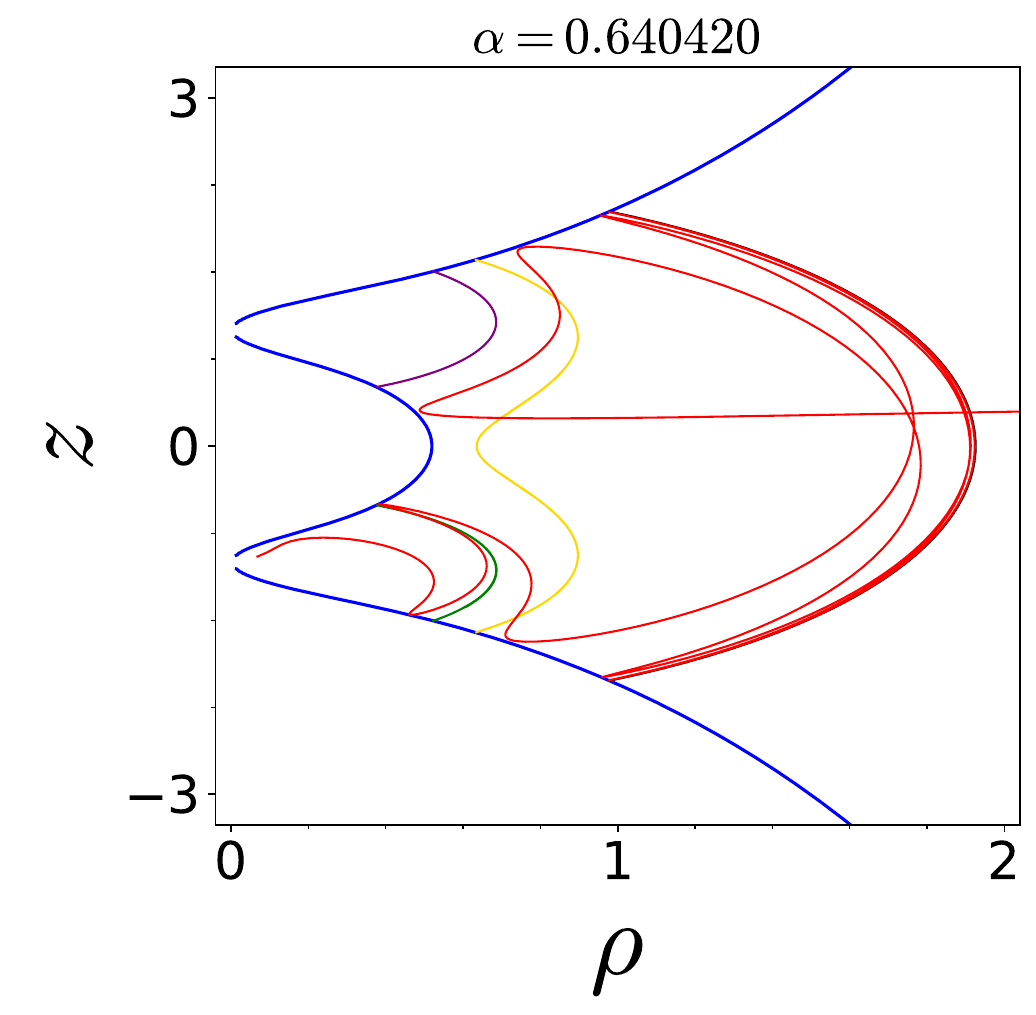} 
    \includegraphics[width=0.15\textwidth]{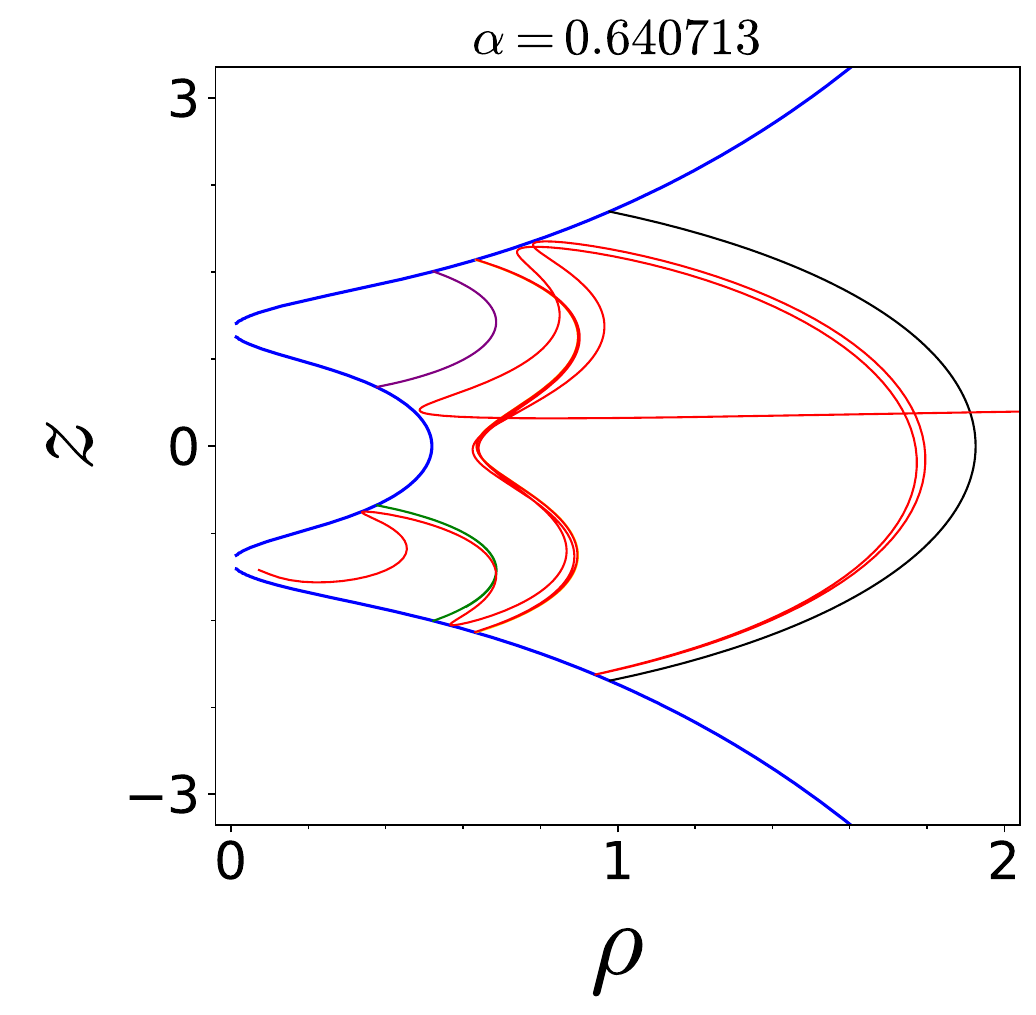} 
  \caption{Top: Two photons launched with impact parameters $\alpha\approx 5.194319M$ and $\alpha\approx 0.645129M$ respectively. The first one resonates with a $O^{0_{-}}_0$ and a $O^{1_{+}}_0$ FPO, while the second with all three FPOs. Bottom: Two photons launched with impact parameters $\alpha\approx 0.640M$ but with a difference in their 4th decimal. The first one resonates with two FPOs while the second one with three.} 
  \label{fig:diffcombsmp}
\end{center}
\end{figure}

The MP di-hole is another example of a spacetime describing a prolate configuration. The difference with the previous two cases is that this spacetime is not rotating, therefore it is symmetric with respect to the axis of symmetry of the di-hole. The MP spacetime admits a pocket feature for a specific range for the coordinate separation parameter $d$.  This pocket, just as in the HT spacetime, acts as a \textit{randomizing} region that leads to rich phenomenology for null geodesics \cite{Shipley:2019kfq}. We focus on the case where the system is an open Hamiltonian one, that is for a separation parameter $d=2M$ (\cref{fig:mpfpos}), where we do not have the formation of a pocket. We set the two masses at $M=1$ units of length and the two BHs are located at $z_+=M$ and $z_-=-M$ respectively. The potential function $h_+$ and the separatrix for an impact parameter $b=4M$ along with the FPOs are shown in \cref{fig:mpfpos}. Light rays that are launched with different $\alpha$-impact parameters and resonate with different combinations of the FPOs are shown in \cref{fig:diffcombsmp}. 

\begin{figure}[h] 
\begin{center}
    \includegraphics[width=0.18\textwidth]{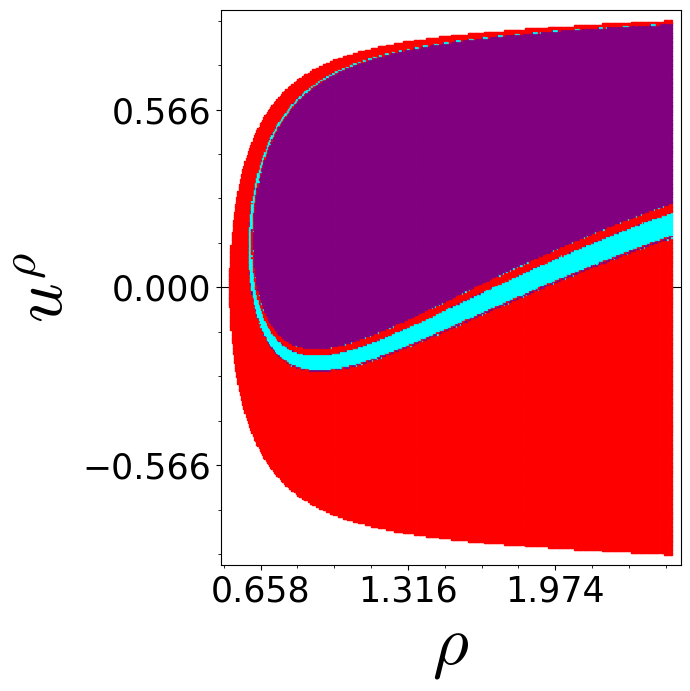} 
    \includegraphics[width=0.18\textwidth]{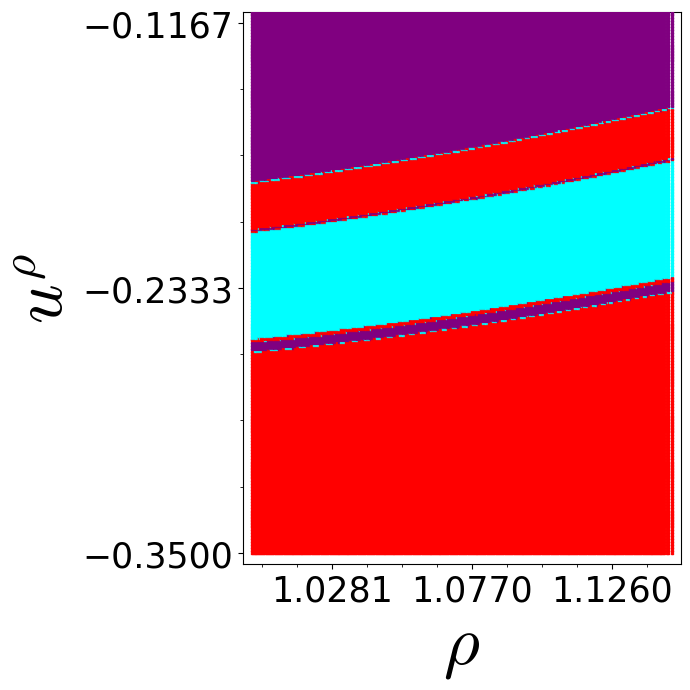} 
    
    \includegraphics[width=0.18\textwidth]{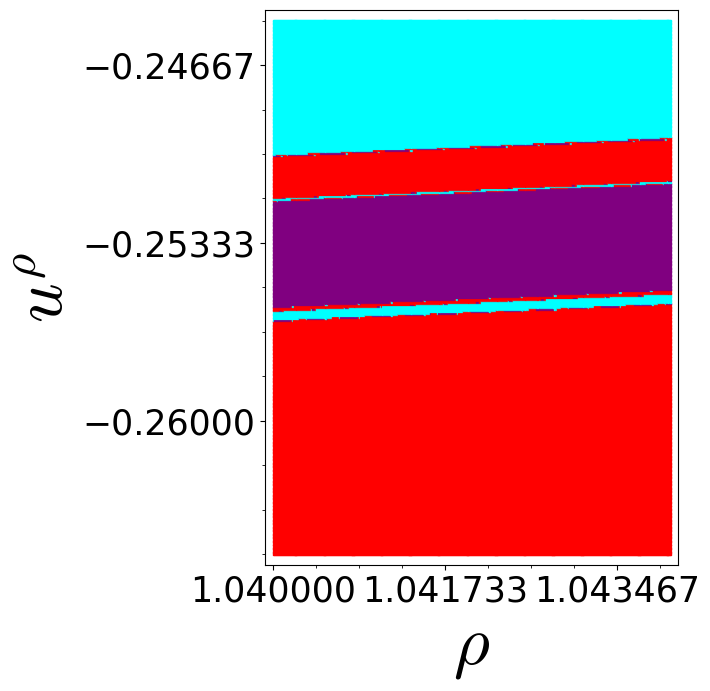} 
    \includegraphics[width=0.18\textwidth]{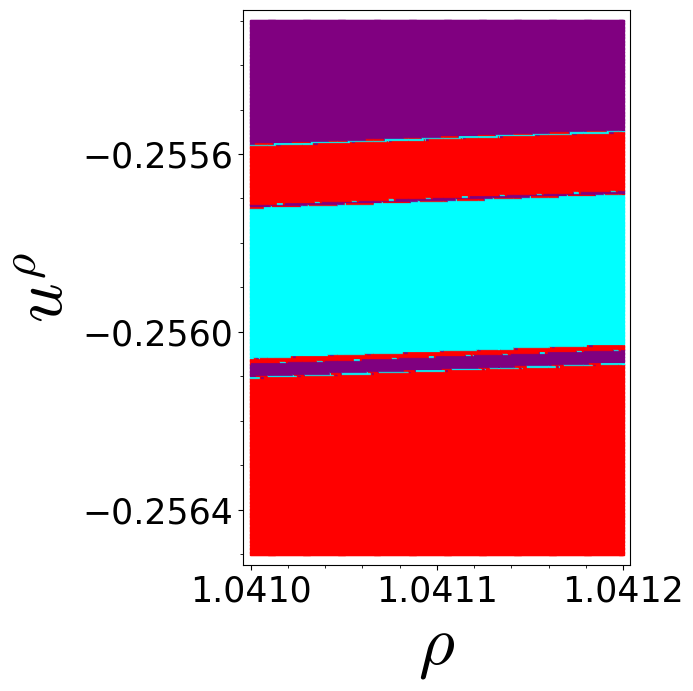} 
  \caption{Exit basin diagrams for the MP spacetime discussed. 
  Purple (cyan) basins correspond to light rays that fall into the object through the upper (lower) escape and red basins to ones that escape to infinity. The diagram is self-similar with a fractal dimension $D_s=1.81397$.} 
  \label{fig:exbsMP}
\end{center}
\end{figure}

\begin{figure}[H]
\begin{center}
    \includegraphics[width=0.2\textwidth]{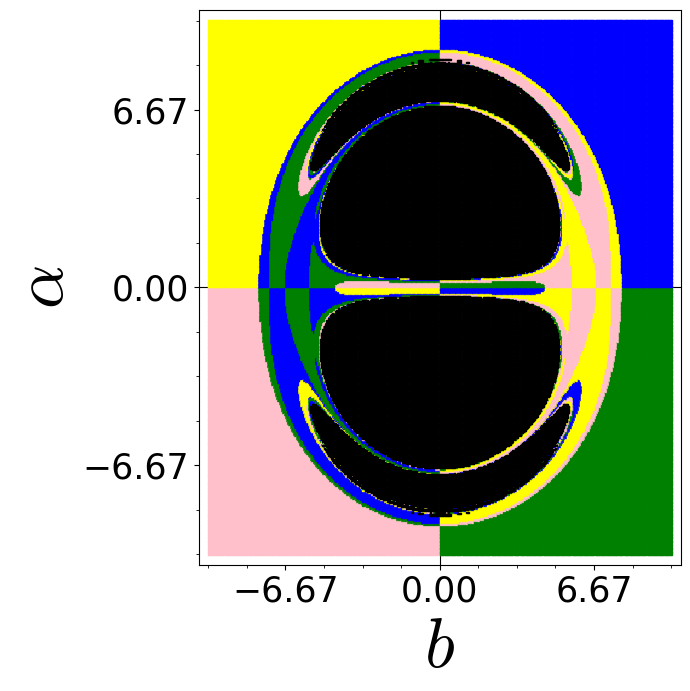} \\
    \includegraphics[width=0.2\textwidth]{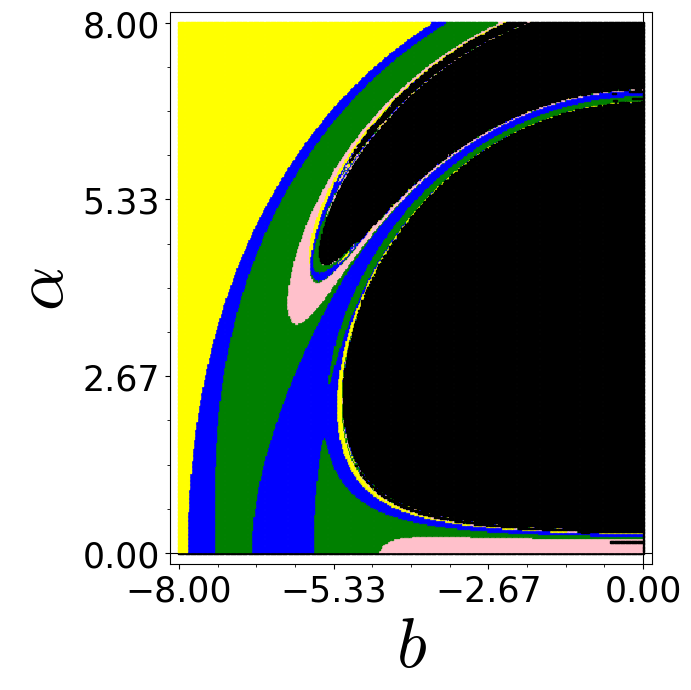} 
    \includegraphics[width=0.2\textwidth]{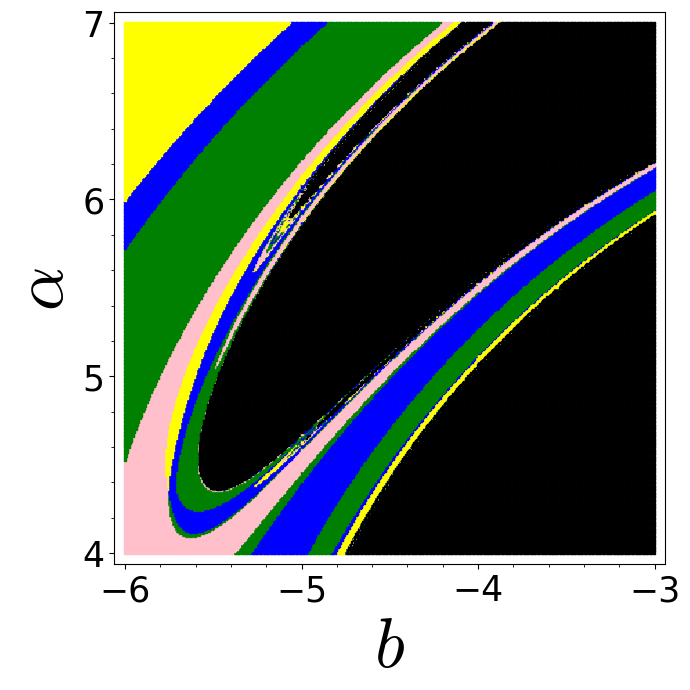}\\
    
    \includegraphics[width=0.2\textwidth]{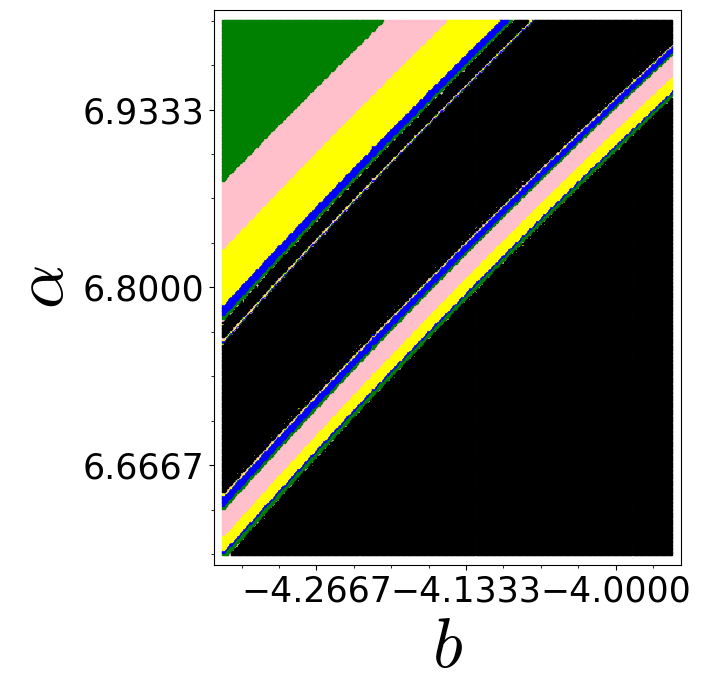} 
    \includegraphics[width=0.2\textwidth]{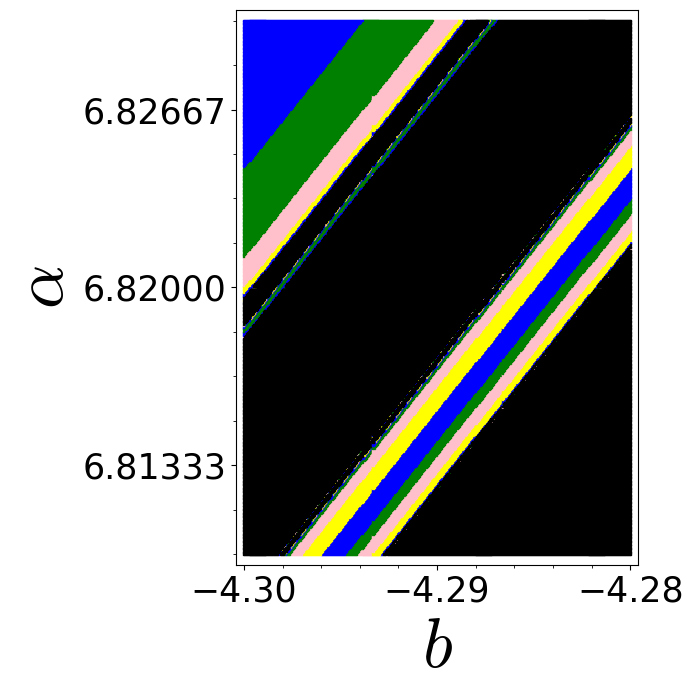}
    \caption{Shadow of the MP di-hole discussed. 
    The shadow has self-similar fractal eyebrows that correspond to different resonances of the light rays with the FPOs.}
    \label{fig:MP_shadows_black}
\end{center}
\end{figure}

The exit basins for $b=4M$ and $d=2M$, highly resemble those of the HT spacetime and are presented in \cref{fig:exbsMP}. Similarly to the HT and JP spacetimes (\cref{fig:exbsHT,fig:exbsJP}) the basins are self-similar. 
We find the fractal dimension to be $D_s=1.81397$ with a goodness of fit $R^2= 0.99951$. The shadow of a MP di-hole with the same parameters is presented in \cref{fig:MP_shadows_black}. It shares the same qualitative features with the HT and JP cases with the shadow being surrounded by self-similar eyebrow features. Since in this case, the space between the two BHs is available to photons, the shadow is split in two halves symmetric with respect to the equatorial plane. Implementing the box-counting method we find that the shadow has a fractal dimension $D_s=1.80970$ with a goodness of fit $R^2=0.99867$.

\begin{figure}[H]
\begin{center}
    \includegraphics[width=0.2\textwidth]{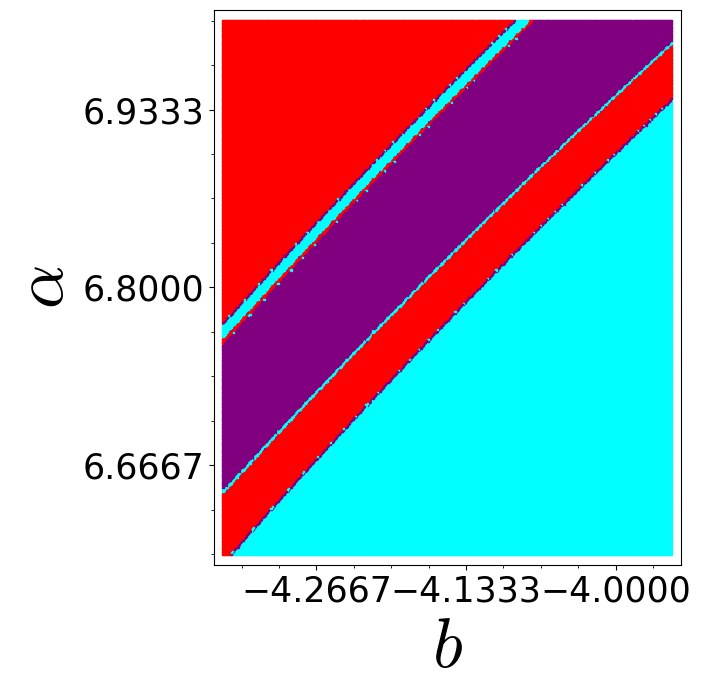} 
    \includegraphics[width=0.2\textwidth]{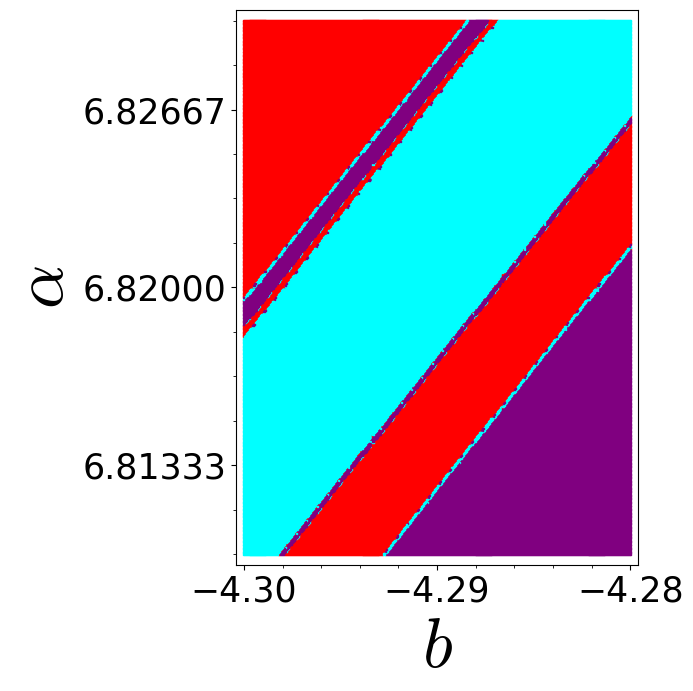}
    \caption{Shadow of the MP di-hole discussed. 
    The light rays are color-coded as purple(cyan) if they fall into the surface of the object through the upper(lower) exit and red if they reach infinity.}
    \label{fig:MP_shadows_clr}
\end{center}
\end{figure}

As in the previous two cases, \cref{fig:MP_shadows_clr} demonstrates how the eyebrow-like features form from the change of photon orbits between different FPO resonances and escapes.

%%%%------------------------
\subsection{Fractal shadows with eyebrow-like formations}
%%%%------------------------

At this point, we can briefly summarize our results. In all three spacetime cases we investigated we have assumed configurations with prolate structure. The spacetime around these objects is therefore also prolate with a positive quadrupole deformation from the quadrupole of a Kerr BH (the Kerr quadrupole is $Q=-\chi^2M^3$, with the negative sign characterizing oblate objects). This allows for the bifurcation of the equatorial light ring and the formation of two off-equatorial light-rings that form two exits towards the central object that photons can follow.

\begin{figure}[h]
\begin{center}
\includegraphics[width=0.22\textwidth]{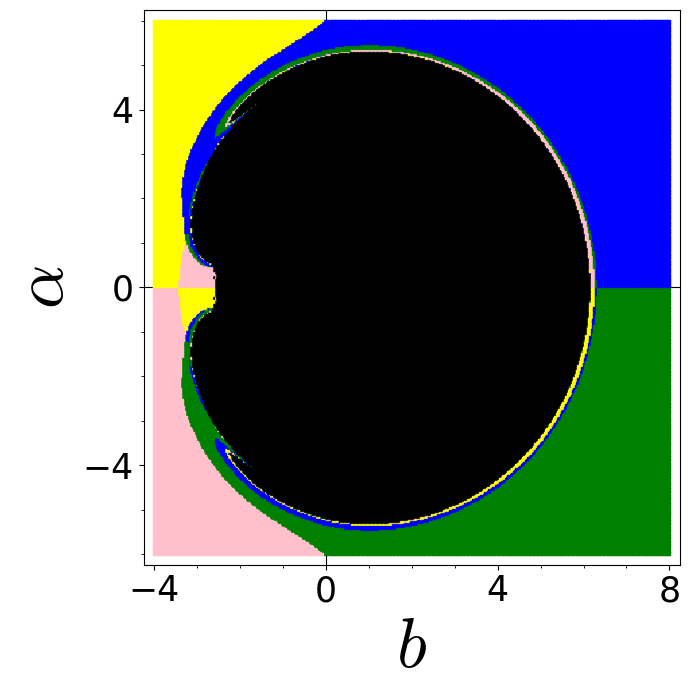} 
    \includegraphics[width=0.22\textwidth]{JP_shad_1prdblk.png}
    \caption{The shadows of a HT spacetime with $\delta q=1$ and $\chi=0.6$ (left) and a JP spacetime with $a=0.85M$ and $\epsilon_3=1$ (right). The shadows are clearly non-Kerr-like. In the HT case, the shift from $\chi=0.35$ to $\chi=0.6$ makes the eyebrow-like formation even more visible.}
    \label{fig:shadows_summary}
\end{center}
\end{figure}

These spacetimes admit several FPOs associated with the different exits of the open Hamiltonian system and general photon orbits can resonate with these FPOs. As a photon orbit shifts through the parameter space it can change from one resonance to another cycling between different combinations of FPOs that result in the photons exiting from different exits. This causes the formation of fractal structures in the shadow of the compact objects such as the eyebrow-like formations we can see in \cref{fig:shadows_summary}. It is worth noting that such prolate configurations may also arise in scenarios inspired by quantum gravity and result in similar shadow structures \cite{Eichhorn_2021,Eichhorn:2021etc}.

%%%%%%%%%%%%%%%%%%%%%%%%%%%
\section{Accretion Illuminated Shadows}
\label{sec:accretion}
%%%%%%%%%%%%%%%%%%%%%%%%%%%

The discussion thus far has been on the properties of the ``mathematical'' shadows of the non-Kerr spacetimes. We will now consider a more astrophysically relevant scenario, where the compact object is illuminated by the accreting matter that surrounds it. As a non-Kerr spacetime, we will use the JP spacetime and we will compare the shadows between Kerr and non-Kerr objects. 
 
In this case, instead of an isotropically illuminated object, an un-realisable illumination background for compact objects in nature, the compact object will be illuminated by the emission of a disk(torus)-like structure around it. 
We will assume a torus located at a radial distance between $r=700M$ and $r=1000M$ from the central object. In astrophysical situations, such a disk would be located between the outer edge of the inner accretion disk and the inner edge of the broad line region (BLR) of a SMBH at the center of a galaxy, a transitional zone that may contribute significantly to the formation of broad emission lines (BELs) \cite{Carroll_Ostlie_2017}. 

Such a disk, being located far from the SMBH, leaves the strongly lensed region free from any emission contamination and has longer dynamical time scales yielding slower emission variability, making it easier to discern any features of interest. 
In short, this compact-object illumination scenario is better suited for observing the finer features of the shadow that are of interest to us, bypassing some of the complications that astrophysics could bring \cite{Glampedakis:2021oie,Volkel:2020xlc,Gralla:2020pra,Ozel:2021ayr,Lara:2021zth}. In this case these features are the fractal structures of the shadow while in other cases they could be the higher-order light-rings \cite{Olmo:2023lil}. Furthermore, we assume that the emission from the disk is monochromatic emission at some rest frame frequency $\nu_0$ (for the astrophysical possibility of this scenario see \cite{Kostaros2024,Thi2024}). 

In cylindrical coordinates ($\varpi,\theta,z)$, we will assume a torus with the boundaries $700M<\varpi<1000M$ and $-85M<z<85M$. We will also assume a density profile, which also serves as our local rest-frame emissivity $j_0$, that follows a simple $j_0\sim r^{-2}$ power law and a velocity profile for the fluid that follows
\begin{equation}
v^{\phi}=\left(\frac{\sqrt{M}}{(r\sin\theta)^{3/2}+a\sqrt{M}}\right),
\label{eq:velocityprofSL}
\end{equation}
which describes an almost Keplerian disk.

In order to produce the image of the BH, we will have to integrate the radiative transfer equations.  
The Lorentz Invariant intensity $\mathcal{I}_\nu=I_\nu/\nu^3$ of every light ray can be calculated through the radiative transfer equation \cite{Younsi2012,Fuerst_2004,Fuerst_2007},
\begin{equation}
    \frac{d\mathcal{I}_\nu}{d\lambda}=\gamma^{-1}\left(\frac{j_0}{\nu^3_{0}}\right),
\end{equation}
where $\gamma=\nu/\nu_0$, with $\nu$ being the frequency and $"0"$ denoting quantities in the local rest frame. We have assumed zero absorption for simplicity. At every point of the geodesic, this is transformed to the specific intensity using
\begin{equation}
\begin{split}
    dI_\nu&=d\mathcal{I}_\nu \nu_{obs}^3=\gamma^{-1}\left(\frac{j_0}{\nu_0^3}\right)\nu^3_{obs}d\lambda\\
    &=\gamma^2 j_0 d\lambda,
\end{split}
\end{equation}
which is the contribution to the intensity from a differential segment of the geodesic $d\lambda$. Since we assume that the emission in the emitter's frame is given by the delta function $\delta(\nu_{em}-\nu_{0})$, transforming the energy from the emitter's frame to the observer's frame introduces an additional $\gamma$ factor. This transformation has to take place just before outputting the results (at the observer's screen) since the intensity is affected by the act of going from the emission frame to the observer's frame \cite{Fuerst_2004}. In this way, for every light ray we obtain a specific intensity distribution at the observer. We use the specific intensity $I_\nu$ at every point along the geodesic, paired with its respective $\nu$, to calculate direct intensity images through
\begin{equation}
    dI=I_\nu d\nu
\end{equation}
for every point on the observer's screen.

We will focus here only on the optically thick case where emission comes from the torus' surface alone, so the analysis is greatly simplified since for every light ray there is intensity at only one frequency since every point on the observer's screen is connected to only one fluid element on the surface of the disk. We use three different compact objects, a Kerr BH rotating at the Thorne limit with $a=0.998M$ and $M=1$ \cite{1974ApJThorne}, a Kerr BH with $a=0.85M$ and $M=1$ and a non-Kerr JP object with $a=0.85M$, $\epsilon_3=1$, and $M=1$. The torus-like structure will be the same for all three cases and it will lie on the equatorial plane of the compact object. The observer will be placed at various angles with respect to the axis of rotation of the compact object.

\begin{figure}[h]
\begin{center}
    \includegraphics[width=0.155\textwidth]{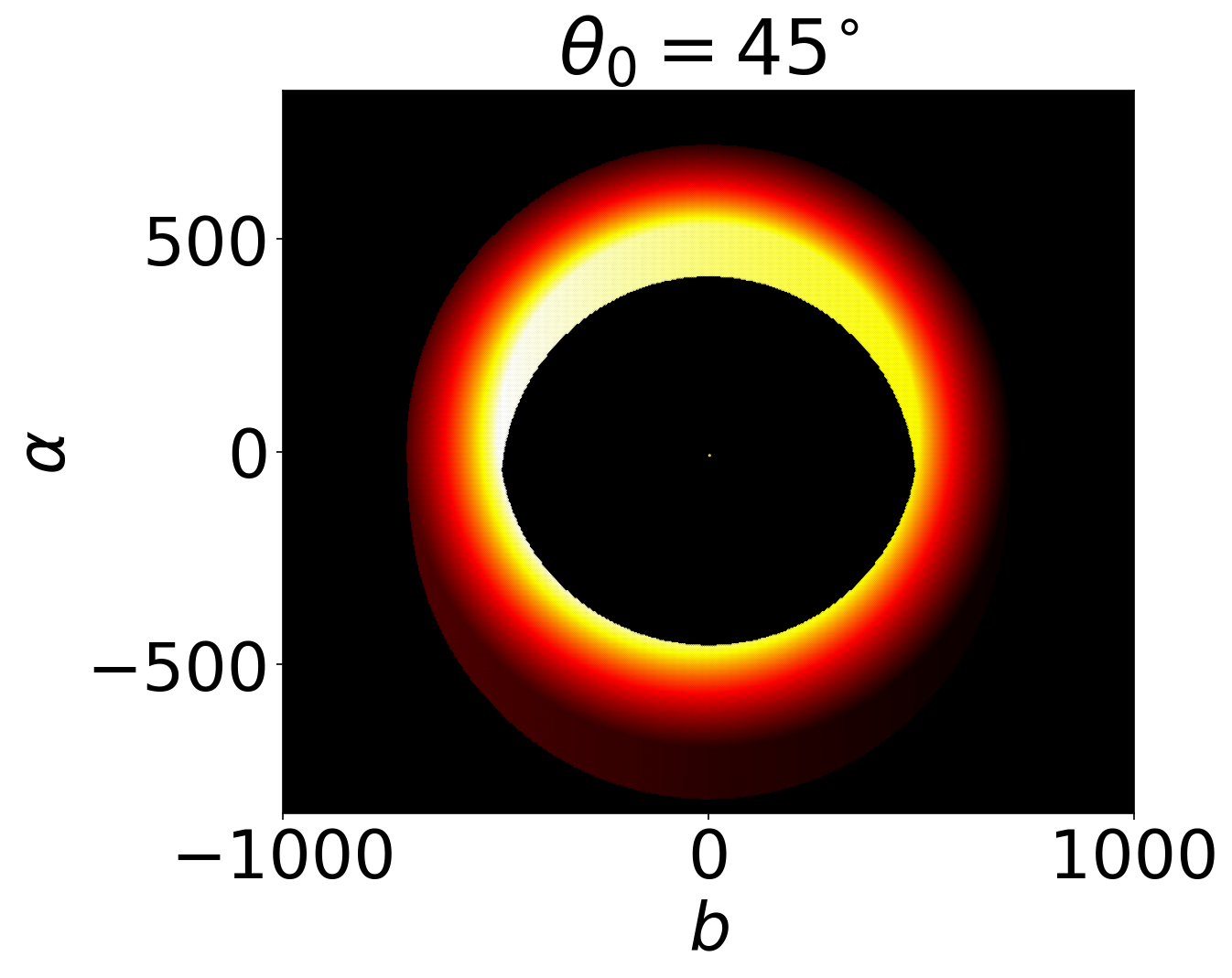}
    \includegraphics[width=0.155\textwidth]{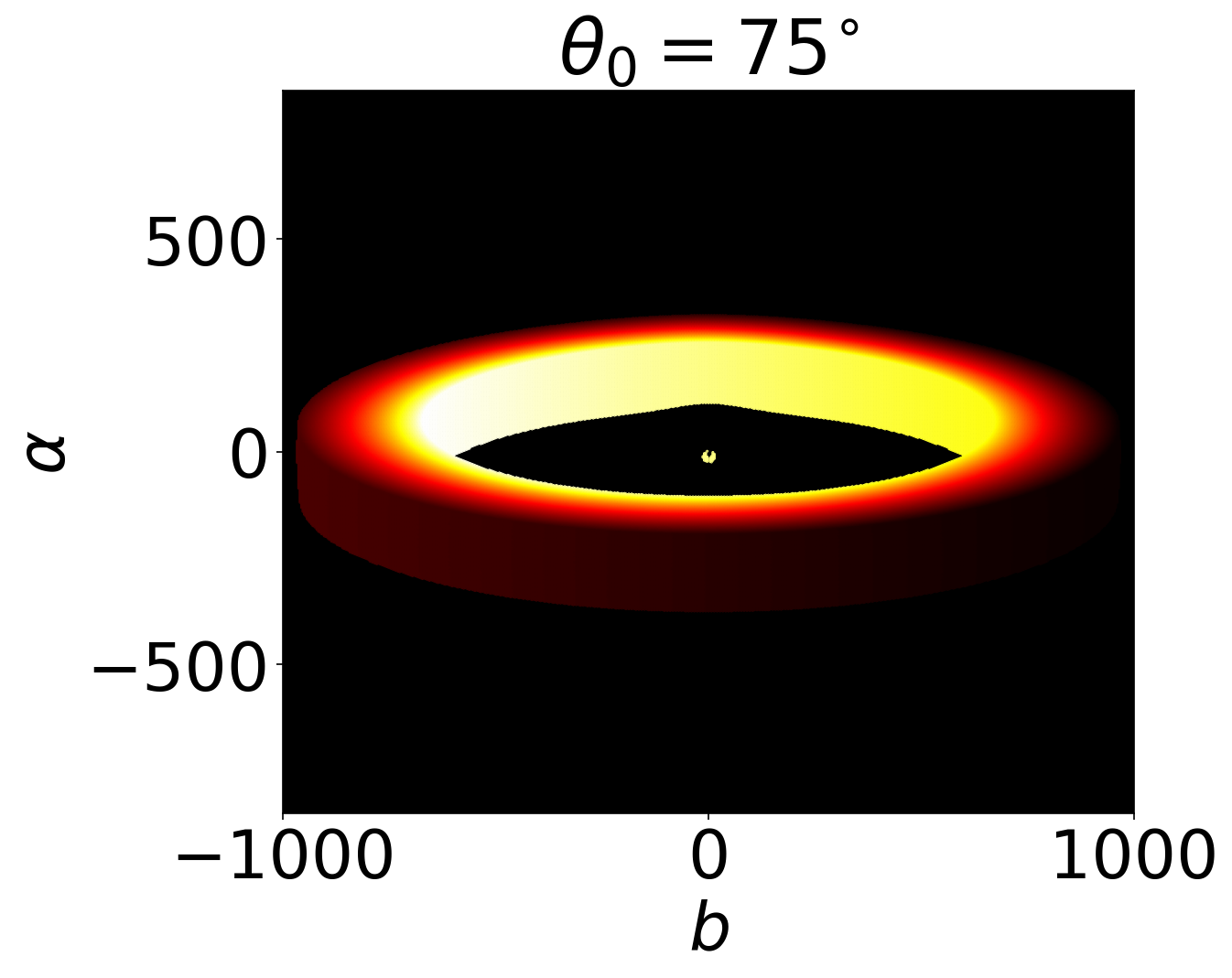} 
    \includegraphics[width=0.155\textwidth]{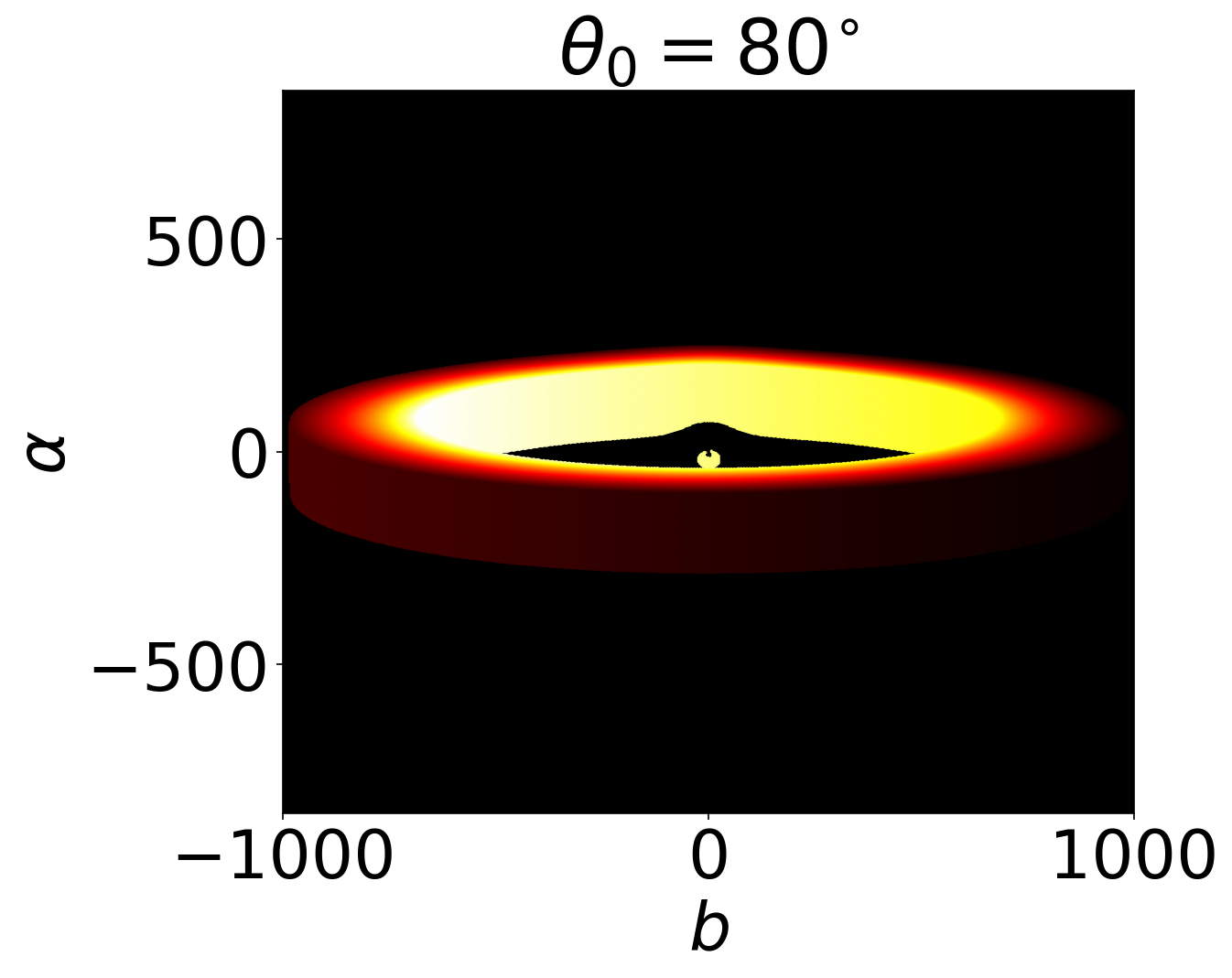}
    
    \caption{Direct intensity images of an optically thick torus located at $700M<\varpi<1000M$ and $-85M<z<85M$, for the viewing inclinations (from the rotation axis) $\theta_0=45,75^{\circ},80^{\circ}$ (left to right).}
    \label{fig:TT_bigscale}
\end{center}
\end{figure}

We show in \cref{fig:TT_bigscale} the direct intensity images of the torus at viewing angles $\theta_0=45^{\circ},75^{\circ},80^{\circ}$ (left to right) with a $600\times600$ resolution. At this scale, these images are the same across all three spacetime setups. As the observer's viewing angle approaches the equatorial plane, the visibility of the inner more dense regions of the torus increases. These regions are the brightest due to Doppler beaming since there is more material moving towards the observer. In the images for the angles  $\theta_0=75^{\circ}, 80^{\circ}$ one can also see the lensing effect of the compact object on the appearance of the torus, where a small deformation appears near the center (more visible for the $80^{\circ}$).

\begin{figure}[h]
\begin{center}
    \includegraphics[width=0.155\textwidth]{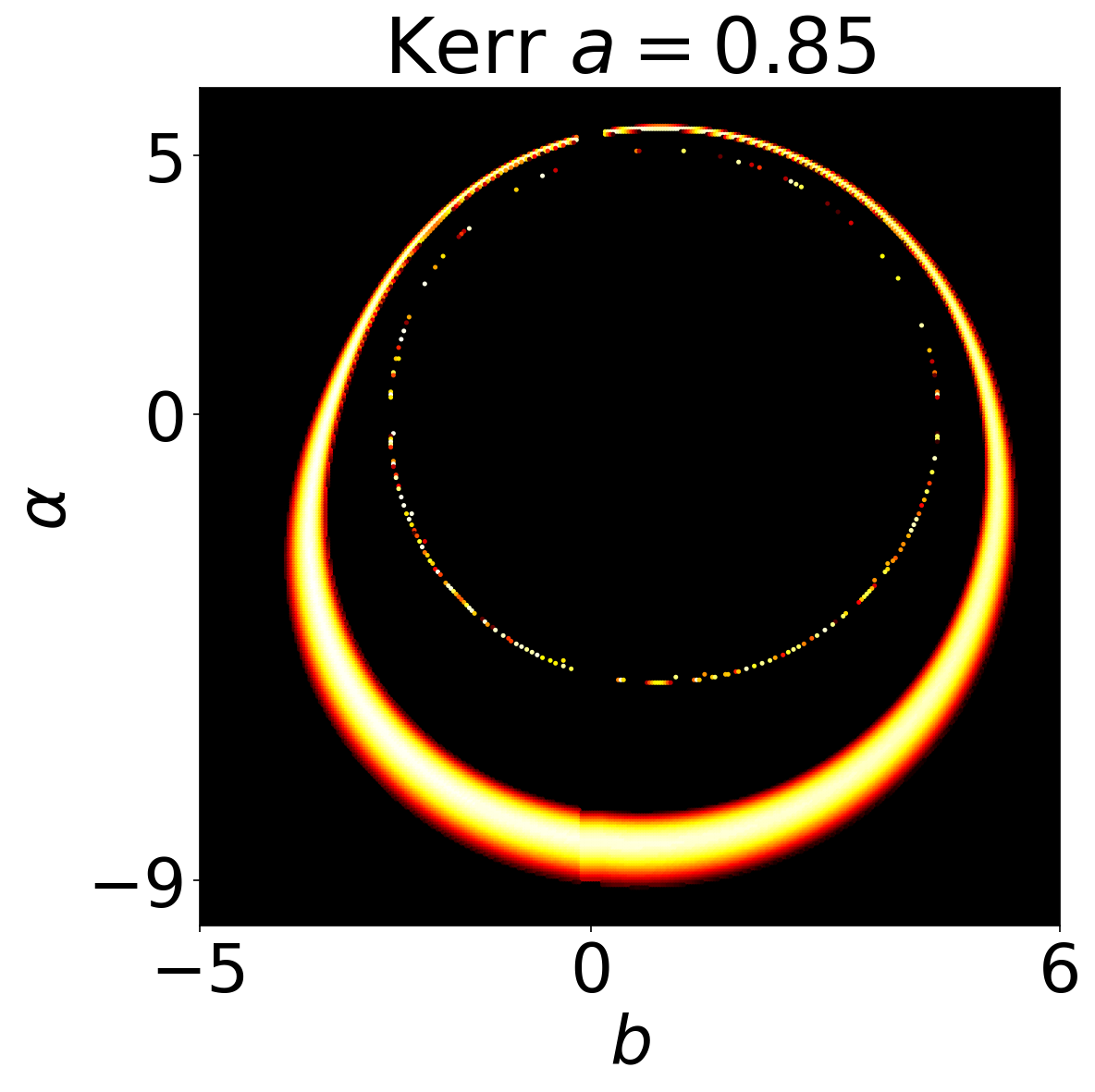}
    \includegraphics[width=0.155\textwidth]{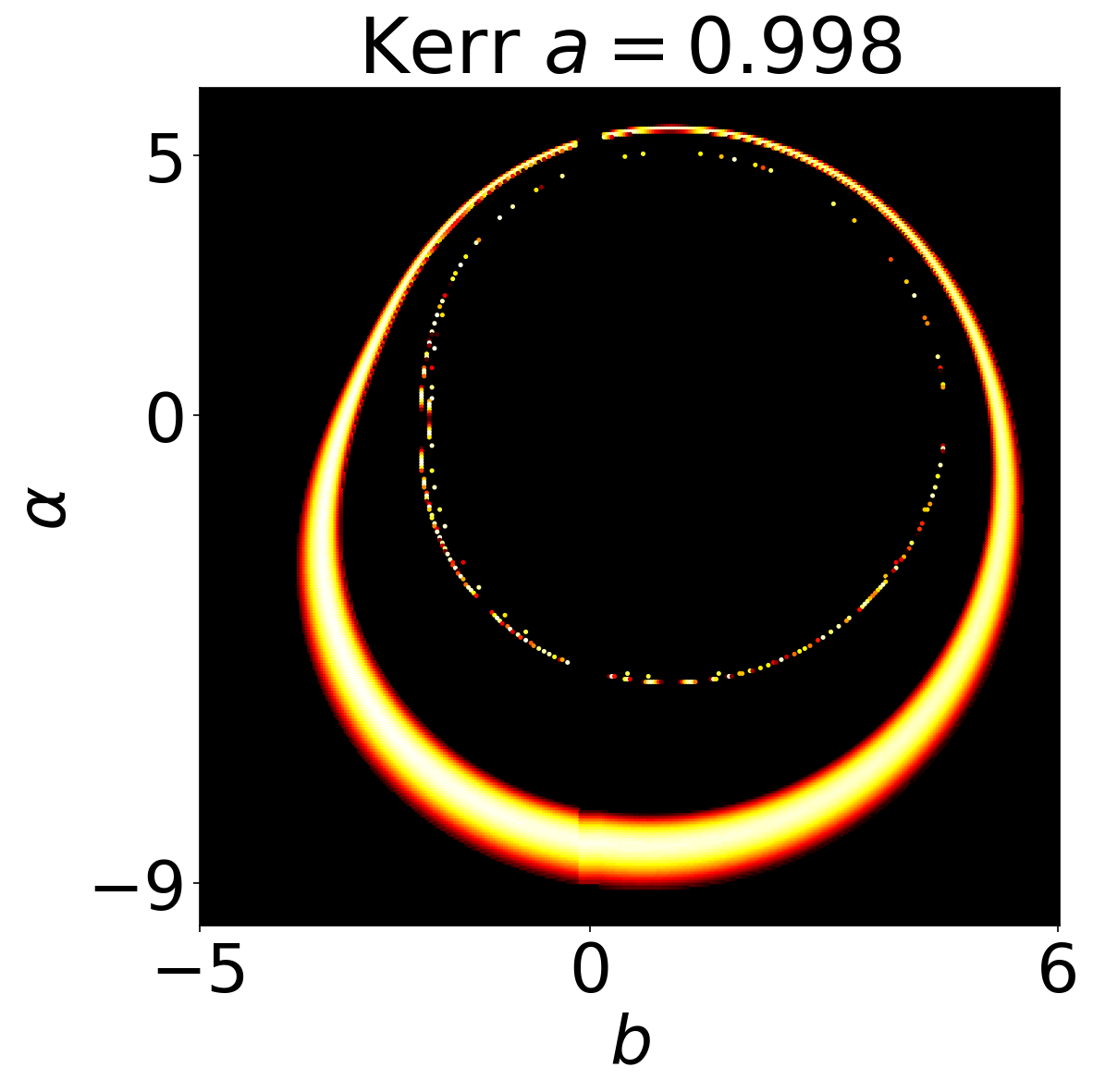} 
    \includegraphics[width=0.155\textwidth]{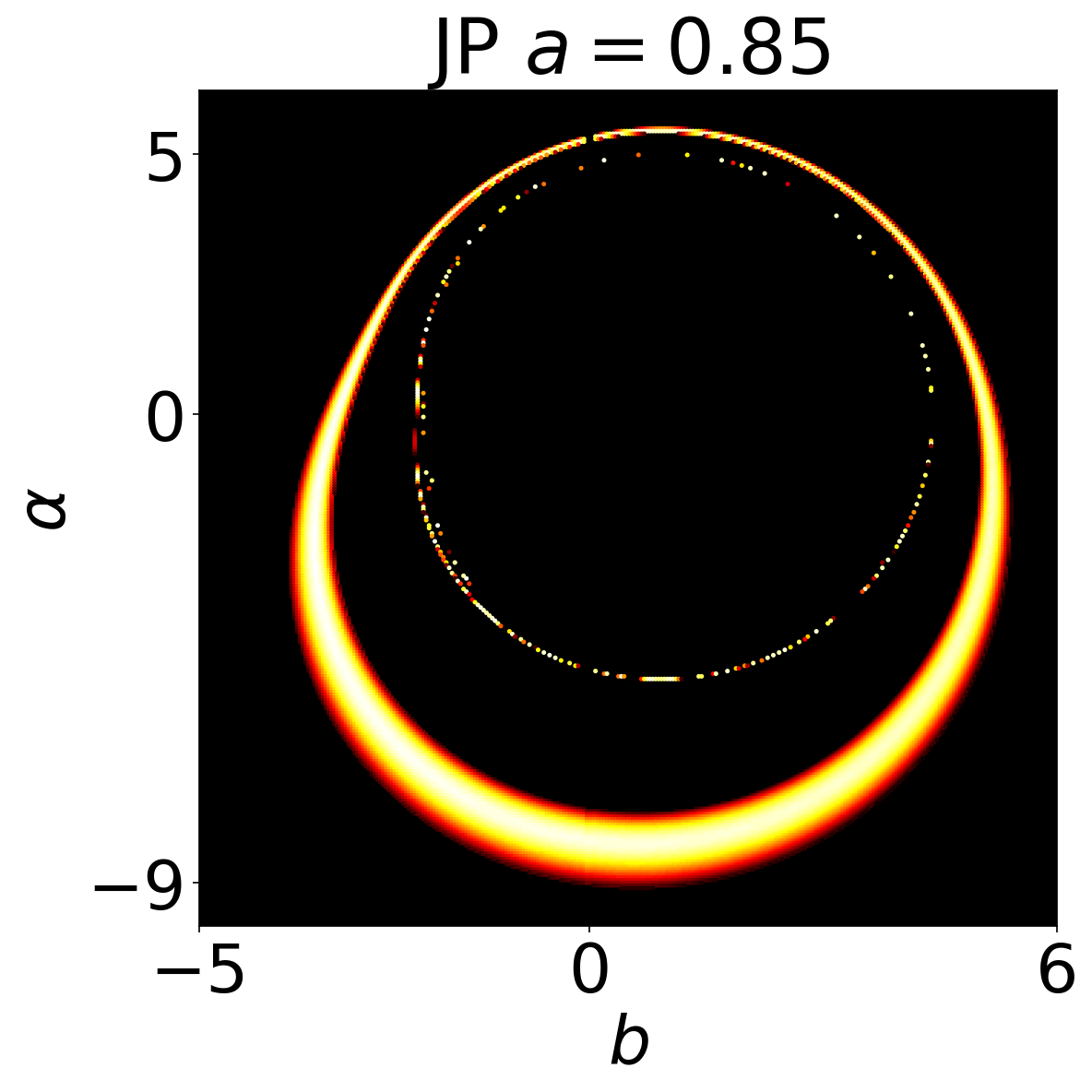}

    \includegraphics[width=0.155\textwidth]{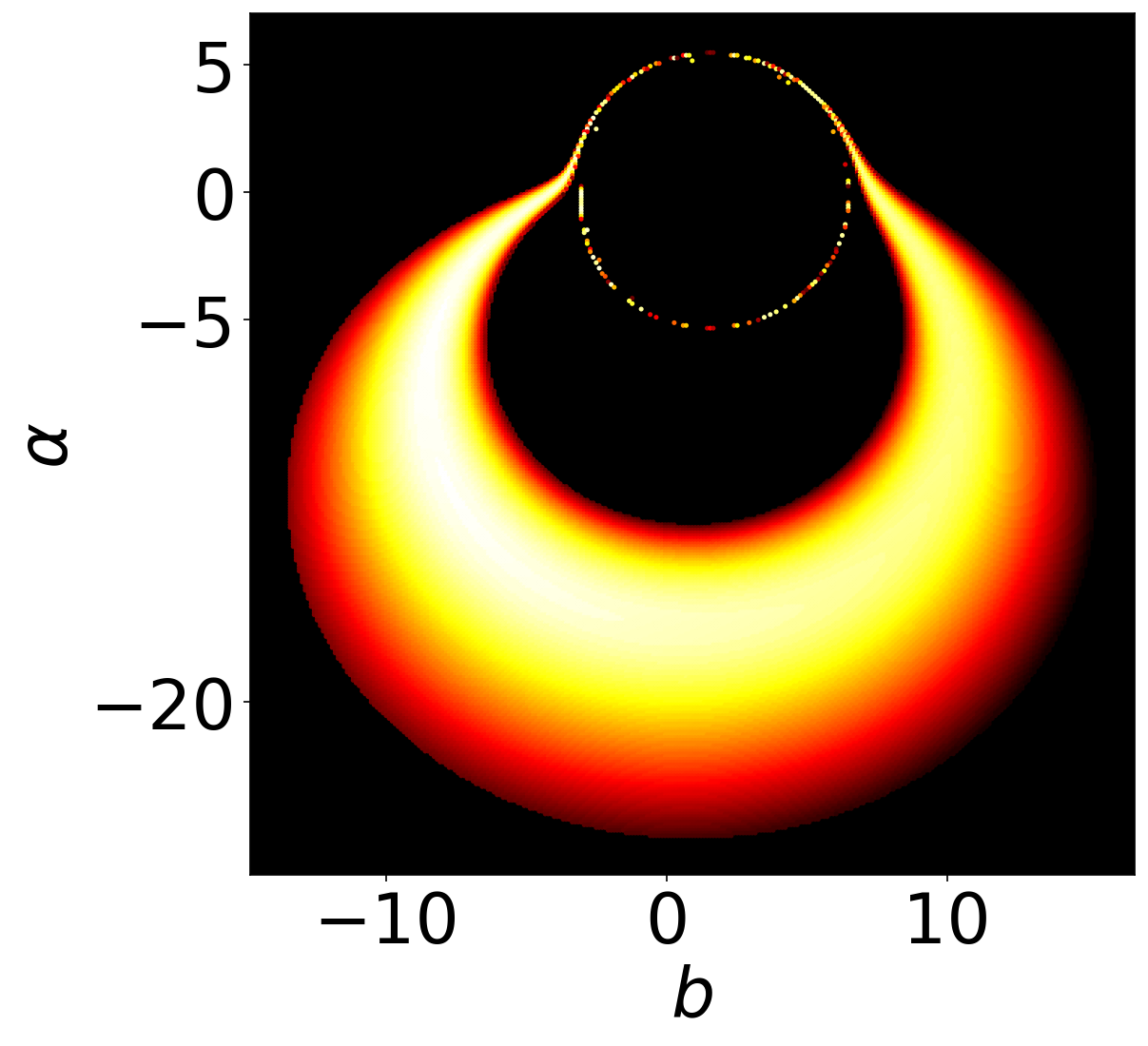}
    \includegraphics[width=0.155\textwidth]{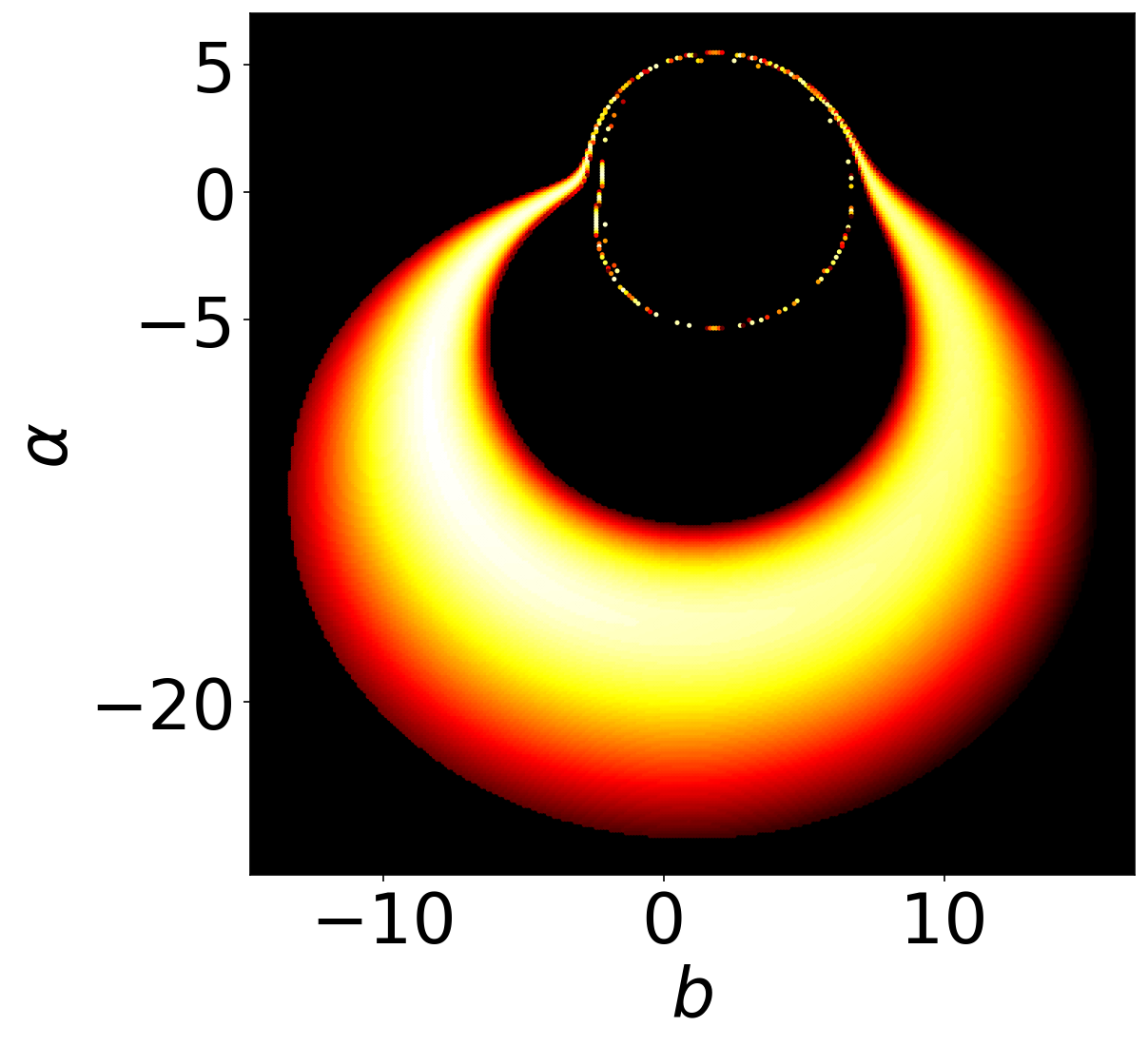} 
    \includegraphics[width=0.155\textwidth]{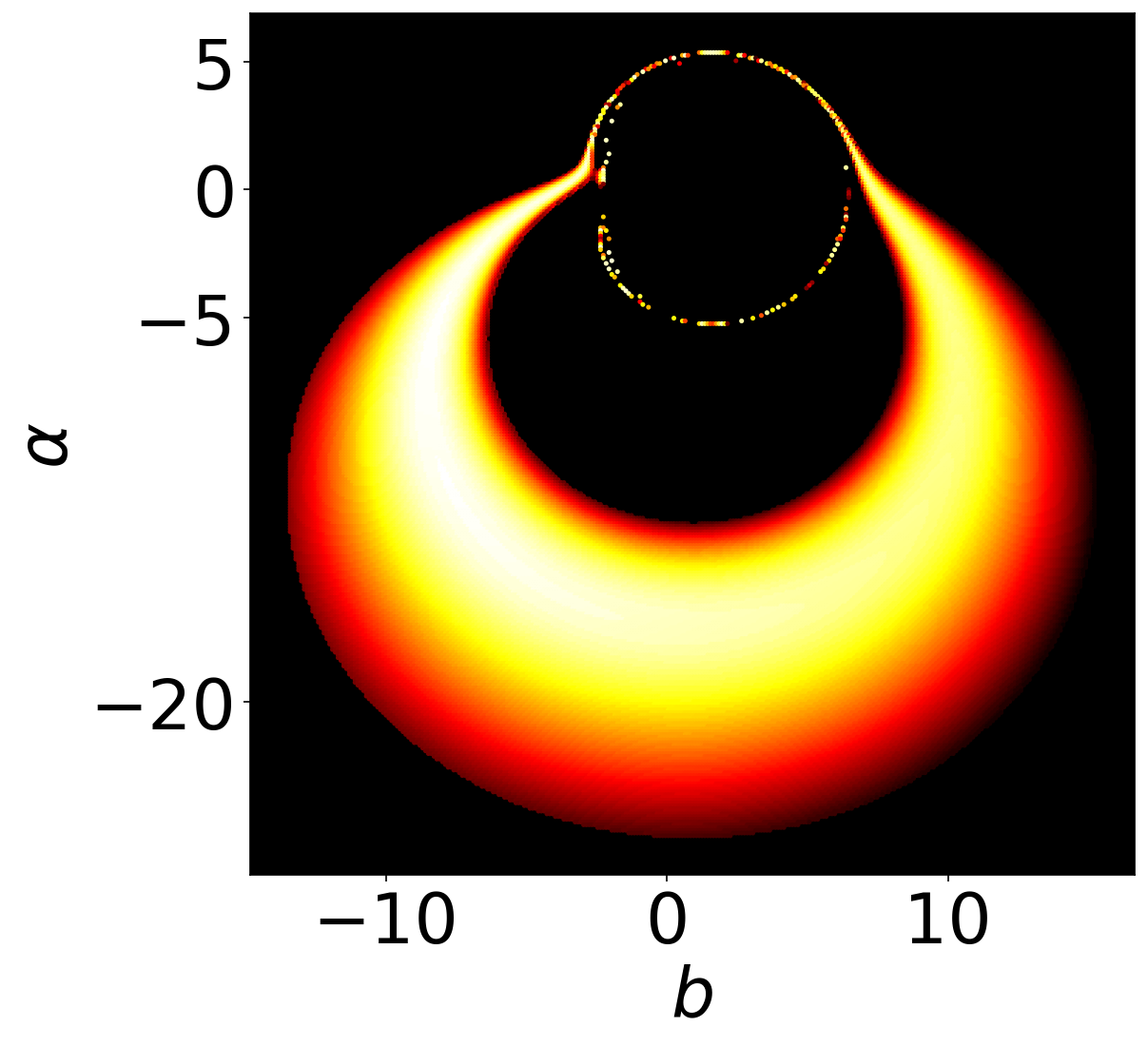}

    \includegraphics[width=0.155\textwidth]{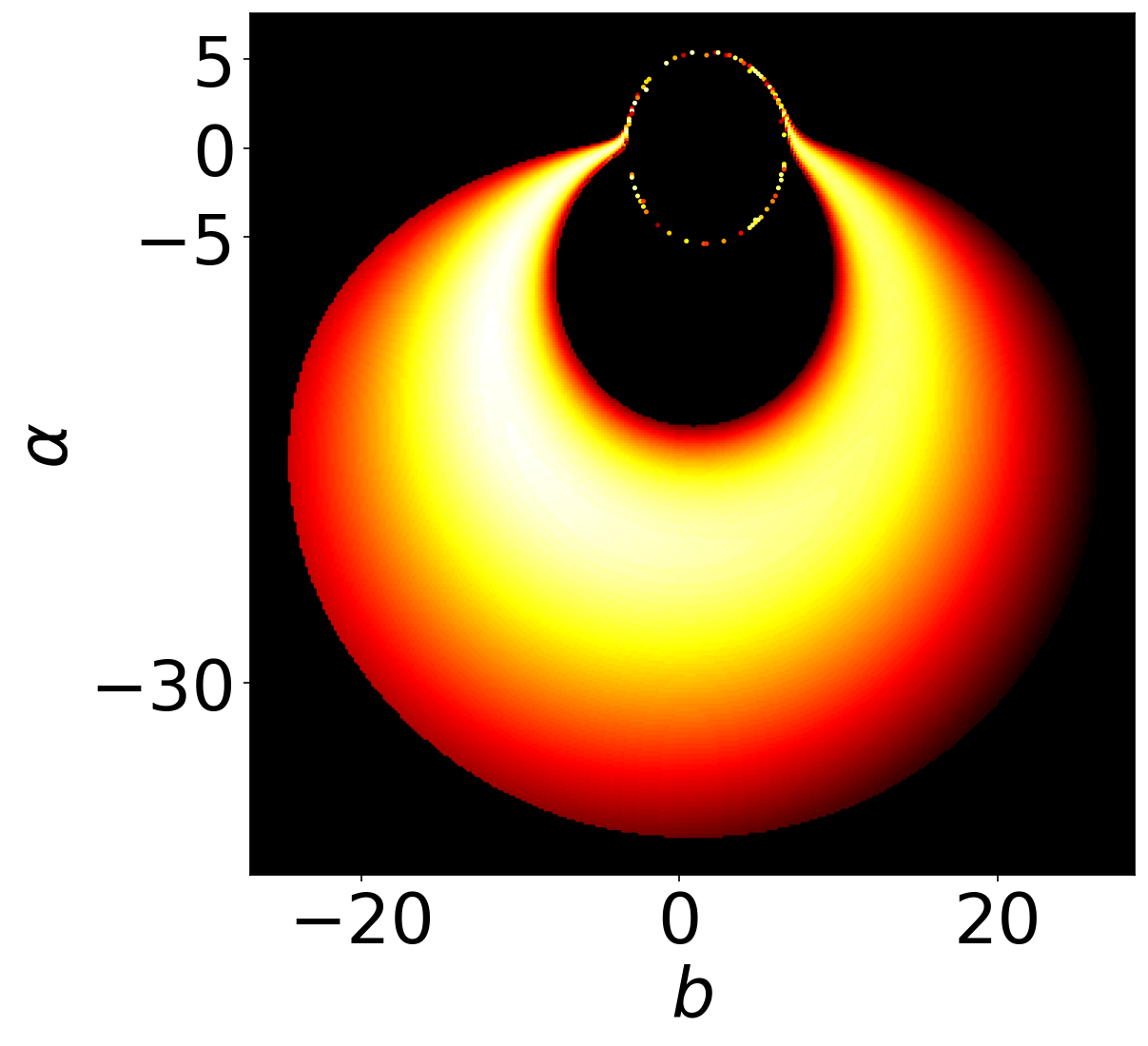}
    \includegraphics[width=0.155\textwidth]{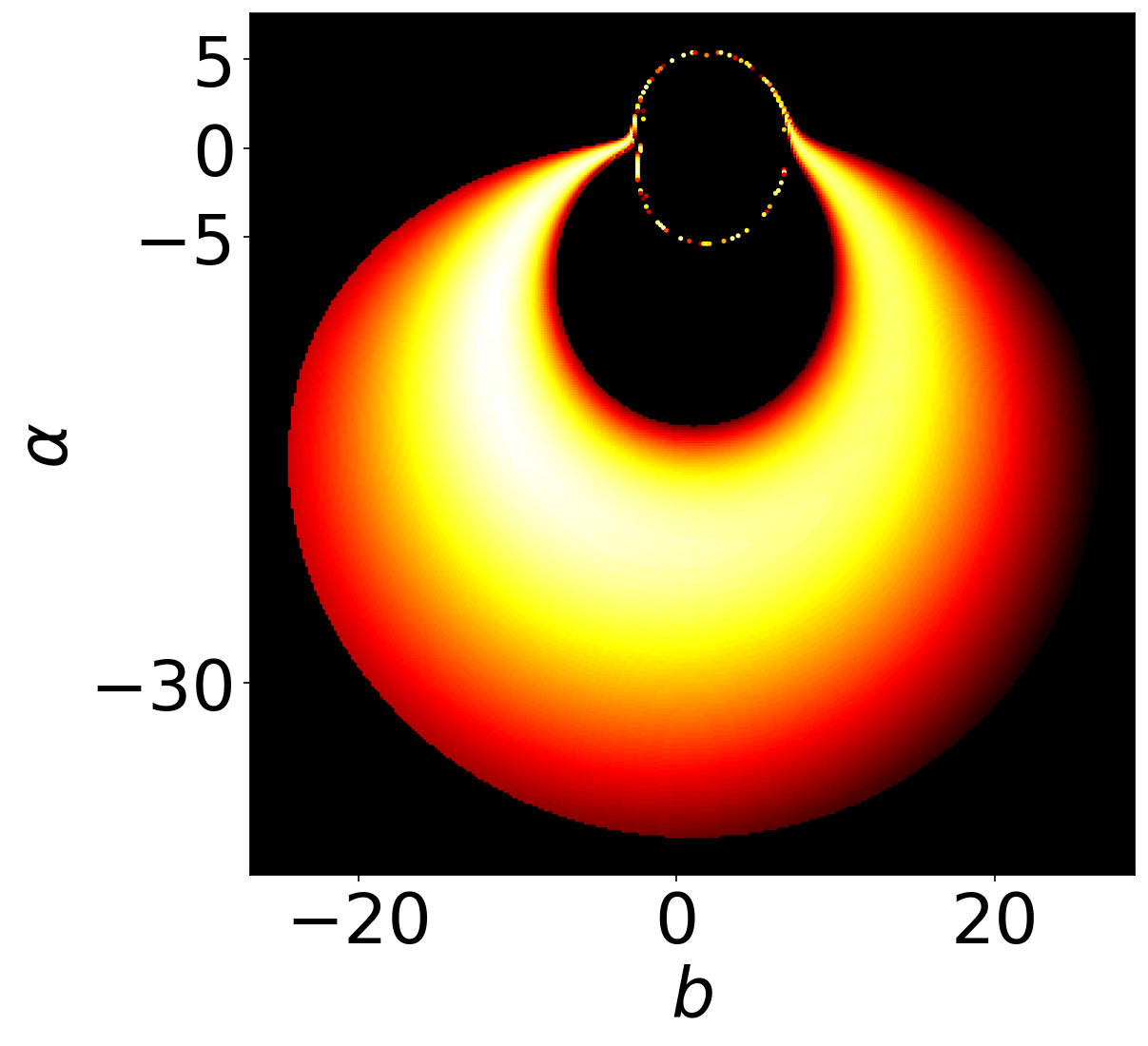} 
    \includegraphics[width=0.155\textwidth]{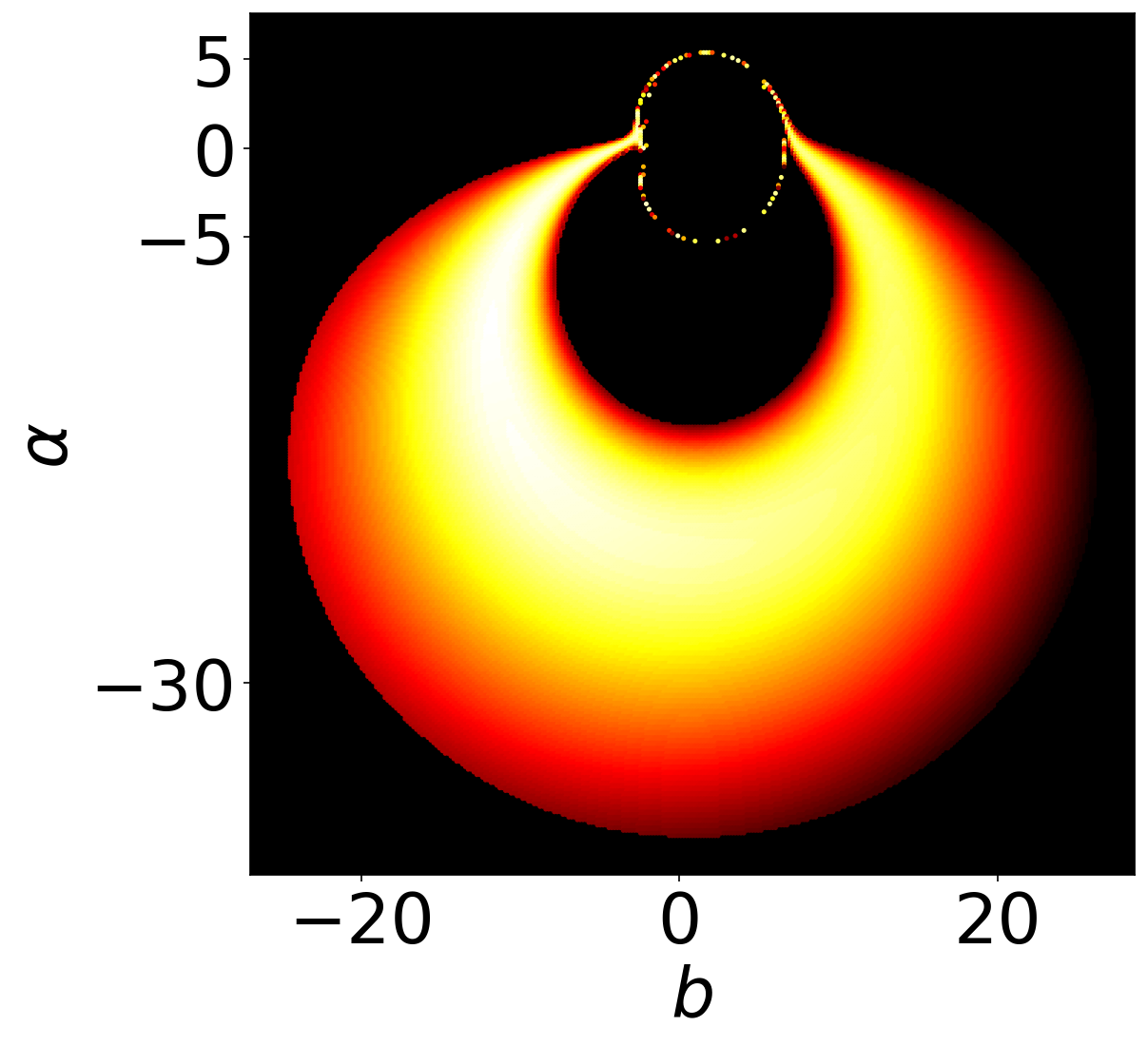}
    \caption{Direct intensity images of the strongly lensed region in the center of the optically thick torus discussed.
    The viewing inclinations are again $\theta_0=45^{\circ},75^{\circ},80^{\circ}$ (top to bottom) for three spacetime setups: a Kerr BH with $a=0.85M$, a Kerr BH with $a=0.998M$ and a JP object with $a=0.85M$, $\epsilon_3=1$ (left to right) }
    \label{fig:TT_midscale}
\end{center}
\end{figure}

We continue by zooming-in on the central and strongly lensed region for which we present in \cref{fig:TT_midscale} direct intensity images for the three spacetimes (left to right) at the $\theta_0=45^{\circ},75^{\circ},80^{\circ}$ viewing angles (top to bottom). There are two characteristic rings in each image, the larger one that is the lensed image of the torus and a smaller one that is the light-ring around the object's shadow. Both are created by photons that leave the torus' inner surface, get lensed by the central object and reach the observer's screen. The light-rings in particular are formed by the photons that orbit the central object several times. 

The lensed images (larger ring) of the torus are the same across the three spacetimes for all viewing angles. 
This is due to the fact that the disk is relatively far from the object. At such distance, the spacetimes are too similar (i.e., Schwarzschild-like) to produce any significant differences in the structure of the disk. These spacetime-induced differences in the disk structure is what would be visible in either the direct image or the larger lensed image of the torus, therefore these images cannot tell the spacetimes apart. 
As the viewing inclination increases the lensed photons seem to be more influenced by the relativistic beaming 
effect since we observe an increased asymmetry in the surface's brightness. Additionally, the lensed torus appears larger and elongated at larger angles, which would make it brighter and easier to observe.

\begin{figure}[H]
\begin{center}
    \includegraphics[width=0.155\textwidth]{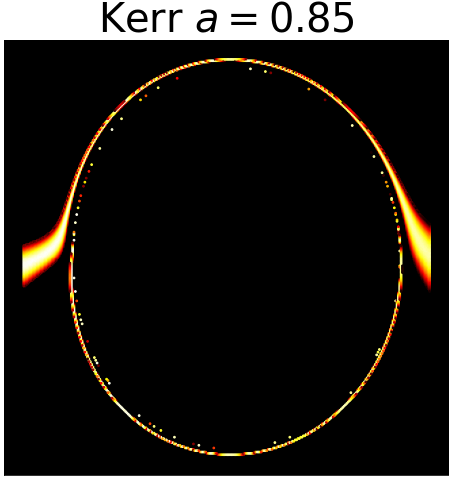}
    \includegraphics[width=0.155\textwidth]{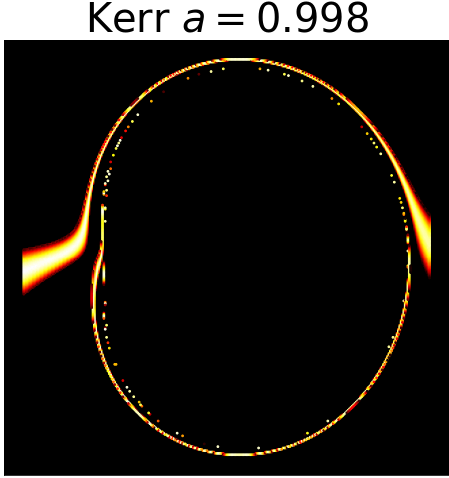} 
    \includegraphics[width=0.155\textwidth]{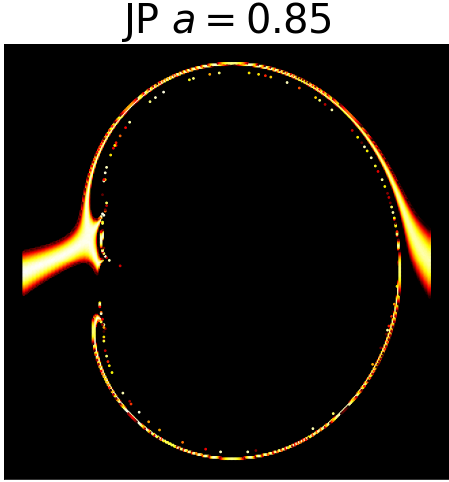}
    \caption{Zoom-ins on the photon ring viewed at the observer's angle $\theta_0=80^{\circ}$, for three different spacetimes, a Kerr BH with a spin parameter $a=0.85M$, a Kerr BH with a spin parameter $a=0.998M$ and a JP object with $a=0.85M$ and $\epsilon_3=1$  (left to right). The impact parameter ranges are $b\in [-4.5M,7.5M]$ and $\alpha \in [-5.5M,5.5M]$.}
    \label{fig:TT_ringscale}
\end{center}
\end{figure}

To examine the possible differences between the three spacetimes we zoom-in further to the light-ring's scale where one approaches the mathematical shadow of the object (this is a little smaller than the scale of the EHT images \cite{Akiyama_2019,EHT_Sgr1}). We do this for a viewing angle of $\theta_0=80^{\circ}$ and the resulting direct intensity images of the three objects are presented in \cref{fig:TT_ringscale}. At this scale, there are visible differences between the Kerr BHs and the JP object. In the JP spacetime the lensed torus exhibits a bifurcation-like behavior, approaching both the shadow's cusp and top eyebrow feature. The light-ring gets disconnected, wrapping itself around the bottom eyebrow feature {\it in a clear smoking gun signal of a non-Kerr spacetime.} We note here that our analysis refers to stationary and isolated compact objects.
 
This also demonstrates the advantage of the distant-disk illumination of the central object since the spacetime-related differences are easier to identify as compared to illumination by disks close to the central object. In the latter cases the large ``noise'' of the accretion itself (due to e.g. the inevitable stochasticity of emission caused by the local excitation conditions, turbulent velocity fields, MHD effects) can ``mask'' the interesting features we study here. This is simply because such physical ``noise'' will be vastly higher in the strongly turbulent, very hot, highly-variable, and highly-ionized accretion disk close to the central object.

\begin{figure}[h]
\centering
\includegraphics[width=0.5\textwidth]{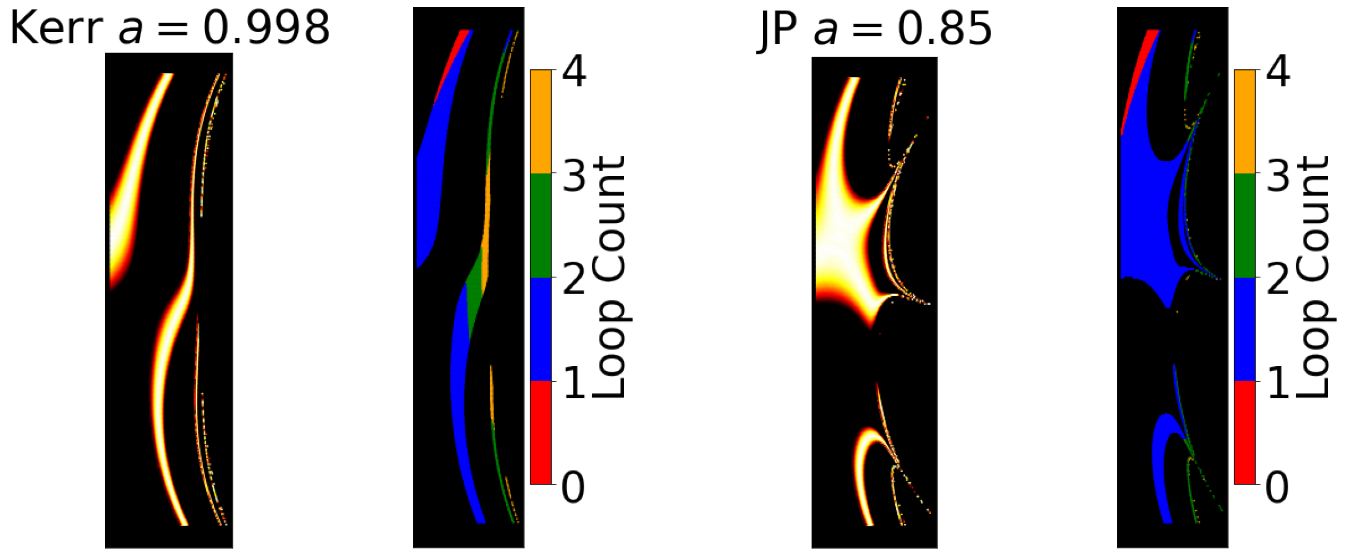}
\caption{Zoom-in images of the prograde side of the Kerr $a=0.998M$ BH (left) and JP $a=0.85M$, $\epsilon_3=1$ object (right) and the orders of the geodesics that create them. Color code up to $n<4$: red for $[0,1)$ windings; blue for $[1,2)$ windings; green for $[2,3)$ windings; orange for $[3,4)$ windings.}
\label{fig:TT_superzoomscale}
\end{figure}

In the ranges of impact parameters and image resolution we have studied thus far, we have merely captured the lowest order light-ring, mainly produced by $n\lesssim1$ geodesics (where $n$ is the number of windings a light ray does around the central object). The light-ring though is created by a sequence of subrings made of photons that perform an increasing number of windings. Each subring of the set that constitutes the light-ring is exponentially narrower than the last, requiring exponentially finer resolution in order to resolve the set \cite{Gralla_2020}. To capture the image of photons with more windings, we zoom-in in the ranges $b\in [-1.8M,-2.8M]$ and $\alpha \in [-2.5M,2.5M]$ with a $300\times 300$ resolution. The direct intensity images and the different orders of the geodesics that create them are presented in \cref{fig:TT_superzoomscale}. In the Kerr case, the subrings are successive and nearly circular as they approach the BH's shadow. This circularity is clearly broken in the JP spacetime where the shadow of the object has a richer structure. The $n>3$ subrings are thinner, approaching the main cusps and self-similar eyebrows of the JP objects mathematical shadow. In this work, we will not  further pursue these higher-order features, but these higher-order rings are worth further study.

As a final note, we should mention that the images at the scale of the light-ring and the shadow, show that there are favorable observation angles where the disk illuminates the features of interest. One could use cases like these, if they present themselves, to better study the higher-order rings and look for finer fractal features.

%%%%%%%%%%%%%%%%%%%%%%%%%%%%%%
\section{Conclusions and future work}
\label{sec:conclusions}
%%%%%%%%%%%%%%%%%%%%%%%%%%%%%%

In this work, we studied the photon orbits of three non-Kerr spacetimes: the slowly rotating Hartle~\&~Thorne spacetime, the deformed Kerr metric by Johannsen and Psaltis, and the static Majumdar-Papapetrou di-hole. We focused on configurations of prolate compact objects that form off-equatorial light-rings and allow for three escapes to the photon geodesics, leaving these spacetimes to be open Hamiltonian systems. Each spacetime specifically admits two non-equatorial light-rings, which is a generic feature, and multiple FPOs, whose number and formation depend on the parameters of the configuration. These FPOs are dynamically connected and photons that are launched from infinity can resonate with them either individually or in any combination. 

To further study these properties of the photon orbits we evaluated exit basin diagrams in every case. The exit basin diagrams of each spacetime are self-similar and form eyebrows on top of eyebrows that correspond to different types of resonances of the photons with the FPOs. This behavior is inherited by the respective mathematical shadows which exhibit self-similar, fractal, eyebrow features. These fractal features are therefore demonstrated to be products of the resonances and the multiple available escapes between which photon orbits shift.

In order to apply our findings to a more realistic astrophysical scenario and explore the phenomenology, we carried out a comparative analysis between the shadows of three objects illuminated by an accretion disk. The three objects we compare are, a non-Kerr Johannsen-Psaltis object, considered as a typical example of the non-Kerr objects we are interested in, and two Kerr BHs of different spins. 
The illuminating torus is placed in the transitional zone between the inner disk and the BLR and is assumed to be optically thick.

The torus appears with three main images on the observer's screen. The first is a large direct image with very little lensing taking place. The second is a strongly lensed image coming from the compact object and is at a scale comparable to but larger than the scale of the light-ring. Finally, the third main image is a very strongly lensed image of the accretion disk that forms the light-ring that defines the edge of the shadow. 

Both the direct images of the torus and the strongly lensed images are the same across all spacetimes, showing no obvious differences. This is mainly due to the fact that the disk is relatively far from the compact object to a region where all spacetimes are quite similar. 
To identify possible differences between the images of the three objects one needs to look at the third and very strongly lensed image that resolves the scale of the light ring. 
At the inclination of 80 degrees, the emission from the disk illuminates the JP spacetime's fractal features clearly showing it to be a non-Kerr spacetime. 

In this work, we have tried to broaden our scope of black hole shadow analysis through the systematic exploration of the dynamics of photon geodesics within both a mathematical and an astrophysical framework. The demonstration of the existence of self-similar structures across different stationary and axisymmetric prolate non-Kerr objects, suggests a universal aspect to chaotic scattering that goes beyond the existence of a dynamical pocket \cite{Sengo_2023,Kostaros_2022,Cunha2016PhRvD,Shipley_2016} and possibly a universal aspect to the formation of fractal structures in the shadow of these objects. Furthermore, our astrophysical application indicates that such structures in the image of the shadow can be observable under favorable conditions (for the geometry of the illuminating accretion disk and its relation to the observer) with future instrument capabilities. 

In a future direction for this work we will perform a more detailed investigation of how small prolate deviations from Kerr can be identified by the formation of fractal features in the shadows. 
The parameter space of different spins and quadrupole deformations will be systematically explored for the type and magnitude of effects that can be produced and may be accessible with current and future observational capabilities of instruments like the EHT \cite{Ayzenberg:2023hfw}. 
This at first may seem impossible given the complications of the accretion disk physics and the corresponding uncertainties of the radiation field illuminating the central object. Nevertheless, given that the expected phases of any symmetric object (like the shadow of not too rapidly rotating Kerr SMBH) in Fourier space (the measurement domain of interferometers) will be zero, any deviation from symmetry, betraying a non-Kerr fractal shadow, could be in principle detectable via the detection of non-zero features in the phases, which are the most (emission structure)-sensitive measurables in interferometric imaging. Higher resolution and denser u-v coverage would be necessary for approaching such a goal, but future EHT capabilities (e.g operation at 345\,GHz, addition of space-based baselines) do aim towards such a direction \cite{Gurvits:2019ioq,Fromm:2021flr}.

\section*{Acknowledgements}

The authors would like to thank %
T. Apostolatos for useful discussions throughout the development of this work. The numerical integration of geodesics was carried out using SageMath's built-in integrator \cite{sagemath}. All results presented in this work have been produced using the Aristotle University of Thessaloniki (AUTh) High Performance Infrastructure and Resources. The publication of the article in OA mode was financially supported by HEAL-Link.

\bibliography{bibliography.bib}

\begin{thebibliography}{118}%
\makeatletter
\providecommand \@ifxundefined [1]{%
 \@ifx{#1\undefined}
}%
\providecommand \@ifnum [1]{%
 \ifnum #1\expandafter \@firstoftwo
 \else \expandafter \@secondoftwo
 \fi
}%
\providecommand \@ifx [1]{%
 \ifx #1\expandafter \@firstoftwo
 \else \expandafter \@secondoftwo
 \fi
}%
\providecommand \natexlab [1]{#1}%
\providecommand \enquote  [1]{``#1''}%
\providecommand \bibnamefont  [1]{#1}%
\providecommand \bibfnamefont [1]{#1}%
\providecommand \citenamefont [1]{#1}%
\providecommand \href@noop [0]{\@secondoftwo}%
\providecommand \href [0]{\begingroup \@sanitize@url \@href}%
\providecommand \@href[1]{\@@startlink{#1}\@@href}%
\providecommand \@@href[1]{\endgroup#1\@@endlink}%
\providecommand \@sanitize@url [0]{\catcode `\\12\catcode `\$12\catcode
  `\&12\catcode `\#12\catcode `\^12\catcode `\_12\catcode `\%12\relax}%
\providecommand \@@startlink[1]{}%
\providecommand \@@endlink[0]{}%
\providecommand \url  [0]{\begingroup\@sanitize@url \@url }%
\providecommand \@url [1]{\endgroup\@href {#1}{\urlprefix }}%
\providecommand \urlprefix  [0]{URL }%
\providecommand \Eprint [0]{\href }%
\providecommand \doibase [0]{http://dx.doi.org/}%
\providecommand \selectlanguage [0]{\@gobble}%
\providecommand \bibinfo  [0]{\@secondoftwo}%
\providecommand \bibfield  [0]{\@secondoftwo}%
\providecommand \translation [1]{[#1]}%
\providecommand \BibitemOpen [0]{}%
\providecommand \bibitemStop [0]{}%
\providecommand \bibitemNoStop [0]{.\EOS\space}%
\providecommand \EOS [0]{\spacefactor3000\relax}%
\providecommand \BibitemShut  [1]{\csname bibitem#1\endcsname}%
\let\auto@bib@innerbib\@empty
%</preamble>
\bibitem [{\citenamefont {Collaboration}(2019{\natexlab{a}})}]{Akiyama_2019}%
  \BibitemOpen
  \bibfield  {author} {\bibinfo {author} {\bibfnamefont {T.~E. H.~T.}\
  \bibnamefont {Collaboration}},\ }\href {\doibase 10.3847/2041-8213/ab0ec7}
  {\bibfield  {journal} {\bibinfo  {journal} {The Astrophysical Journal
  Letters}\ }\textbf {\bibinfo {volume} {875}},\ \bibinfo {pages} {L1}
  (\bibinfo {year} {2019}{\natexlab{a}})}\BibitemShut {NoStop}%
\bibitem [{\citenamefont {Collaboration}(2019{\natexlab{b}})}]{EHT_M87_2}%
  \BibitemOpen
  \bibfield  {author} {\bibinfo {author} {\bibfnamefont {T.~E. H.~T.}\
  \bibnamefont {Collaboration}},\ }\href {\doibase 10.3847/2041-8213/ab0c96}
  {\bibfield  {journal} {\bibinfo  {journal} {The Astrophysical Journal
  Letters}\ }\textbf {\bibinfo {volume} {875}},\ \bibinfo {eid} {L2} (\bibinfo
  {year} {2019}{\natexlab{b}})},\ \Eprint {http://arxiv.org/abs/1906.11239}
  {arXiv:1906.11239 [astro-ph.IM]} \BibitemShut {NoStop}%
\bibitem [{\citenamefont {Collaboration}(2019{\natexlab{c}})}]{EHT_M87_3}%
  \BibitemOpen
  \bibfield  {author} {\bibinfo {author} {\bibfnamefont {T.~E. H.~T.}\
  \bibnamefont {Collaboration}},\ }\href {\doibase 10.3847/2041-8213/ab0c57}
  {\bibfield  {journal} {\bibinfo  {journal} {The Astrophysical Journal
  Letters}\ }\textbf {\bibinfo {volume} {875}},\ \bibinfo {eid} {L3} (\bibinfo
  {year} {2019}{\natexlab{c}})},\ \Eprint {http://arxiv.org/abs/1906.11240}
  {arXiv:1906.11240 [astro-ph.GA]} \BibitemShut {NoStop}%
\bibitem [{\citenamefont {Collaboration}(2019{\natexlab{d}})}]{EHT_M87_4}%
  \BibitemOpen
  \bibfield  {author} {\bibinfo {author} {\bibfnamefont {T.~E. H.~T.}\
  \bibnamefont {Collaboration}},\ }\href {\doibase 10.3847/2041-8213/ab0e85}
  {\bibfield  {journal} {\bibinfo  {journal} {The Astrophysical Journal
  Letters}\ }\textbf {\bibinfo {volume} {875}},\ \bibinfo {eid} {L4} (\bibinfo
  {year} {2019}{\natexlab{d}})},\ \Eprint {http://arxiv.org/abs/1906.11241}
  {arXiv:1906.11241 [astro-ph.GA]} \BibitemShut {NoStop}%
\bibitem [{\citenamefont {Collaboration}(2019{\natexlab{e}})}]{EHT_M87_5}%
  \BibitemOpen
  \bibfield  {author} {\bibinfo {author} {\bibfnamefont {T.~E. H.~T.}\
  \bibnamefont {Collaboration}},\ }\href {\doibase 10.3847/2041-8213/ab0f43}
  {\bibfield  {journal} {\bibinfo  {journal} {The Astrophysical Journal
  Letters}\ }\textbf {\bibinfo {volume} {875}},\ \bibinfo {eid} {L5} (\bibinfo
  {year} {2019}{\natexlab{e}})},\ \Eprint {http://arxiv.org/abs/1906.11242}
  {arXiv:1906.11242 [astro-ph.GA]} \BibitemShut {NoStop}%
\bibitem [{\citenamefont {Collaboration}(2019{\natexlab{f}})}]{EHT_M87_6}%
  \BibitemOpen
  \bibfield  {author} {\bibinfo {author} {\bibfnamefont {T.~E. H.~T.}\
  \bibnamefont {Collaboration}},\ }\href {\doibase 10.3847/2041-8213/ab1141}
  {\bibfield  {journal} {\bibinfo  {journal} {The Astrophysical Journal
  Letters}\ }\textbf {\bibinfo {volume} {875}},\ \bibinfo {eid} {L6} (\bibinfo
  {year} {2019}{\natexlab{f}})},\ \Eprint {http://arxiv.org/abs/1906.11243}
  {arXiv:1906.11243 [astro-ph.GA]} \BibitemShut {NoStop}%
\bibitem [{\citenamefont {Collaboration}(2022{\natexlab{a}})}]{EHT_Sgr1}%
  \BibitemOpen
  \bibfield  {author} {\bibinfo {author} {\bibfnamefont {T.~E. H.~T.}\
  \bibnamefont {Collaboration}},\ }\href {\doibase 10.3847/2041-8213/ac6674}
  {\bibfield  {journal} {\bibinfo  {journal} {The Astrophysical Journal
  Letters}\ }\textbf {\bibinfo {volume} {930}},\ \bibinfo {eid} {L12} (\bibinfo
  {year} {2022}{\natexlab{a}})}\BibitemShut {NoStop}%
\bibitem [{\citenamefont {Collaboration}(2022{\natexlab{b}})}]{EHT_Sgr2}%
  \BibitemOpen
  \bibfield  {author} {\bibinfo {author} {\bibfnamefont {T.~E. H.~T.}\
  \bibnamefont {Collaboration}},\ }\href {\doibase 10.3847/2041-8213/ac6675}
  {\bibfield  {journal} {\bibinfo  {journal} {The Astrophysical Journal
  Letters}\ }\textbf {\bibinfo {volume} {930}},\ \bibinfo {eid} {L13} (\bibinfo
  {year} {2022}{\natexlab{b}})}\BibitemShut {NoStop}%
\bibitem [{\citenamefont {Collaboration}(2022{\natexlab{c}})}]{EHT_Sgr3}%
  \BibitemOpen
  \bibfield  {author} {\bibinfo {author} {\bibfnamefont {T.~E. H.~T.}\
  \bibnamefont {Collaboration}},\ }\href {\doibase 10.3847/2041-8213/ac6429}
  {\bibfield  {journal} {\bibinfo  {journal} {The Astrophysical Journal
  Letters}\ }\textbf {\bibinfo {volume} {930}},\ \bibinfo {eid} {L14} (\bibinfo
  {year} {2022}{\natexlab{c}})}\BibitemShut {NoStop}%
\bibitem [{\citenamefont {Collaboration}(2022{\natexlab{d}})}]{EHT_Sgr4}%
  \BibitemOpen
  \bibfield  {author} {\bibinfo {author} {\bibfnamefont {T.~E. H.~T.}\
  \bibnamefont {Collaboration}},\ }\href {\doibase 10.3847/2041-8213/ac6736}
  {\bibfield  {journal} {\bibinfo  {journal} {The Astrophysical Journal
  Letters}\ }\textbf {\bibinfo {volume} {930}},\ \bibinfo {eid} {L15} (\bibinfo
  {year} {2022}{\natexlab{d}})}\BibitemShut {NoStop}%
\bibitem [{\citenamefont {Collaboration}(2022{\natexlab{e}})}]{EHT_Sgr5}%
  \BibitemOpen
  \bibfield  {author} {\bibinfo {author} {\bibfnamefont {T.~E. H.~T.}\
  \bibnamefont {Collaboration}},\ }\href {\doibase 10.3847/2041-8213/ac6672}
  {\bibfield  {journal} {\bibinfo  {journal} {The Astrophysical Journal
  Letters}\ }\textbf {\bibinfo {volume} {930}},\ \bibinfo {eid} {L16} (\bibinfo
  {year} {2022}{\natexlab{e}})}\BibitemShut {NoStop}%
\bibitem [{\citenamefont {Collaboration}(2022{\natexlab{f}})}]{EHT_Sgr6}%
  \BibitemOpen
  \bibfield  {author} {\bibinfo {author} {\bibfnamefont {T.~E. H.~T.}\
  \bibnamefont {Collaboration}},\ }\href {\doibase 10.3847/2041-8213/ac6756}
  {\bibfield  {journal} {\bibinfo  {journal} {The Astrophysical Journal
  Letters}\ }\textbf {\bibinfo {volume} {930}},\ \bibinfo {eid} {L17} (\bibinfo
  {year} {2022}{\natexlab{f}})}\BibitemShut {NoStop}%
\bibitem [{\citenamefont {Collaboration}(2024)}]{EHT_M87_7}%
  \BibitemOpen
  \bibfield  {author} {\bibinfo {author} {\bibfnamefont {T.~E. H.~T.}\
  \bibnamefont {Collaboration}},\ }\href {\doibase 10.1051/0004-6361/202347932}
  {\bibfield  {journal} {\bibinfo  {journal} {{A\&A}}\ }\textbf {\bibinfo
  {volume} {681}},\ \bibinfo {eid} {A79} (\bibinfo {year} {2024})}\BibitemShut
  {NoStop}%
\bibitem [{\citenamefont {Glampedakis}\ \emph {et~al.}(2017)\citenamefont
  {Glampedakis}, \citenamefont {Pappas}, \citenamefont {Silva},\ and\
  \citenamefont {Berti}}]{Glampedakis:2017dvb}%
  \BibitemOpen
  \bibfield  {author} {\bibinfo {author} {\bibfnamefont {K.}~\bibnamefont
  {Glampedakis}}, \bibinfo {author} {\bibfnamefont {G.}~\bibnamefont {Pappas}},
  \bibinfo {author} {\bibfnamefont {H.~O.}\ \bibnamefont {Silva}}, \ and\
  \bibinfo {author} {\bibfnamefont {E.}~\bibnamefont {Berti}},\ }\href
  {\doibase 10.1103/PhysRevD.96.064054} {\bibfield  {journal} {\bibinfo
  {journal} {Phys. Rev. D}\ }\textbf {\bibinfo {volume} {96}},\ \bibinfo
  {pages} {064054} (\bibinfo {year} {2017})},\ \Eprint
  {http://arxiv.org/abs/1706.07658} {arXiv:1706.07658 [gr-qc]} \BibitemShut
  {NoStop}%
\bibitem [{\citenamefont {Glampedakis}\ and\ \citenamefont
  {Pappas}(2018)}]{Glampedakis_2018}%
  \BibitemOpen
  \bibfield  {author} {\bibinfo {author} {\bibfnamefont {K.}~\bibnamefont
  {Glampedakis}}\ and\ \bibinfo {author} {\bibfnamefont {G.}~\bibnamefont
  {Pappas}},\ }\href {\doibase 10.1103/physrevd.97.041502} {\bibfield
  {journal} {\bibinfo  {journal} {Physical Review D}\ }\textbf {\bibinfo
  {volume} {97}} (\bibinfo {year} {2018}),\
  10.1103/physrevd.97.041502}\BibitemShut {NoStop}%
\bibitem [{\citenamefont {Glampedakis}\ and\ \citenamefont
  {Silva}(2019)}]{Glampedakis:2019dqh}%
  \BibitemOpen
  \bibfield  {author} {\bibinfo {author} {\bibfnamefont {K.}~\bibnamefont
  {Glampedakis}}\ and\ \bibinfo {author} {\bibfnamefont {H.~O.}\ \bibnamefont
  {Silva}},\ }\href {\doibase 10.1103/PhysRevD.100.044040} {\bibfield
  {journal} {\bibinfo  {journal} {Phys. Rev. D}\ }\textbf {\bibinfo {volume}
  {100}},\ \bibinfo {pages} {044040} (\bibinfo {year} {2019})},\ \Eprint
  {http://arxiv.org/abs/1906.05455} {arXiv:1906.05455 [gr-qc]} \BibitemShut
  {NoStop}%
\bibitem [{\citenamefont {Cardoso}\ \emph {et~al.}(2019)\citenamefont
  {Cardoso}, \citenamefont {Kimura}, \citenamefont {Maselli}, \citenamefont
  {Berti}, \citenamefont {Macedo},\ and\ \citenamefont
  {McManus}}]{Cardoso:2019mqo}%
  \BibitemOpen
  \bibfield  {author} {\bibinfo {author} {\bibfnamefont {V.}~\bibnamefont
  {Cardoso}}, \bibinfo {author} {\bibfnamefont {M.}~\bibnamefont {Kimura}},
  \bibinfo {author} {\bibfnamefont {A.}~\bibnamefont {Maselli}}, \bibinfo
  {author} {\bibfnamefont {E.}~\bibnamefont {Berti}}, \bibinfo {author}
  {\bibfnamefont {C.~F.~B.}\ \bibnamefont {Macedo}}, \ and\ \bibinfo {author}
  {\bibfnamefont {R.}~\bibnamefont {McManus}},\ }\href {\doibase
  10.1103/PhysRevD.99.104077} {\bibfield  {journal} {\bibinfo  {journal} {Phys.
  Rev. D}\ }\textbf {\bibinfo {volume} {99}},\ \bibinfo {pages} {104077}
  (\bibinfo {year} {2019})},\ \Eprint {http://arxiv.org/abs/1901.01265}
  {arXiv:1901.01265 [gr-qc]} \BibitemShut {NoStop}%
\bibitem [{\citenamefont {McManus}\ \emph {et~al.}(2019)\citenamefont
  {McManus}, \citenamefont {Berti}, \citenamefont {Macedo}, \citenamefont
  {Kimura}, \citenamefont {Maselli},\ and\ \citenamefont
  {Cardoso}}]{McManus:2019ulj}%
  \BibitemOpen
  \bibfield  {author} {\bibinfo {author} {\bibfnamefont {R.}~\bibnamefont
  {McManus}}, \bibinfo {author} {\bibfnamefont {E.}~\bibnamefont {Berti}},
  \bibinfo {author} {\bibfnamefont {C.~F.~B.}\ \bibnamefont {Macedo}}, \bibinfo
  {author} {\bibfnamefont {M.}~\bibnamefont {Kimura}}, \bibinfo {author}
  {\bibfnamefont {A.}~\bibnamefont {Maselli}}, \ and\ \bibinfo {author}
  {\bibfnamefont {V.}~\bibnamefont {Cardoso}},\ }\href {\doibase
  10.1103/PhysRevD.100.044061} {\bibfield  {journal} {\bibinfo  {journal}
  {Phys. Rev. D}\ }\textbf {\bibinfo {volume} {100}},\ \bibinfo {pages}
  {044061} (\bibinfo {year} {2019})},\ \Eprint
  {http://arxiv.org/abs/1906.05155} {arXiv:1906.05155 [gr-qc]} \BibitemShut
  {NoStop}%
\bibitem [{\citenamefont {Silva}\ and\ \citenamefont
  {Glampedakis}(2020)}]{Silva:2019scu}%
  \BibitemOpen
  \bibfield  {author} {\bibinfo {author} {\bibfnamefont {H.~O.}\ \bibnamefont
  {Silva}}\ and\ \bibinfo {author} {\bibfnamefont {K.}~\bibnamefont
  {Glampedakis}},\ }\href {\doibase 10.1103/PhysRevD.101.044051} {\bibfield
  {journal} {\bibinfo  {journal} {Phys. Rev. D}\ }\textbf {\bibinfo {volume}
  {101}},\ \bibinfo {pages} {044051} (\bibinfo {year} {2020})},\ \Eprint
  {http://arxiv.org/abs/1912.09286} {arXiv:1912.09286 [gr-qc]} \BibitemShut
  {NoStop}%
\bibitem [{\citenamefont {V\"olkel}\ \emph {et~al.}(2021)\citenamefont
  {V\"olkel}, \citenamefont {Barausse}, \citenamefont {Franchini},\ and\
  \citenamefont {Broderick}}]{Volkel:2020xlc}%
  \BibitemOpen
  \bibfield  {author} {\bibinfo {author} {\bibfnamefont {S.~H.}\ \bibnamefont
  {V\"olkel}}, \bibinfo {author} {\bibfnamefont {E.}~\bibnamefont {Barausse}},
  \bibinfo {author} {\bibfnamefont {N.}~\bibnamefont {Franchini}}, \ and\
  \bibinfo {author} {\bibfnamefont {A.~E.}\ \bibnamefont {Broderick}},\ }\href
  {\doibase 10.1088/1361-6382/ac27ed} {\bibfield  {journal} {\bibinfo
  {journal} {Class. Quant. Grav.}\ }\textbf {\bibinfo {volume} {38}},\ \bibinfo
  {pages} {21LT01} (\bibinfo {year} {2021})},\ \Eprint
  {http://arxiv.org/abs/2011.06812} {arXiv:2011.06812 [gr-qc]} \BibitemShut
  {NoStop}%
\bibitem [{\citenamefont {Glampedakis}\ and\ \citenamefont
  {Pappas}(2021)}]{Glampedakis:2021oie}%
  \BibitemOpen
  \bibfield  {author} {\bibinfo {author} {\bibfnamefont {K.}~\bibnamefont
  {Glampedakis}}\ and\ \bibinfo {author} {\bibfnamefont {G.}~\bibnamefont
  {Pappas}},\ }\href {\doibase 10.1103/PhysRevD.104.L081503} {\bibfield
  {journal} {\bibinfo  {journal} {Phys. Rev. D}\ }\textbf {\bibinfo {volume}
  {104}},\ \bibinfo {pages} {L081503} (\bibinfo {year} {2021})},\ \Eprint
  {http://arxiv.org/abs/2102.13573} {arXiv:2102.13573 [gr-qc]} \BibitemShut
  {NoStop}%
\bibitem [{\citenamefont {Lima}\ \emph {et~al.}(2021)\citenamefont {Lima},
  \citenamefont {Crispino}, \citenamefont {Cunha},\ and\ \citenamefont
  {Herdeiro}}]{Lima:2021las}%
  \BibitemOpen
  \bibfield  {author} {\bibinfo {author} {\bibfnamefont {H.~C.~D.}\
  \bibnamefont {Lima}, \bibfnamefont {Junior.}}, \bibinfo {author}
  {\bibfnamefont {L.~C.~B.}\ \bibnamefont {Crispino}}, \bibinfo {author}
  {\bibfnamefont {P.~V.~P.}\ \bibnamefont {Cunha}}, \ and\ \bibinfo {author}
  {\bibfnamefont {C.~A.~R.}\ \bibnamefont {Herdeiro}},\ }\href {\doibase
  10.1103/PhysRevD.103.084040} {\bibfield  {journal} {\bibinfo  {journal}
  {Phys. Rev. D}\ }\textbf {\bibinfo {volume} {103}},\ \bibinfo {pages}
  {084040} (\bibinfo {year} {2021})},\ \Eprint
  {http://arxiv.org/abs/2102.07034} {arXiv:2102.07034 [gr-qc]} \BibitemShut
  {NoStop}%
\bibitem [{\citenamefont {Bryant}\ \emph {et~al.}(2021)\citenamefont {Bryant},
  \citenamefont {Silva}, \citenamefont {Yagi},\ and\ \citenamefont
  {Glampedakis}}]{Bryant:2021xdh}%
  \BibitemOpen
  \bibfield  {author} {\bibinfo {author} {\bibfnamefont {A.}~\bibnamefont
  {Bryant}}, \bibinfo {author} {\bibfnamefont {H.~O.}\ \bibnamefont {Silva}},
  \bibinfo {author} {\bibfnamefont {K.}~\bibnamefont {Yagi}}, \ and\ \bibinfo
  {author} {\bibfnamefont {K.}~\bibnamefont {Glampedakis}},\ }\href {\doibase
  10.1103/PhysRevD.104.044051} {\bibfield  {journal} {\bibinfo  {journal}
  {Phys. Rev. D}\ }\textbf {\bibinfo {volume} {104}},\ \bibinfo {pages}
  {044051} (\bibinfo {year} {2021})},\ \Eprint
  {http://arxiv.org/abs/2106.09657} {arXiv:2106.09657 [gr-qc]} \BibitemShut
  {NoStop}%
\bibitem [{\citenamefont {Konoplya}\ \emph {et~al.}(2020)\citenamefont
  {Konoplya}, \citenamefont {Schee},\ and\ \citenamefont
  {Ovchinnikov}}]{konoplya2020shadow}%
  \BibitemOpen
  \bibfield  {author} {\bibinfo {author} {\bibfnamefont {R.~A.}\ \bibnamefont
  {Konoplya}}, \bibinfo {author} {\bibfnamefont {J.}~\bibnamefont {Schee}}, \
  and\ \bibinfo {author} {\bibfnamefont {D.}~\bibnamefont {Ovchinnikov}},\
  }\href@noop {} {\  (\bibinfo {year} {2020})},\ \Eprint
  {http://arxiv.org/abs/2008.04118} {arXiv:2008.04118 [gr-qc]} \BibitemShut
  {NoStop}%
\bibitem [{\citenamefont {Guo}\ and\ \citenamefont {Li}(2020)}]{Guo_2020}%
  \BibitemOpen
  \bibfield  {author} {\bibinfo {author} {\bibfnamefont {M.}~\bibnamefont
  {Guo}}\ and\ \bibinfo {author} {\bibfnamefont {P.-C.}\ \bibnamefont {Li}},\
  }\href {\doibase 10.1140/epjc/s10052-020-8164-7} {\bibfield  {journal}
  {\bibinfo  {journal} {The European Physical Journal C}\ }\textbf {\bibinfo
  {volume} {80}} (\bibinfo {year} {2020}),\
  10.1140/epjc/s10052-020-8164-7}\BibitemShut {NoStop}%
\bibitem [{\citenamefont {Contreras}\ \emph {et~al.}(2020)\citenamefont
  {Contreras}, \citenamefont {Rinc\'on}, \citenamefont {Panotopoulos},
  \citenamefont {Bargue\~no},\ and\ \citenamefont {Koch}}]{Contreras_2020}%
  \BibitemOpen
  \bibfield  {author} {\bibinfo {author} {\bibfnamefont {E.}~\bibnamefont
  {Contreras}}, \bibinfo {author} {\bibfnamefont {A.}~\bibnamefont {Rinc\'on}},
  \bibinfo {author} {\bibfnamefont {G.}~\bibnamefont {Panotopoulos}}, \bibinfo
  {author} {\bibfnamefont {P.}~\bibnamefont {Bargue\~no}}, \ and\ \bibinfo
  {author} {\bibfnamefont {B.}~\bibnamefont {Koch}},\ }\href {\doibase
  10.1103/PhysRevD.101.064053} {\bibfield  {journal} {\bibinfo  {journal}
  {Phys. Rev. D}\ }\textbf {\bibinfo {volume} {101}},\ \bibinfo {pages}
  {064053} (\bibinfo {year} {2020})}\BibitemShut {NoStop}%
\bibitem [{\citenamefont {Wang}\ \emph {et~al.}(2019)\citenamefont {Wang},
  \citenamefont {Xu},\ and\ \citenamefont {Wei}}]{Wang_2019}%
  \BibitemOpen
  \bibfield  {author} {\bibinfo {author} {\bibfnamefont {H.-M.}\ \bibnamefont
  {Wang}}, \bibinfo {author} {\bibfnamefont {Y.-M.}\ \bibnamefont {Xu}}, \ and\
  \bibinfo {author} {\bibfnamefont {S.-W.}\ \bibnamefont {Wei}},\ }\href
  {\doibase 10.1088/1475-7516/2019/03/046} {\bibfield  {journal} {\bibinfo
  {journal} {Journal of Cosmology and Astroparticle Physics}\ }\textbf
  {\bibinfo {volume} {2019}},\ \bibinfo {pages} {046–046} (\bibinfo {year}
  {2019})}\BibitemShut {NoStop}%
\bibitem [{\citenamefont {Wang}\ \emph {et~al.}(2020)\citenamefont {Wang},
  \citenamefont {Chen}, \citenamefont {Wang},\ and\ \citenamefont
  {Jing}}]{Wang_2020}%
  \BibitemOpen
  \bibfield  {author} {\bibinfo {author} {\bibfnamefont {M.}~\bibnamefont
  {Wang}}, \bibinfo {author} {\bibfnamefont {S.}~\bibnamefont {Chen}}, \bibinfo
  {author} {\bibfnamefont {J.}~\bibnamefont {Wang}}, \ and\ \bibinfo {author}
  {\bibfnamefont {J.}~\bibnamefont {Jing}},\ }\href {\doibase
  10.1140/epjc/s10052-020-7641-3} {\bibfield  {journal} {\bibinfo  {journal}
  {The European Physical Journal C}\ }\textbf {\bibinfo {volume} {80}}
  (\bibinfo {year} {2020}),\ 10.1140/epjc/s10052-020-7641-3}\BibitemShut
  {NoStop}%
\bibitem [{\citenamefont {Ayzenberg}\ and\ \citenamefont
  {Yunes}(2018)}]{Ayzenberg_2018}%
  \BibitemOpen
  \bibfield  {author} {\bibinfo {author} {\bibfnamefont {D.}~\bibnamefont
  {Ayzenberg}}\ and\ \bibinfo {author} {\bibfnamefont {N.}~\bibnamefont
  {Yunes}},\ }\href {\doibase 10.1088/1361-6382/aae87b} {\bibfield  {journal}
  {\bibinfo  {journal} {Classical and Quantum Gravity}\ }\textbf {\bibinfo
  {volume} {35}},\ \bibinfo {pages} {235002} (\bibinfo {year}
  {2018})}\BibitemShut {NoStop}%
\bibitem [{\citenamefont {Cunha}\ \emph
  {et~al.}(2017{\natexlab{a}})\citenamefont {Cunha}, \citenamefont {Herdeiro},\
  and\ \citenamefont {Radu}}]{FPOs}%
  \BibitemOpen
  \bibfield  {author} {\bibinfo {author} {\bibfnamefont {P.~V.~P.}\
  \bibnamefont {Cunha}}, \bibinfo {author} {\bibfnamefont {C.~A.~R.}\
  \bibnamefont {Herdeiro}}, \ and\ \bibinfo {author} {\bibfnamefont
  {E.}~\bibnamefont {Radu}},\ }\href {\doibase 10.1103/PhysRevD.96.024039}
  {\bibfield  {journal} {\bibinfo  {journal} {Phys. Rev. D}\ }\textbf {\bibinfo
  {volume} {96}},\ \bibinfo {pages} {024039} (\bibinfo {year}
  {2017}{\natexlab{a}})}\BibitemShut {NoStop}%
\bibitem [{\citenamefont {{Cunha}}\ \emph
  {et~al.}(2016{\natexlab{a}})\citenamefont {{Cunha}}, \citenamefont
  {{Herdeiro}}, \citenamefont {{Radu}},\ and\ \citenamefont
  {{R{\'u}narsson}}}]{Cunha2016IJMPD}%
  \BibitemOpen
  \bibfield  {author} {\bibinfo {author} {\bibfnamefont {P.~V.~P.}\
  \bibnamefont {{Cunha}}}, \bibinfo {author} {\bibfnamefont {C.~A.~R.}\
  \bibnamefont {{Herdeiro}}}, \bibinfo {author} {\bibfnamefont
  {E.}~\bibnamefont {{Radu}}}, \ and\ \bibinfo {author} {\bibfnamefont {H.~F.}\
  \bibnamefont {{R{\'u}narsson}}},\ }\href {\doibase 10.1142/S0218271816410212}
  {\bibfield  {journal} {\bibinfo  {journal} {Int. J. Mod. Phys. D}\ }\textbf
  {\bibinfo {volume} {25}},\ \bibinfo {eid} {1641021} (\bibinfo {year}
  {2016}{\natexlab{a}})}\BibitemShut {NoStop}%
\bibitem [{\citenamefont {{Cunha}}\ and\ \citenamefont
  {{Herdeiro}}(2018)}]{Cunha2018GReGr}%
  \BibitemOpen
  \bibfield  {author} {\bibinfo {author} {\bibfnamefont {P.~V.~P.}\
  \bibnamefont {{Cunha}}}\ and\ \bibinfo {author} {\bibfnamefont {C.~A.~R.}\
  \bibnamefont {{Herdeiro}}},\ }\href {\doibase 10.1007/s10714-018-2361-9}
  {\bibfield  {journal} {\bibinfo  {journal} {Gen. Relativ. Gravit.}\ }\textbf
  {\bibinfo {volume} {50}},\ \bibinfo {eid} {42} (\bibinfo {year}
  {2018})}\BibitemShut {NoStop}%
\bibitem [{\citenamefont {Kostaros}\ and\ \citenamefont
  {Pappas}(2022)}]{Kostaros_2022}%
  \BibitemOpen
  \bibfield  {author} {\bibinfo {author} {\bibfnamefont {K.}~\bibnamefont
  {Kostaros}}\ and\ \bibinfo {author} {\bibfnamefont {G.}~\bibnamefont
  {Pappas}},\ }\href {\doibase 10.1088/1361-6382/ac7028} {\bibfield  {journal}
  {\bibinfo  {journal} {Classical and Quantum Gravity}\ }\textbf {\bibinfo
  {volume} {39}},\ \bibinfo {pages} {134001} (\bibinfo {year}
  {2022})}\BibitemShut {NoStop}%
\bibitem [{\citenamefont {Glampedakis}\ and\ \citenamefont
  {Pappas}(2023)}]{Glampedakis:2023eek}%
  \BibitemOpen
  \bibfield  {author} {\bibinfo {author} {\bibfnamefont {K.}~\bibnamefont
  {Glampedakis}}\ and\ \bibinfo {author} {\bibfnamefont {G.}~\bibnamefont
  {Pappas}},\ }\href {\doibase 10.1103/PhysRevD.107.064001} {\bibfield
  {journal} {\bibinfo  {journal} {Phys. Rev. D}\ }\textbf {\bibinfo {volume}
  {107}},\ \bibinfo {pages} {064001} (\bibinfo {year} {2023})},\ \Eprint
  {http://arxiv.org/abs/2302.06140} {arXiv:2302.06140 [gr-qc]} \BibitemShut
  {NoStop}%
\bibitem [{\citenamefont {Gralla}\ \emph {et~al.}(2020)\citenamefont {Gralla},
  \citenamefont {Lupsasca},\ and\ \citenamefont {Marrone}}]{Gralla_2020}%
  \BibitemOpen
  \bibfield  {author} {\bibinfo {author} {\bibfnamefont {S.~E.}\ \bibnamefont
  {Gralla}}, \bibinfo {author} {\bibfnamefont {A.}~\bibnamefont {Lupsasca}}, \
  and\ \bibinfo {author} {\bibfnamefont {D.~P.}\ \bibnamefont {Marrone}},\
  }\href {\doibase 10.1103/physrevd.102.124004} {\bibfield  {journal} {\bibinfo
   {journal} {Physical Review D}\ }\textbf {\bibinfo {volume} {102}} (\bibinfo
  {year} {2020}),\ 10.1103/physrevd.102.124004}\BibitemShut {NoStop}%
\bibitem [{\citenamefont {Younsi}\ \emph {et~al.}(2023)\citenamefont {Younsi},
  \citenamefont {Psaltis},\ and\ \citenamefont {\"Ozel}}]{Younsi:2021dxe}%
  \BibitemOpen
  \bibfield  {author} {\bibinfo {author} {\bibfnamefont {Z.}~\bibnamefont
  {Younsi}}, \bibinfo {author} {\bibfnamefont {D.}~\bibnamefont {Psaltis}}, \
  and\ \bibinfo {author} {\bibfnamefont {F.}~\bibnamefont {\"Ozel}},\ }\href
  {\doibase 10.3847/1538-4357/aca58a} {\bibfield  {journal} {\bibinfo
  {journal} {Astrophys. J.}\ }\textbf {\bibinfo {volume} {942}},\ \bibinfo
  {pages} {47} (\bibinfo {year} {2023})},\ \Eprint
  {http://arxiv.org/abs/2111.01752} {arXiv:2111.01752 [astro-ph.HE]}
  \BibitemShut {NoStop}%
\bibitem [{\citenamefont {Bauer}\ \emph {et~al.}(2022)\citenamefont {Bauer},
  \citenamefont {C\'ardenas-Avenda\~no}, \citenamefont {Gammie},\ and\
  \citenamefont {Yunes}}]{Bauer:2021atk}%
  \BibitemOpen
  \bibfield  {author} {\bibinfo {author} {\bibfnamefont {A.~M.}\ \bibnamefont
  {Bauer}}, \bibinfo {author} {\bibfnamefont {A.}~\bibnamefont
  {C\'ardenas-Avenda\~no}}, \bibinfo {author} {\bibfnamefont {C.~F.}\
  \bibnamefont {Gammie}}, \ and\ \bibinfo {author} {\bibfnamefont
  {N.}~\bibnamefont {Yunes}},\ }\href {\doibase 10.3847/1538-4357/ac3a03}
  {\bibfield  {journal} {\bibinfo  {journal} {Astrophys. J.}\ }\textbf
  {\bibinfo {volume} {925}},\ \bibinfo {pages} {119} (\bibinfo {year}
  {2022})},\ \Eprint {http://arxiv.org/abs/2111.02178} {arXiv:2111.02178
  [gr-qc]} \BibitemShut {NoStop}%
\bibitem [{\citenamefont {Gralla}(2020)}]{Gralla:2020nwp}%
  \BibitemOpen
  \bibfield  {author} {\bibinfo {author} {\bibfnamefont {S.~E.}\ \bibnamefont
  {Gralla}},\ }\href {\doibase 10.1103/PhysRevD.102.044017} {\bibfield
  {journal} {\bibinfo  {journal} {Phys. Rev. D}\ }\textbf {\bibinfo {volume}
  {102}},\ \bibinfo {pages} {044017} (\bibinfo {year} {2020})},\ \Eprint
  {http://arxiv.org/abs/2005.03856} {arXiv:2005.03856 [astro-ph.HE]}
  \BibitemShut {NoStop}%
\bibitem [{\citenamefont {Medeiros}\ \emph {et~al.}(2020)\citenamefont
  {Medeiros}, \citenamefont {Psaltis},\ and\ \citenamefont
  {\"Ozel}}]{Medeiros:2019cde}%
  \BibitemOpen
  \bibfield  {author} {\bibinfo {author} {\bibfnamefont {L.}~\bibnamefont
  {Medeiros}}, \bibinfo {author} {\bibfnamefont {D.}~\bibnamefont {Psaltis}}, \
  and\ \bibinfo {author} {\bibfnamefont {F.}~\bibnamefont {\"Ozel}},\ }\href
  {\doibase 10.3847/1538-4357/ab8bd1} {\bibfield  {journal} {\bibinfo
  {journal} {Astrophys. J.}\ }\textbf {\bibinfo {volume} {896}},\ \bibinfo
  {pages} {7} (\bibinfo {year} {2020})}\BibitemShut {NoStop}%
\bibitem [{\citenamefont {Gralla}\ \emph {et~al.}(2019)\citenamefont {Gralla},
  \citenamefont {Holz},\ and\ \citenamefont {Wald}}]{Gralla:2019xty}%
  \BibitemOpen
  \bibfield  {author} {\bibinfo {author} {\bibfnamefont {S.~E.}\ \bibnamefont
  {Gralla}}, \bibinfo {author} {\bibfnamefont {D.~E.}\ \bibnamefont {Holz}}, \
  and\ \bibinfo {author} {\bibfnamefont {R.~M.}\ \bibnamefont {Wald}},\ }\href
  {\doibase 10.1103/PhysRevD.100.024018} {\bibfield  {journal} {\bibinfo
  {journal} {Phys. Rev. D}\ }\textbf {\bibinfo {volume} {100}},\ \bibinfo
  {pages} {024018} (\bibinfo {year} {2019})},\ \Eprint
  {http://arxiv.org/abs/1906.00873} {arXiv:1906.00873 [astro-ph.HE]}
  \BibitemShut {NoStop}%
\bibitem [{\citenamefont {Cunha}\ \emph
  {et~al.}(2017{\natexlab{b}})\citenamefont {Cunha}, \citenamefont {Berti},\
  and\ \citenamefont {Herdeiro}}]{Cunha:2017qtt}%
  \BibitemOpen
  \bibfield  {author} {\bibinfo {author} {\bibfnamefont {P.~V.~P.}\
  \bibnamefont {Cunha}}, \bibinfo {author} {\bibfnamefont {E.}~\bibnamefont
  {Berti}}, \ and\ \bibinfo {author} {\bibfnamefont {C.~A.~R.}\ \bibnamefont
  {Herdeiro}},\ }\href {\doibase 10.1103/PhysRevLett.119.251102} {\bibfield
  {journal} {\bibinfo  {journal} {Phys. Rev. Lett.}\ }\textbf {\bibinfo
  {volume} {119}},\ \bibinfo {pages} {251102} (\bibinfo {year}
  {2017}{\natexlab{b}})},\ \Eprint {http://arxiv.org/abs/1708.04211}
  {arXiv:1708.04211 [gr-qc]} \BibitemShut {NoStop}%
\bibitem [{\citenamefont {Cunha}\ \emph
  {et~al.}(2017{\natexlab{c}})\citenamefont {Cunha}, \citenamefont {Herdeiro},\
  and\ \citenamefont {Radu}}]{Cunha2017PhysRevD}%
  \BibitemOpen
  \bibfield  {author} {\bibinfo {author} {\bibfnamefont {P.~V.~P.}\
  \bibnamefont {Cunha}}, \bibinfo {author} {\bibfnamefont {C.~A.~R.}\
  \bibnamefont {Herdeiro}}, \ and\ \bibinfo {author} {\bibfnamefont
  {E.}~\bibnamefont {Radu}},\ }\href {\doibase 10.1103/PhysRevD.96.024039}
  {\bibfield  {journal} {\bibinfo  {journal} {Phys. Rev. D}\ }\textbf {\bibinfo
  {volume} {96}},\ \bibinfo {pages} {024039} (\bibinfo {year}
  {2017}{\natexlab{c}})}\BibitemShut {NoStop}%
\bibitem [{\citenamefont {Johannsen}\ and\ \citenamefont
  {Psaltis}(2010)}]{Johannsen_2010}%
  \BibitemOpen
  \bibfield  {author} {\bibinfo {author} {\bibfnamefont {T.}~\bibnamefont
  {Johannsen}}\ and\ \bibinfo {author} {\bibfnamefont {D.}~\bibnamefont
  {Psaltis}},\ }\href {\doibase 10.1088/0004-637x/718/1/446} {\bibfield
  {journal} {\bibinfo  {journal} {The Astrophysical Journal}\ }\textbf
  {\bibinfo {volume} {718}},\ \bibinfo {pages} {446} (\bibinfo {year}
  {2010})}\BibitemShut {NoStop}%
\bibitem [{\citenamefont {Patil}\ \emph {et~al.}(2017)\citenamefont {Patil},
  \citenamefont {Mishra},\ and\ \citenamefont {Narasimha}}]{Patil_2017}%
  \BibitemOpen
  \bibfield  {author} {\bibinfo {author} {\bibfnamefont {M.}~\bibnamefont
  {Patil}}, \bibinfo {author} {\bibfnamefont {P.}~\bibnamefont {Mishra}}, \
  and\ \bibinfo {author} {\bibfnamefont {D.}~\bibnamefont {Narasimha}},\ }\href
  {\doibase 10.1103/physrevd.95.024026} {\bibfield  {journal} {\bibinfo
  {journal} {Physical Review D}\ }\textbf {\bibinfo {volume} {95}} (\bibinfo
  {year} {2017}),\ 10.1103/physrevd.95.024026}\BibitemShut {NoStop}%
\bibitem [{\citenamefont {Kocherlakota}\ \emph
  {et~al.}(2024{\natexlab{a}})\citenamefont {Kocherlakota}, \citenamefont
  {Rezzolla}, \citenamefont {Roy},\ and\ \citenamefont
  {Wielgus}}]{Kocherlakota:2023qgo}%
  \BibitemOpen
  \bibfield  {author} {\bibinfo {author} {\bibfnamefont {P.}~\bibnamefont
  {Kocherlakota}}, \bibinfo {author} {\bibfnamefont {L.}~\bibnamefont
  {Rezzolla}}, \bibinfo {author} {\bibfnamefont {R.}~\bibnamefont {Roy}}, \
  and\ \bibinfo {author} {\bibfnamefont {M.}~\bibnamefont {Wielgus}},\ }\href
  {\doibase 10.1103/PhysRevD.109.064064} {\bibfield  {journal} {\bibinfo
  {journal} {Phys. Rev. D}\ }\textbf {\bibinfo {volume} {109}},\ \bibinfo
  {pages} {064064} (\bibinfo {year} {2024}{\natexlab{a}})},\ \Eprint
  {http://arxiv.org/abs/2307.16841} {arXiv:2307.16841 [gr-qc]} \BibitemShut
  {NoStop}%
\bibitem [{\citenamefont {Kocherlakota}\ \emph
  {et~al.}(2024{\natexlab{b}})\citenamefont {Kocherlakota}, \citenamefont
  {Rezzolla}, \citenamefont {Roy},\ and\ \citenamefont
  {Wielgus}}]{Kocherlakota:2024hyq}%
  \BibitemOpen
  \bibfield  {author} {\bibinfo {author} {\bibfnamefont {P.}~\bibnamefont
  {Kocherlakota}}, \bibinfo {author} {\bibfnamefont {L.}~\bibnamefont
  {Rezzolla}}, \bibinfo {author} {\bibfnamefont {R.}~\bibnamefont {Roy}}, \
  and\ \bibinfo {author} {\bibfnamefont {M.}~\bibnamefont {Wielgus}},\
  }\href@noop {} {\  (\bibinfo {year} {2024}{\natexlab{b}})},\ \Eprint
  {http://arxiv.org/abs/2403.08862} {arXiv:2403.08862 [astro-ph.HE]}
  \BibitemShut {NoStop}%
\bibitem [{\citenamefont {Kocherlakota}\ and\ \citenamefont
  {Rezzolla}(2022)}]{Kocherlakota:2022jnz}%
  \BibitemOpen
  \bibfield  {author} {\bibinfo {author} {\bibfnamefont {P.}~\bibnamefont
  {Kocherlakota}}\ and\ \bibinfo {author} {\bibfnamefont {L.}~\bibnamefont
  {Rezzolla}},\ }\href {\doibase 10.1093/mnras/stac891} {\bibfield  {journal}
  {\bibinfo  {journal} {Mon. Not. Roy. Astron. Soc.}\ }\textbf {\bibinfo
  {volume} {513}},\ \bibinfo {pages} {1229} (\bibinfo {year} {2022})},\ \Eprint
  {http://arxiv.org/abs/2201.05641} {arXiv:2201.05641 [gr-qc]} \BibitemShut
  {NoStop}%
\bibitem [{\citenamefont {Manko}\ and\ \citenamefont
  {Novikov}(1992)}]{Manko_1992}%
  \BibitemOpen
  \bibfield  {author} {\bibinfo {author} {\bibfnamefont {V.~S.}\ \bibnamefont
  {Manko}}\ and\ \bibinfo {author} {\bibfnamefont {I.~D.}\ \bibnamefont
  {Novikov}},\ }\href {\doibase 10.1088/0264-9381/9/11/013} {\bibfield
  {journal} {\bibinfo  {journal} {Classical and Quantum Gravity}\ }\textbf
  {\bibinfo {volume} {9}},\ \bibinfo {pages} {2477} (\bibinfo {year}
  {1992})}\BibitemShut {NoStop}%
\bibitem [{\citenamefont {Friedberg}\ \emph {et~al.}(1987)\citenamefont
  {Friedberg}, \citenamefont {Lee},\ and\ \citenamefont
  {Pang}}]{friedberg1987}%
  \BibitemOpen
  \bibfield  {author} {\bibinfo {author} {\bibfnamefont {R.}~\bibnamefont
  {Friedberg}}, \bibinfo {author} {\bibfnamefont {T.~D.}\ \bibnamefont {Lee}},
  \ and\ \bibinfo {author} {\bibfnamefont {Y.}~\bibnamefont {Pang}},\ }\href
  {\doibase 10.1103/PhysRevD.35.3640} {\bibfield  {journal} {\bibinfo
  {journal} {Phys. Rev. D}\ }\textbf {\bibinfo {volume} {35}},\ \bibinfo
  {pages} {3640} (\bibinfo {year} {1987})}\BibitemShut {NoStop}%
\bibitem [{\citenamefont {Mazur}\ and\ \citenamefont
  {Mottola}(2023)}]{Mazur:2001fv}%
  \BibitemOpen
  \bibfield  {author} {\bibinfo {author} {\bibfnamefont {P.~O.}\ \bibnamefont
  {Mazur}}\ and\ \bibinfo {author} {\bibfnamefont {E.}~\bibnamefont
  {Mottola}},\ }\href {\doibase 10.3390/universe9020088} {\bibfield  {journal}
  {\bibinfo  {journal} {Universe}\ }\textbf {\bibinfo {volume} {9}},\ \bibinfo
  {pages} {88} (\bibinfo {year} {2023})},\ \Eprint
  {http://arxiv.org/abs/gr-qc/0109035} {arXiv:gr-qc/0109035} \BibitemShut
  {NoStop}%
\bibitem [{\citenamefont {Barcel\'o}\ \emph {et~al.}(2008)\citenamefont
  {Barcel\'o}, \citenamefont {Liberati}, \citenamefont {Sonego},\ and\
  \citenamefont {Visser}}]{barcelo2008}%
  \BibitemOpen
  \bibfield  {author} {\bibinfo {author} {\bibfnamefont {C.}~\bibnamefont
  {Barcel\'o}}, \bibinfo {author} {\bibfnamefont {S.}~\bibnamefont {Liberati}},
  \bibinfo {author} {\bibfnamefont {S.}~\bibnamefont {Sonego}}, \ and\ \bibinfo
  {author} {\bibfnamefont {M.}~\bibnamefont {Visser}},\ }\href {\doibase
  10.1103/PhysRevD.77.044032} {\bibfield  {journal} {\bibinfo  {journal} {Phys.
  Rev. D}\ }\textbf {\bibinfo {volume} {77}},\ \bibinfo {pages} {044032}
  (\bibinfo {year} {2008})}\BibitemShut {NoStop}%
\bibitem [{\citenamefont {Yagi}\ and\ \citenamefont
  {Yunes}(2016)}]{Yagi:2016ejg}%
  \BibitemOpen
  \bibfield  {author} {\bibinfo {author} {\bibfnamefont {K.}~\bibnamefont
  {Yagi}}\ and\ \bibinfo {author} {\bibfnamefont {N.}~\bibnamefont {Yunes}},\
  }\href {\doibase 10.1088/0264-9381/33/9/095005} {\bibfield  {journal}
  {\bibinfo  {journal} {Class. Quant. Grav.}\ }\textbf {\bibinfo {volume}
  {33}},\ \bibinfo {pages} {095005} (\bibinfo {year} {2016})},\ \Eprint
  {http://arxiv.org/abs/1601.02171} {arXiv:1601.02171 [gr-qc]} \BibitemShut
  {NoStop}%
\bibitem [{\citenamefont {Psaltis}\ \emph {et~al.}(2008)\citenamefont
  {Psaltis}, \citenamefont {Perrodin}, \citenamefont {Dienes},\ and\
  \citenamefont {Mocioiu}}]{psaltis2008}%
  \BibitemOpen
  \bibfield  {author} {\bibinfo {author} {\bibfnamefont {D.}~\bibnamefont
  {Psaltis}}, \bibinfo {author} {\bibfnamefont {D.}~\bibnamefont {Perrodin}},
  \bibinfo {author} {\bibfnamefont {K.~R.}\ \bibnamefont {Dienes}}, \ and\
  \bibinfo {author} {\bibfnamefont {I.}~\bibnamefont {Mocioiu}},\ }\href
  {\doibase 10.1103/PhysRevLett.100.091101} {\bibfield  {journal} {\bibinfo
  {journal} {Phys. Rev. Lett.}\ }\textbf {\bibinfo {volume} {100}},\ \bibinfo
  {pages} {091101} (\bibinfo {year} {2008})}\BibitemShut {NoStop}%
\bibitem [{\citenamefont {Herdeiro}\ and\ \citenamefont
  {Radu}(2015)}]{Herdeiro:2015waa}%
  \BibitemOpen
  \bibfield  {author} {\bibinfo {author} {\bibfnamefont {C.~A.~R.}\
  \bibnamefont {Herdeiro}}\ and\ \bibinfo {author} {\bibfnamefont
  {E.}~\bibnamefont {Radu}},\ }\href {\doibase 10.1142/S0218271815420146}
  {\bibfield  {journal} {\bibinfo  {journal} {Int. J. Mod. Phys. D}\ }\textbf
  {\bibinfo {volume} {24}},\ \bibinfo {pages} {1542014} (\bibinfo {year}
  {2015})},\ \Eprint {http://arxiv.org/abs/1504.08209} {arXiv:1504.08209
  [gr-qc]} \BibitemShut {NoStop}%
\bibitem [{\citenamefont {Eichhorn}\ and\ \citenamefont
  {Held}(2021{\natexlab{a}})}]{Eichhorn_2021}%
  \BibitemOpen
  \bibfield  {author} {\bibinfo {author} {\bibfnamefont {A.}~\bibnamefont
  {Eichhorn}}\ and\ \bibinfo {author} {\bibfnamefont {A.}~\bibnamefont
  {Held}},\ }\href {\doibase 10.1088/1475-7516/2021/05/073} {\bibfield
  {journal} {\bibinfo  {journal} {Journal of Cosmology and Astroparticle
  Physics}\ }\textbf {\bibinfo {volume} {2021}},\ \bibinfo {pages} {073}
  (\bibinfo {year} {2021}{\natexlab{a}})}\BibitemShut {NoStop}%
\bibitem [{\citenamefont {Barack}\ \emph {et~al.}(2019)\citenamefont {Barack}
  \emph {et~al.}}]{Barack:2018yly}%
  \BibitemOpen
  \bibfield  {author} {\bibinfo {author} {\bibfnamefont {L.}~\bibnamefont
  {Barack}} \emph {et~al.},\ }\href {\doibase 10.1088/1361-6382/ab0587}
  {\bibfield  {journal} {\bibinfo  {journal} {Class. Quant. Grav.}\ }\textbf
  {\bibinfo {volume} {36}},\ \bibinfo {pages} {143001} (\bibinfo {year}
  {2019})},\ \Eprint {http://arxiv.org/abs/1806.05195} {arXiv:1806.05195
  [gr-qc]} \BibitemShut {NoStop}%
\bibitem [{\citenamefont {Berti}\ \emph {et~al.}(2015)\citenamefont {Berti}
  \emph {et~al.}}]{Berti:2015itd}%
  \BibitemOpen
  \bibfield  {author} {\bibinfo {author} {\bibfnamefont {E.}~\bibnamefont
  {Berti}} \emph {et~al.},\ }\href {\doibase 10.1088/0264-9381/32/24/243001}
  {\bibfield  {journal} {\bibinfo  {journal} {Class. Quant. Grav.}\ }\textbf
  {\bibinfo {volume} {32}},\ \bibinfo {pages} {243001} (\bibinfo {year}
  {2015})},\ \Eprint {http://arxiv.org/abs/1501.07274} {arXiv:1501.07274
  [gr-qc]} \BibitemShut {NoStop}%
\bibitem [{\citenamefont {Abbott}\ \emph {et~al.}(2016)\citenamefont {Abbott}
  \emph {et~al.}}]{LIGOScientific:2016lio}%
  \BibitemOpen
  \bibfield  {author} {\bibinfo {author} {\bibfnamefont {B.~P.}\ \bibnamefont
  {Abbott}} \emph {et~al.} (\bibinfo {collaboration} {LIGO Scientific,
  Virgo}),\ }\href {\doibase 10.1103/PhysRevLett.116.221101} {\bibfield
  {journal} {\bibinfo  {journal} {Phys. Rev. Lett.}\ }\textbf {\bibinfo
  {volume} {116}},\ \bibinfo {pages} {221101} (\bibinfo {year} {2016})},\
  \bibinfo {note} {[Erratum: Phys.Rev.Lett. 121, 129902 (2018)]},\ \Eprint
  {http://arxiv.org/abs/1602.03841} {arXiv:1602.03841 [gr-qc]} \BibitemShut
  {NoStop}%
\bibitem [{\citenamefont {Visser}\ and\ \citenamefont
  {Wiltshire}(2004)}]{Visser:2003ge}%
  \BibitemOpen
  \bibfield  {author} {\bibinfo {author} {\bibfnamefont {M.}~\bibnamefont
  {Visser}}\ and\ \bibinfo {author} {\bibfnamefont {D.~L.}\ \bibnamefont
  {Wiltshire}},\ }\href {\doibase 10.1088/0264-9381/21/4/027} {\bibfield
  {journal} {\bibinfo  {journal} {Class. Quant. Grav.}\ }\textbf {\bibinfo
  {volume} {21}},\ \bibinfo {pages} {1135} (\bibinfo {year} {2004})},\ \Eprint
  {http://arxiv.org/abs/gr-qc/0310107} {arXiv:gr-qc/0310107} \BibitemShut
  {NoStop}%
\bibitem [{\citenamefont {Eby}\ \emph {et~al.}(2016)\citenamefont {Eby},
  \citenamefont {Kouvaris}, \citenamefont {Nielsen},\ and\ \citenamefont
  {Wijewardhana}}]{Eby_2016}%
  \BibitemOpen
  \bibfield  {author} {\bibinfo {author} {\bibfnamefont {J.}~\bibnamefont
  {Eby}}, \bibinfo {author} {\bibfnamefont {C.}~\bibnamefont {Kouvaris}},
  \bibinfo {author} {\bibfnamefont {N.~G.}\ \bibnamefont {Nielsen}}, \ and\
  \bibinfo {author} {\bibfnamefont {L.~C.~R.}\ \bibnamefont {Wijewardhana}},\
  }\href {\doibase 10.1007/jhep02(2016)028} {\bibfield  {journal} {\bibinfo
  {journal} {Journal of High Energy Physics}\ }\textbf {\bibinfo {volume}
  {2016}} (\bibinfo {year} {2016}),\ 10.1007/jhep02(2016)028}\BibitemShut
  {NoStop}%
\bibitem [{\citenamefont {Schunck}\ and\ \citenamefont
  {Mielke}(2003)}]{Schunck:2003kk}%
  \BibitemOpen
  \bibfield  {author} {\bibinfo {author} {\bibfnamefont {F.~E.}\ \bibnamefont
  {Schunck}}\ and\ \bibinfo {author} {\bibfnamefont {E.~W.}\ \bibnamefont
  {Mielke}},\ }\href {\doibase 10.1088/0264-9381/20/20/201} {\bibfield
  {journal} {\bibinfo  {journal} {Class. Quant. Grav.}\ }\textbf {\bibinfo
  {volume} {20}},\ \bibinfo {pages} {R301} (\bibinfo {year} {2003})},\ \Eprint
  {http://arxiv.org/abs/0801.0307} {arXiv:0801.0307 [astro-ph]} \BibitemShut
  {NoStop}%
\bibitem [{\citenamefont {Cardoso}\ and\ \citenamefont
  {Pani}(2019)}]{darkobjcardoso}%
  \BibitemOpen
  \bibfield  {author} {\bibinfo {author} {\bibfnamefont {V.}~\bibnamefont
  {Cardoso}}\ and\ \bibinfo {author} {\bibfnamefont {P.}~\bibnamefont {Pani}},\
  }\href {\doibase 10.1007/s41114-019-0020-4} {\bibfield  {journal} {\bibinfo
  {journal} {Living Reviews in Relativity}\ }\textbf {\bibinfo {volume} {22}}
  (\bibinfo {year} {2019}),\ 10.1007/s41114-019-0020-4}\BibitemShut {NoStop}%
\bibitem [{\citenamefont {Glampedakis}\ and\ \citenamefont
  {Babak}(2006)}]{Glampedakis:2005cf}%
  \BibitemOpen
  \bibfield  {author} {\bibinfo {author} {\bibfnamefont {K.}~\bibnamefont
  {Glampedakis}}\ and\ \bibinfo {author} {\bibfnamefont {S.}~\bibnamefont
  {Babak}},\ }\href {\doibase 10.1088/0264-9381/23/12/013} {\bibfield
  {journal} {\bibinfo  {journal} {Class. Quant. Grav.}\ }\textbf {\bibinfo
  {volume} {23}},\ \bibinfo {pages} {4167} (\bibinfo {year} {2006})},\ \Eprint
  {http://arxiv.org/abs/gr-qc/0510057} {arXiv:gr-qc/0510057} \BibitemShut
  {NoStop}%
\bibitem [{\citenamefont {Collins}\ and\ \citenamefont
  {Hughes}(2004)}]{Collins:2004ex}%
  \BibitemOpen
  \bibfield  {author} {\bibinfo {author} {\bibfnamefont {N.~A.}\ \bibnamefont
  {Collins}}\ and\ \bibinfo {author} {\bibfnamefont {S.~A.}\ \bibnamefont
  {Hughes}},\ }\href {\doibase 10.1103/PhysRevD.69.124022} {\bibfield
  {journal} {\bibinfo  {journal} {Phys. Rev. D}\ }\textbf {\bibinfo {volume}
  {69}},\ \bibinfo {pages} {124022} (\bibinfo {year} {2004})},\ \Eprint
  {http://arxiv.org/abs/gr-qc/0402063} {arXiv:gr-qc/0402063} \BibitemShut
  {NoStop}%
\bibitem [{\citenamefont {Vigeland}\ \emph {et~al.}(2011)\citenamefont
  {Vigeland}, \citenamefont {Yunes},\ and\ \citenamefont
  {Stein}}]{Vigeland:2011ji}%
  \BibitemOpen
  \bibfield  {author} {\bibinfo {author} {\bibfnamefont {S.}~\bibnamefont
  {Vigeland}}, \bibinfo {author} {\bibfnamefont {N.}~\bibnamefont {Yunes}}, \
  and\ \bibinfo {author} {\bibfnamefont {L.}~\bibnamefont {Stein}},\ }\href
  {\doibase 10.1103/PhysRevD.83.104027} {\bibfield  {journal} {\bibinfo
  {journal} {Phys. Rev. D}\ }\textbf {\bibinfo {volume} {83}},\ \bibinfo
  {pages} {104027} (\bibinfo {year} {2011})},\ \Eprint
  {http://arxiv.org/abs/1102.3706} {arXiv:1102.3706 [gr-qc]} \BibitemShut
  {NoStop}%
\bibitem [{\citenamefont {Johannsen}\ and\ \citenamefont
  {Psaltis}(2011)}]{Johannsen:2011dh}%
  \BibitemOpen
  \bibfield  {author} {\bibinfo {author} {\bibfnamefont {T.}~\bibnamefont
  {Johannsen}}\ and\ \bibinfo {author} {\bibfnamefont {D.}~\bibnamefont
  {Psaltis}},\ }\href {\doibase 10.1103/PhysRevD.83.124015} {\bibfield
  {journal} {\bibinfo  {journal} {Phys. Rev. D}\ }\textbf {\bibinfo {volume}
  {83}},\ \bibinfo {pages} {124015} (\bibinfo {year} {2011})},\ \Eprint
  {http://arxiv.org/abs/1105.3191} {arXiv:1105.3191 [gr-qc]} \BibitemShut
  {NoStop}%
\bibitem [{\citenamefont {Johannsen}(2013)}]{Johannsen:2013szh}%
  \BibitemOpen
  \bibfield  {author} {\bibinfo {author} {\bibfnamefont {T.}~\bibnamefont
  {Johannsen}},\ }\href {\doibase 10.1103/PhysRevD.88.044002} {\bibfield
  {journal} {\bibinfo  {journal} {Phys. Rev. D}\ }\textbf {\bibinfo {volume}
  {88}},\ \bibinfo {pages} {044002} (\bibinfo {year} {2013})},\ \Eprint
  {http://arxiv.org/abs/1501.02809} {arXiv:1501.02809 [gr-qc]} \BibitemShut
  {NoStop}%
\bibitem [{\citenamefont {Rezzolla}\ and\ \citenamefont
  {Zhidenko}(2014)}]{Rezzolla:2014mua}%
  \BibitemOpen
  \bibfield  {author} {\bibinfo {author} {\bibfnamefont {L.}~\bibnamefont
  {Rezzolla}}\ and\ \bibinfo {author} {\bibfnamefont {A.}~\bibnamefont
  {Zhidenko}},\ }\href {\doibase 10.1103/PhysRevD.90.084009} {\bibfield
  {journal} {\bibinfo  {journal} {Phys. Rev. D}\ }\textbf {\bibinfo {volume}
  {90}},\ \bibinfo {pages} {084009} (\bibinfo {year} {2014})},\ \Eprint
  {http://arxiv.org/abs/1407.3086} {arXiv:1407.3086 [gr-qc]} \BibitemShut
  {NoStop}%
\bibitem [{\citenamefont {Cardoso}\ \emph {et~al.}(2014)\citenamefont
  {Cardoso}, \citenamefont {Pani},\ and\ \citenamefont
  {Rico}}]{Cardoso:2014rha}%
  \BibitemOpen
  \bibfield  {author} {\bibinfo {author} {\bibfnamefont {V.}~\bibnamefont
  {Cardoso}}, \bibinfo {author} {\bibfnamefont {P.}~\bibnamefont {Pani}}, \
  and\ \bibinfo {author} {\bibfnamefont {J.}~\bibnamefont {Rico}},\ }\href
  {\doibase 10.1103/PhysRevD.89.064007} {\bibfield  {journal} {\bibinfo
  {journal} {Phys. Rev. D}\ }\textbf {\bibinfo {volume} {89}},\ \bibinfo
  {pages} {064007} (\bibinfo {year} {2014})},\ \Eprint
  {http://arxiv.org/abs/1401.0528} {arXiv:1401.0528 [gr-qc]} \BibitemShut
  {NoStop}%
\bibitem [{\citenamefont {Konoplya}\ \emph {et~al.}(2016)\citenamefont
  {Konoplya}, \citenamefont {Rezzolla},\ and\ \citenamefont
  {Zhidenko}}]{Konoplya:2016jvv}%
  \BibitemOpen
  \bibfield  {author} {\bibinfo {author} {\bibfnamefont {R.}~\bibnamefont
  {Konoplya}}, \bibinfo {author} {\bibfnamefont {L.}~\bibnamefont {Rezzolla}},
  \ and\ \bibinfo {author} {\bibfnamefont {A.}~\bibnamefont {Zhidenko}},\
  }\href {\doibase 10.1103/PhysRevD.93.064015} {\bibfield  {journal} {\bibinfo
  {journal} {Phys. Rev. D}\ }\textbf {\bibinfo {volume} {93}},\ \bibinfo
  {pages} {064015} (\bibinfo {year} {2016})},\ \Eprint
  {http://arxiv.org/abs/1602.02378} {arXiv:1602.02378 [gr-qc]} \BibitemShut
  {NoStop}%
\bibitem [{\citenamefont {Papadopoulos}\ and\ \citenamefont
  {Kokkotas}(2018)}]{Papadopoulos:2018nvd}%
  \BibitemOpen
  \bibfield  {author} {\bibinfo {author} {\bibfnamefont {G.~O.}\ \bibnamefont
  {Papadopoulos}}\ and\ \bibinfo {author} {\bibfnamefont {K.~D.}\ \bibnamefont
  {Kokkotas}},\ }\href {\doibase 10.1088/1361-6382/aad7f4} {\bibfield
  {journal} {\bibinfo  {journal} {Class. Quant. Grav.}\ }\textbf {\bibinfo
  {volume} {35}},\ \bibinfo {pages} {185014} (\bibinfo {year} {2018})},\
  \Eprint {http://arxiv.org/abs/1807.08594} {arXiv:1807.08594 [gr-qc]}
  \BibitemShut {NoStop}%
\bibitem [{\citenamefont {Carson}\ and\ \citenamefont
  {Yagi}(2020)}]{Carson:2020dez}%
  \BibitemOpen
  \bibfield  {author} {\bibinfo {author} {\bibfnamefont {Z.}~\bibnamefont
  {Carson}}\ and\ \bibinfo {author} {\bibfnamefont {K.}~\bibnamefont {Yagi}},\
  }\href {\doibase 10.1103/PhysRevD.101.084030} {\bibfield  {journal} {\bibinfo
   {journal} {Phys. Rev. D}\ }\textbf {\bibinfo {volume} {101}},\ \bibinfo
  {pages} {084030} (\bibinfo {year} {2020})},\ \Eprint
  {http://arxiv.org/abs/2002.01028} {arXiv:2002.01028 [gr-qc]} \BibitemShut
  {NoStop}%
\bibitem [{\citenamefont {Yagi}\ \emph {et~al.}(2024)\citenamefont {Yagi},
  \citenamefont {Lomuscio}, \citenamefont {Lowrey},\ and\ \citenamefont
  {Carson}}]{Yagi:2023eap}%
  \BibitemOpen
  \bibfield  {author} {\bibinfo {author} {\bibfnamefont {K.}~\bibnamefont
  {Yagi}}, \bibinfo {author} {\bibfnamefont {S.}~\bibnamefont {Lomuscio}},
  \bibinfo {author} {\bibfnamefont {T.}~\bibnamefont {Lowrey}}, \ and\ \bibinfo
  {author} {\bibfnamefont {Z.}~\bibnamefont {Carson}},\ }\href {\doibase
  10.1103/PhysRevD.109.044017} {\bibfield  {journal} {\bibinfo  {journal}
  {Phys. Rev. D}\ }\textbf {\bibinfo {volume} {109}},\ \bibinfo {pages}
  {044017} (\bibinfo {year} {2024})},\ \Eprint
  {http://arxiv.org/abs/2311.08659} {arXiv:2311.08659 [gr-qc]} \BibitemShut
  {NoStop}%
\bibitem [{\citenamefont {Shipley}(2019)}]{Shipley:2019kfq}%
  \BibitemOpen
  \bibfield  {author} {\bibinfo {author} {\bibfnamefont {J.~O.}\ \bibnamefont
  {Shipley}},\ }\emph {\bibinfo {title} {{Strong-field gravitational lensing by
  black holes}}},\ \href@noop {} {Ph.D. thesis},\ \bibinfo  {school} {Sheffield
  U.} (\bibinfo {year} {2019}),\ \Eprint {http://arxiv.org/abs/1909.04691}
  {arXiv:1909.04691 [gr-qc]} \BibitemShut {NoStop}%
\bibitem [{\citenamefont {Wang}\ \emph {et~al.}(2018)\citenamefont {Wang},
  \citenamefont {Chen},\ and\ \citenamefont {Jing}}]{WangPhysRevD2018}%
  \BibitemOpen
  \bibfield  {author} {\bibinfo {author} {\bibfnamefont {M.}~\bibnamefont
  {Wang}}, \bibinfo {author} {\bibfnamefont {S.}~\bibnamefont {Chen}}, \ and\
  \bibinfo {author} {\bibfnamefont {J.}~\bibnamefont {Jing}},\ }\href {\doibase
  10.1103/PhysRevD.98.104040} {\bibfield  {journal} {\bibinfo  {journal} {Phys.
  Rev. D}\ }\textbf {\bibinfo {volume} {98}},\ \bibinfo {pages} {104040}
  (\bibinfo {year} {2018})}\BibitemShut {NoStop}%
\bibitem [{\citenamefont {Sengo}\ \emph {et~al.}(2023)\citenamefont {Sengo},
  \citenamefont {Cunha}, \citenamefont {Herdeiro},\ and\ \citenamefont
  {Radu}}]{Sengo_2023}%
  \BibitemOpen
  \bibfield  {author} {\bibinfo {author} {\bibfnamefont {I.}~\bibnamefont
  {Sengo}}, \bibinfo {author} {\bibfnamefont {P.~V.}\ \bibnamefont {Cunha}},
  \bibinfo {author} {\bibfnamefont {C.~A.}\ \bibnamefont {Herdeiro}}, \ and\
  \bibinfo {author} {\bibfnamefont {E.}~\bibnamefont {Radu}},\ }\href {\doibase
  10.1088/1475-7516/2023/01/047} {\bibfield  {journal} {\bibinfo  {journal}
  {Journal of Cosmology and Astroparticle Physics}\ }\textbf {\bibinfo {volume}
  {2023}},\ \bibinfo {pages} {047} (\bibinfo {year} {2023})}\BibitemShut
  {NoStop}%
\bibitem [{\citenamefont {Long}\ \emph {et~al.}(2020)\citenamefont {Long},
  \citenamefont {Chen}, \citenamefont {Wang},\ and\ \citenamefont
  {Jing}}]{Wang_long2020shadow}%
  \BibitemOpen
  \bibfield  {author} {\bibinfo {author} {\bibfnamefont {F.}~\bibnamefont
  {Long}}, \bibinfo {author} {\bibfnamefont {S.}~\bibnamefont {Chen}}, \bibinfo
  {author} {\bibfnamefont {M.}~\bibnamefont {Wang}}, \ and\ \bibinfo {author}
  {\bibfnamefont {J.}~\bibnamefont {Jing}},\ }\href@noop {} {\enquote {\bibinfo
  {title} {Shadow of a disformal kerr black hole in quadratic dhost
  theories},}\ } (\bibinfo {year} {2020}),\ \Eprint
  {http://arxiv.org/abs/2009.07508} {arXiv:2009.07508 [gr-qc]} \BibitemShut
  {NoStop}%
\bibitem [{\citenamefont {Wang}\ \emph {et~al.}(2022)\citenamefont {Wang},
  \citenamefont {Chen},\ and\ \citenamefont {Jing}}]{Wang:2022kvg}%
  \BibitemOpen
  \bibfield  {author} {\bibinfo {author} {\bibfnamefont {M.}~\bibnamefont
  {Wang}}, \bibinfo {author} {\bibfnamefont {S.}~\bibnamefont {Chen}}, \ and\
  \bibinfo {author} {\bibfnamefont {J.}~\bibnamefont {Jing}},\ }\href {\doibase
  10.1088/1572-9494/ac6e5c} {\bibfield  {journal} {\bibinfo  {journal} {Commun.
  Theor. Phys.}\ }\textbf {\bibinfo {volume} {74}},\ \bibinfo {pages} {097401}
  (\bibinfo {year} {2022})},\ \Eprint {http://arxiv.org/abs/2205.05855}
  {arXiv:2205.05855 [gr-qc]} \BibitemShut {NoStop}%
\bibitem [{\citenamefont {Dettmann}\ \emph {et~al.}(1994)\citenamefont
  {Dettmann}, \citenamefont {Frankel},\ and\ \citenamefont
  {Cornish}}]{Dettmann:1994dj}%
  \BibitemOpen
  \bibfield  {author} {\bibinfo {author} {\bibfnamefont {C.~P.}\ \bibnamefont
  {Dettmann}}, \bibinfo {author} {\bibfnamefont {N.~E.}\ \bibnamefont
  {Frankel}}, \ and\ \bibinfo {author} {\bibfnamefont {N.~J.}\ \bibnamefont
  {Cornish}},\ }\href {\doibase 10.1103/PhysRevD.50.R618} {\bibfield  {journal}
  {\bibinfo  {journal} {Phys. Rev. D}\ }\textbf {\bibinfo {volume} {50}},\
  \bibinfo {pages} {R618} (\bibinfo {year} {1994})},\ \Eprint
  {http://arxiv.org/abs/gr-qc/9402027} {arXiv:gr-qc/9402027} \BibitemShut
  {NoStop}%
\bibitem [{\citenamefont {Dettmann}\ \emph {et~al.}(1995)\citenamefont
  {Dettmann}, \citenamefont {Frankel},\ and\ \citenamefont
  {Cornish}}]{Dettmann:1995ex}%
  \BibitemOpen
  \bibfield  {author} {\bibinfo {author} {\bibfnamefont {C.~P.}\ \bibnamefont
  {Dettmann}}, \bibinfo {author} {\bibfnamefont {N.~E.}\ \bibnamefont
  {Frankel}}, \ and\ \bibinfo {author} {\bibfnamefont {N.~J.}\ \bibnamefont
  {Cornish}},\ }\href {\doibase 10.1142/S0218348X9500014X} {\bibfield
  {journal} {\bibinfo  {journal} {Fractals}\ }\textbf {\bibinfo {volume} {3}},\
  \bibinfo {pages} {161} (\bibinfo {year} {1995})},\ \Eprint
  {http://arxiv.org/abs/gr-qc/9502014} {arXiv:gr-qc/9502014} \BibitemShut
  {NoStop}%
\bibitem [{\citenamefont {Sota}\ \emph {et~al.}(1996)\citenamefont {Sota},
  \citenamefont {Suzuki},\ and\ \citenamefont {Maeda}}]{Sota:1995ms}%
  \BibitemOpen
  \bibfield  {author} {\bibinfo {author} {\bibfnamefont {Y.}~\bibnamefont
  {Sota}}, \bibinfo {author} {\bibfnamefont {S.}~\bibnamefont {Suzuki}}, \ and\
  \bibinfo {author} {\bibfnamefont {K.-i.}\ \bibnamefont {Maeda}},\ }\href
  {\doibase 10.1088/0264-9381/13/5/034} {\bibfield  {journal} {\bibinfo
  {journal} {Class. Quant. Grav.}\ }\textbf {\bibinfo {volume} {13}},\ \bibinfo
  {pages} {1241} (\bibinfo {year} {1996})},\ \Eprint
  {http://arxiv.org/abs/gr-qc/9505036} {arXiv:gr-qc/9505036} \BibitemShut
  {NoStop}%
\bibitem [{\citenamefont {Gueron}\ and\ \citenamefont
  {Letelier}(2001)}]{Gueron:2001ey}%
  \BibitemOpen
  \bibfield  {author} {\bibinfo {author} {\bibfnamefont {E.}~\bibnamefont
  {Gueron}}\ and\ \bibinfo {author} {\bibfnamefont {P.~S.}\ \bibnamefont
  {Letelier}},\ }\href {\doibase 10.1103/PhysRevE.63.035201} {\bibfield
  {journal} {\bibinfo  {journal} {Phys. Rev. E}\ }\textbf {\bibinfo {volume}
  {63}},\ \bibinfo {pages} {035201} (\bibinfo {year} {2001})},\ \Eprint
  {http://arxiv.org/abs/astro-ph/0101164} {arXiv:astro-ph/0101164} \BibitemShut
  {NoStop}%
\bibitem [{\citenamefont {Han}(2008)}]{Han:2008zzd}%
  \BibitemOpen
  \bibfield  {author} {\bibinfo {author} {\bibfnamefont {W.-b.}\ \bibnamefont
  {Han}},\ }\href {\doibase 10.1103/PhysRevD.77.123007} {\bibfield  {journal}
  {\bibinfo  {journal} {Phys. Rev. D}\ }\textbf {\bibinfo {volume} {77}},\
  \bibinfo {pages} {123007} (\bibinfo {year} {2008})},\ \Eprint
  {http://arxiv.org/abs/1006.2234} {arXiv:1006.2234 [gr-qc]} \BibitemShut
  {NoStop}%
\bibitem [{\citenamefont {Zelenka}\ and\ \citenamefont
  {Lukes-Gerakopoulos}(2017)}]{Zelenka:2017aqn}%
  \BibitemOpen
  \bibfield  {author} {\bibinfo {author} {\bibfnamefont {O.}~\bibnamefont
  {Zelenka}}\ and\ \bibinfo {author} {\bibfnamefont {G.}~\bibnamefont
  {Lukes-Gerakopoulos}},\ }in\ \href@noop {} {\emph {\bibinfo {booktitle}
  {{Workshop on Black Holes and Neutron Stars}}}}\ (\bibinfo {year} {2017})\
  \Eprint {http://arxiv.org/abs/1711.02442} {arXiv:1711.02442 [gr-qc]}
  \BibitemShut {NoStop}%
\bibitem [{\citenamefont {Destounis}\ and\ \citenamefont
  {Kokkotas}(2023)}]{Destounis:2023cim}%
  \BibitemOpen
  \bibfield  {author} {\bibinfo {author} {\bibfnamefont {K.}~\bibnamefont
  {Destounis}}\ and\ \bibinfo {author} {\bibfnamefont {K.~D.}\ \bibnamefont
  {Kokkotas}},\ }\href {\doibase 10.1007/s10714-023-03170-z} {\bibfield
  {journal} {\bibinfo  {journal} {Gen. Rel. Grav.}\ }\textbf {\bibinfo {volume}
  {55}},\ \bibinfo {pages} {123} (\bibinfo {year} {2023})},\ \Eprint
  {http://arxiv.org/abs/2305.18522} {arXiv:2305.18522 [gr-qc]} \BibitemShut
  {NoStop}%
\bibitem [{\citenamefont {Kopáček}\ \emph {et~al.}(2010)\citenamefont
  {Kopáček}, \citenamefont {Kovář}, \citenamefont {Karas},\ and\
  \citenamefont {Stuchlík}}]{kopacek2010}%
  \BibitemOpen
  \bibfield  {author} {\bibinfo {author} {\bibfnamefont {O.}~\bibnamefont
  {Kopáček}}, \bibinfo {author} {\bibfnamefont {J.}~\bibnamefont {Kovář}},
  \bibinfo {author} {\bibfnamefont {V.}~\bibnamefont {Karas}}, \ and\ \bibinfo
  {author} {\bibfnamefont {Z.}~\bibnamefont {Stuchlík}},\ }\href {\doibase
  10.1063/1.3506071} {\bibfield  {journal} {\bibinfo  {journal} {AIP Conference
  Proceedings}\ }\textbf {\bibinfo {volume} {1283}},\ \bibinfo {pages} {278}
  (\bibinfo {year} {2010})},\ \Eprint
  {http://arxiv.org/abs/https://pubs.aip.org/aip/acp/article-pdf/1283/1/278/12077873/278\_1\_online.pdf}
  {https://pubs.aip.org/aip/acp/article-pdf/1283/1/278/12077873/278\_1\_online.pdf}
  \BibitemShut {NoStop}%
\bibitem [{\citenamefont {Contopoulos}\ \emph {et~al.}(2011)\citenamefont
  {Contopoulos}, \citenamefont {Lukes-Gerakopoulos},\ and\ \citenamefont
  {Apostolatos}}]{CONTOPOULOS_2011}%
  \BibitemOpen
  \bibfield  {author} {\bibinfo {author} {\bibfnamefont {G.}~\bibnamefont
  {Contopoulos}}, \bibinfo {author} {\bibfnamefont {G.}~\bibnamefont
  {Lukes-Gerakopoulos}}, \ and\ \bibinfo {author} {\bibfnamefont {T.~A.}\
  \bibnamefont {Apostolatos}},\ }\href {\doibase 10.1142/S0218127411029768}
  {\bibfield  {journal} {\bibinfo  {journal} {Int. J. Bifurc. Chaos}\ }\textbf
  {\bibinfo {volume} {21}},\ \bibinfo {pages} {2261} (\bibinfo {year}
  {2011})},\ \Eprint {http://arxiv.org/abs/1108.5057} {arXiv:1108.5057 [gr-qc]}
  \BibitemShut {NoStop}%
\bibitem [{\citenamefont {Dubeibe}\ \emph {et~al.}(2007)\citenamefont
  {Dubeibe}, \citenamefont {Pachón},\ and\ \citenamefont
  {Sanabria-Gómez}}]{Dubeibe_2007}%
  \BibitemOpen
  \bibfield  {author} {\bibinfo {author} {\bibfnamefont {F.~L.}\ \bibnamefont
  {Dubeibe}}, \bibinfo {author} {\bibfnamefont {L.~A.}\ \bibnamefont
  {Pachón}}, \ and\ \bibinfo {author} {\bibfnamefont {J.~D.}\ \bibnamefont
  {Sanabria-Gómez}},\ }\href {\doibase 10.1103/physrevd.75.023008} {\bibfield
  {journal} {\bibinfo  {journal} {Physical Review D}\ }\textbf {\bibinfo
  {volume} {75}} (\bibinfo {year} {2007}),\
  10.1103/physrevd.75.023008}\BibitemShut {NoStop}%
\bibitem [{\citenamefont {Igata}\ \emph {et~al.}(2011)\citenamefont {Igata},
  \citenamefont {Ishihara},\ and\ \citenamefont {Takamori}}]{Igata_2011}%
  \BibitemOpen
  \bibfield  {author} {\bibinfo {author} {\bibfnamefont {T.}~\bibnamefont
  {Igata}}, \bibinfo {author} {\bibfnamefont {H.}~\bibnamefont {Ishihara}}, \
  and\ \bibinfo {author} {\bibfnamefont {Y.}~\bibnamefont {Takamori}},\ }\href
  {\doibase 10.1103/physrevd.83.047501} {\bibfield  {journal} {\bibinfo
  {journal} {Physical Review D}\ }\textbf {\bibinfo {volume} {83}} (\bibinfo
  {year} {2011}),\ 10.1103/physrevd.83.047501}\BibitemShut {NoStop}%
\bibitem [{\citenamefont {Eichhorn}\ and\ \citenamefont
  {Held}(2021{\natexlab{b}})}]{Eichhorn:2021etc}%
  \BibitemOpen
  \bibfield  {author} {\bibinfo {author} {\bibfnamefont {A.}~\bibnamefont
  {Eichhorn}}\ and\ \bibinfo {author} {\bibfnamefont {A.}~\bibnamefont
  {Held}},\ }\href {\doibase 10.1140/epjc/s10052-021-09716-2} {\bibfield
  {journal} {\bibinfo  {journal} {Eur. Phys. J. C}\ }\textbf {\bibinfo {volume}
  {81}},\ \bibinfo {pages} {933} (\bibinfo {year} {2021}{\natexlab{b}})},\
  \Eprint {http://arxiv.org/abs/2103.07473} {arXiv:2103.07473 [gr-qc]}
  \BibitemShut {NoStop}%
\bibitem [{\citenamefont {Wald}(1984)}]{Wald:1984rg}%
  \BibitemOpen
  \bibfield  {author} {\bibinfo {author} {\bibfnamefont {R.~M.}\ \bibnamefont
  {Wald}},\ }\href {\doibase 10.7208/chicago/9780226870373.001.0001} {\emph
  {\bibinfo {title} {{General Relativity}}}}\ (\bibinfo  {publisher} {Chicago
  Univ. Pr.},\ \bibinfo {address} {Chicago, USA},\ \bibinfo {year}
  {1984})\BibitemShut {NoStop}%
\bibitem [{\citenamefont {{Hartle}}(1967)}]{1967ApJ...150.1005H}%
  \BibitemOpen
  \bibfield  {author} {\bibinfo {author} {\bibfnamefont {J.~B.}\ \bibnamefont
  {{Hartle}}},\ }\href {\doibase 10.1086/149400} {\bibfield  {journal}
  {\bibinfo  {journal} {\apj}\ }\textbf {\bibinfo {volume} {150}},\ \bibinfo
  {pages} {1005} (\bibinfo {year} {1967})}\BibitemShut {NoStop}%
\bibitem [{\citenamefont {{Hartle}}\ and\ \citenamefont
  {{Thorne}}(1968)}]{hartle2}%
  \BibitemOpen
  \bibfield  {author} {\bibinfo {author} {\bibfnamefont {J.~B.}\ \bibnamefont
  {{Hartle}}}\ and\ \bibinfo {author} {\bibfnamefont {K.~S.}\ \bibnamefont
  {{Thorne}}},\ }\href {\doibase 10.1086/149707} {\bibfield  {journal}
  {\bibinfo  {journal} {\apj}\ }\textbf {\bibinfo {volume} {153}},\ \bibinfo
  {pages} {807} (\bibinfo {year} {1968})}\BibitemShut {NoStop}%
\bibitem [{\citenamefont {Glampedakis}\ and\ \citenamefont
  {Pappas}(2019)}]{Glampedakis:2018blj}%
  \BibitemOpen
  \bibfield  {author} {\bibinfo {author} {\bibfnamefont {K.}~\bibnamefont
  {Glampedakis}}\ and\ \bibinfo {author} {\bibfnamefont {G.}~\bibnamefont
  {Pappas}},\ }\href {\doibase 10.1103/PhysRevD.99.124041} {\bibfield
  {journal} {\bibinfo  {journal} {Phys. Rev. D}\ }\textbf {\bibinfo {volume}
  {99}},\ \bibinfo {pages} {124041} (\bibinfo {year} {2019})},\ \Eprint
  {http://arxiv.org/abs/1806.09333} {arXiv:1806.09333 [gr-qc]} \BibitemShut
  {NoStop}%
\bibitem [{\citenamefont {Papapetrou}(1945)}]{Papapetrou:1947}%
  \BibitemOpen
  \bibfield  {author} {\bibinfo {author} {\bibfnamefont {A.}~\bibnamefont
  {Papapetrou}},\ }\href {http://www.jstor.org/stable/20488481} {\bibfield
  {journal} {\bibinfo  {journal} {Proceedings of the Royal Irish Academy.
  Section A: Mathematical and Physical Sciences}\ }\textbf {\bibinfo {volume}
  {51}},\ \bibinfo {pages} {191} (\bibinfo {year} {1945})}\BibitemShut
  {NoStop}%
\bibitem [{\citenamefont {Majumdar}(1947)}]{Majumdar:1947eu}%
  \BibitemOpen
  \bibfield  {author} {\bibinfo {author} {\bibfnamefont {S.~D.}\ \bibnamefont
  {Majumdar}},\ }\href {\doibase 10.1103/PhysRev.72.390} {\bibfield  {journal}
  {\bibinfo  {journal} {Phys. Rev.}\ }\textbf {\bibinfo {volume} {72}},\
  \bibinfo {pages} {390} (\bibinfo {year} {1947})}\BibitemShut {NoStop}%
\bibitem [{\citenamefont {{Cunha}}\ \emph
  {et~al.}(2016{\natexlab{b}})\citenamefont {{Cunha}}, \citenamefont
  {{Grover}}, \citenamefont {{Herdeiro}}, \citenamefont {{Radu}}, \citenamefont
  {{R{\'u}narsson}},\ and\ \citenamefont {{Wittig}}}]{Cunha2016PhRvD}%
  \BibitemOpen
  \bibfield  {author} {\bibinfo {author} {\bibfnamefont {P.~V.~P.}\
  \bibnamefont {{Cunha}}}, \bibinfo {author} {\bibfnamefont {J.}~\bibnamefont
  {{Grover}}}, \bibinfo {author} {\bibfnamefont {C.}~\bibnamefont
  {{Herdeiro}}}, \bibinfo {author} {\bibfnamefont {E.}~\bibnamefont {{Radu}}},
  \bibinfo {author} {\bibfnamefont {H.}~\bibnamefont {{R{\'u}narsson}}}, \ and\
  \bibinfo {author} {\bibfnamefont {A.}~\bibnamefont {{Wittig}}},\ }\href
  {\doibase 10.1103/PhysRevD.94.104023} {\bibfield  {journal} {\bibinfo
  {journal} {Phys. Rev. D}\ }\textbf {\bibinfo {volume} {94}},\ \bibinfo {eid}
  {104023} (\bibinfo {year} {2016}{\natexlab{b}})}\BibitemShut {NoStop}%
\bibitem [{\citenamefont {Zotos}(2017)}]{Zotos_2017}%
  \BibitemOpen
  \bibfield  {author} {\bibinfo {author} {\bibfnamefont {E.~E.}\ \bibnamefont
  {Zotos}},\ }\href {\doibase 10.1007/s11012-017-0647-8} {\bibfield  {journal}
  {\bibinfo  {journal} {Meccanica}\ }\textbf {\bibinfo {volume} {52}},\
  \bibinfo {pages} {2615} (\bibinfo {year} {2017})}\BibitemShut {NoStop}%
\bibitem [{\citenamefont {Perlick}\ and\ \citenamefont
  {Tsupko}(2022)}]{Perlick_2022}%
  \BibitemOpen
  \bibfield  {author} {\bibinfo {author} {\bibfnamefont {V.}~\bibnamefont
  {Perlick}}\ and\ \bibinfo {author} {\bibfnamefont {O.~Y.}\ \bibnamefont
  {Tsupko}},\ }\href {\doibase 10.1016/j.physrep.2021.10.004} {\bibfield
  {journal} {\bibinfo  {journal} {Physics Reports}\ }\textbf {\bibinfo {volume}
  {947}},\ \bibinfo {pages} {1} (\bibinfo {year} {2022})}\BibitemShut {NoStop}%
\bibitem [{\citenamefont {Bardeen}\ \emph {et~al.}(1972)\citenamefont
  {Bardeen}, \citenamefont {Press},\ and\ \citenamefont
  {Teukolsky}}]{Bardeen:1972fi}%
  \BibitemOpen
  \bibfield  {author} {\bibinfo {author} {\bibfnamefont {J.~M.}\ \bibnamefont
  {Bardeen}}, \bibinfo {author} {\bibfnamefont {W.~H.}\ \bibnamefont {Press}},
  \ and\ \bibinfo {author} {\bibfnamefont {S.~A.}\ \bibnamefont {Teukolsky}},\
  }\href {\doibase 10.1086/151796} {\bibfield  {journal} {\bibinfo  {journal}
  {Astrophys. J.}\ }\textbf {\bibinfo {volume} {178}},\ \bibinfo {pages} {347}
  (\bibinfo {year} {1972})}\BibitemShut {NoStop}%
%%CITATION = ASJOA,178,347;%%
\bibitem [{\citenamefont {Peitgen}\ \emph {et~al.}(2004)\citenamefont
  {Peitgen}, \citenamefont {J{\"u}rgens},\ and\ \citenamefont
  {Saupe}}]{peitgen2004chaos}%
  \BibitemOpen
  \bibfield  {author} {\bibinfo {author} {\bibfnamefont {H.}~\bibnamefont
  {Peitgen}}, \bibinfo {author} {\bibfnamefont {H.}~\bibnamefont
  {J{\"u}rgens}}, \ and\ \bibinfo {author} {\bibfnamefont {D.}~\bibnamefont
  {Saupe}},\ }\href {https://books.google.gr/books?id=jVpS_u0Lg4gC} {\emph
  {\bibinfo {title} {Chaos and Fractals: New Frontiers of Science}}}\ (\bibinfo
   {publisher} {Springer New York},\ \bibinfo {year} {2004})\BibitemShut
  {NoStop}%
\bibitem [{\citenamefont {Mandelbrot}(1967)}]{Mandelbrotbritain}%
  \BibitemOpen
  \bibfield  {author} {\bibinfo {author} {\bibfnamefont {B.}~\bibnamefont
  {Mandelbrot}},\ }\href {http://www.jstor.org/stable/1721427} {\bibfield
  {journal} {\bibinfo  {journal} {Science}\ }\textbf {\bibinfo {volume}
  {156}},\ \bibinfo {pages} {636} (\bibinfo {year} {1967})}\BibitemShut
  {NoStop}%
\bibitem [{\citenamefont {Carroll}\ and\ \citenamefont
  {Ostlie}(2017)}]{Carroll_Ostlie_2017}%
  \BibitemOpen
  \bibfield  {author} {\bibinfo {author} {\bibfnamefont {B.~W.}\ \bibnamefont
  {Carroll}}\ and\ \bibinfo {author} {\bibfnamefont {D.~A.}\ \bibnamefont
  {Ostlie}},\ }\href@noop {} {\emph {\bibinfo {title} {An Introduction to
  Modern Astrophysics}}},\ \bibinfo {edition} {2nd}\ ed.\ (\bibinfo
  {publisher} {Cambridge University Press},\ \bibinfo {year}
  {2017})\BibitemShut {NoStop}%
\bibitem [{\citenamefont {Gralla}(2021)}]{Gralla:2020pra}%
  \BibitemOpen
  \bibfield  {author} {\bibinfo {author} {\bibfnamefont {S.~E.}\ \bibnamefont
  {Gralla}},\ }\href {\doibase 10.1103/PhysRevD.103.024023} {\bibfield
  {journal} {\bibinfo  {journal} {Phys. Rev. D}\ }\textbf {\bibinfo {volume}
  {103}},\ \bibinfo {pages} {024023} (\bibinfo {year} {2021})},\ \Eprint
  {http://arxiv.org/abs/2010.08557} {arXiv:2010.08557 [astro-ph.HE]}
  \BibitemShut {NoStop}%
\bibitem [{\citenamefont {Ozel}\ \emph {et~al.}(2022)\citenamefont {Ozel},
  \citenamefont {Psaltis},\ and\ \citenamefont {Younsi}}]{Ozel:2021ayr}%
  \BibitemOpen
  \bibfield  {author} {\bibinfo {author} {\bibfnamefont {F.}~\bibnamefont
  {Ozel}}, \bibinfo {author} {\bibfnamefont {D.}~\bibnamefont {Psaltis}}, \
  and\ \bibinfo {author} {\bibfnamefont {Z.}~\bibnamefont {Younsi}},\ }\href
  {\doibase 10.3847/1538-4357/ac9fcb} {\bibfield  {journal} {\bibinfo
  {journal} {Astrophys. J.}\ }\textbf {\bibinfo {volume} {941}},\ \bibinfo
  {pages} {88} (\bibinfo {year} {2022})},\ \Eprint
  {http://arxiv.org/abs/2111.01123} {arXiv:2111.01123 [astro-ph.HE]}
  \BibitemShut {NoStop}%
\bibitem [{\citenamefont {Lara}\ \emph {et~al.}(2021)\citenamefont {Lara},
  \citenamefont {V\"olkel},\ and\ \citenamefont {Barausse}}]{Lara:2021zth}%
  \BibitemOpen
  \bibfield  {author} {\bibinfo {author} {\bibfnamefont {G.}~\bibnamefont
  {Lara}}, \bibinfo {author} {\bibfnamefont {S.~H.}\ \bibnamefont {V\"olkel}},
  \ and\ \bibinfo {author} {\bibfnamefont {E.}~\bibnamefont {Barausse}},\
  }\href {\doibase 10.1103/PhysRevD.104.124041} {\bibfield  {journal} {\bibinfo
   {journal} {Phys. Rev. D}\ }\textbf {\bibinfo {volume} {104}},\ \bibinfo
  {pages} {124041} (\bibinfo {year} {2021})},\ \Eprint
  {http://arxiv.org/abs/2110.00026} {arXiv:2110.00026 [gr-qc]} \BibitemShut
  {NoStop}%
\bibitem [{\citenamefont {Olmo}\ \emph {et~al.}(2023)\citenamefont {Olmo},
  \citenamefont {Rosa}, \citenamefont {Rubiera-Garcia},\ and\ \citenamefont
  {Saez-Chillon~Gomez}}]{Olmo:2023lil}%
  \BibitemOpen
  \bibfield  {author} {\bibinfo {author} {\bibfnamefont {G.~J.}\ \bibnamefont
  {Olmo}}, \bibinfo {author} {\bibfnamefont {J.~L.}\ \bibnamefont {Rosa}},
  \bibinfo {author} {\bibfnamefont {D.}~\bibnamefont {Rubiera-Garcia}}, \ and\
  \bibinfo {author} {\bibfnamefont {D.}~\bibnamefont {Saez-Chillon~Gomez}},\
  }\href {\doibase 10.1088/1361-6382/aceacd} {\bibfield  {journal} {\bibinfo
  {journal} {Class. Quant. Grav.}\ }\textbf {\bibinfo {volume} {40}},\ \bibinfo
  {pages} {174002} (\bibinfo {year} {2023})},\ \Eprint
  {http://arxiv.org/abs/2302.12064} {arXiv:2302.12064 [gr-qc]} \BibitemShut
  {NoStop}%
\bibitem [{\citenamefont {Kostaros}\ \emph {et~al.}(ming)\citenamefont
  {Kostaros}, \citenamefont {Papadopoulos},\ and\ \citenamefont
  {Pappas}}]{Kostaros2024}%
  \BibitemOpen
  \bibfield  {author} {\bibinfo {author} {\bibfnamefont {K.}~\bibnamefont
  {Kostaros}}, \bibinfo {author} {\bibfnamefont {P.}~\bibnamefont
  {Papadopoulos}}, \ and\ \bibinfo {author} {\bibfnamefont {G.}~\bibnamefont
  {Pappas}},\ }\href@noop {} {\  (\bibinfo {year} {forthcoming})}\BibitemShut
  {NoStop}%
\bibitem [{\citenamefont {Thi}\ and\ \citenamefont
  {Papadopoulos}(ming)}]{Thi2024}%
  \BibitemOpen
  \bibfield  {author} {\bibinfo {author} {\bibfnamefont {W.-F.}\ \bibnamefont
  {Thi}}\ and\ \bibinfo {author} {\bibfnamefont {P.}~\bibnamefont
  {Papadopoulos}},\ }\href@noop {} {\bibfield  {journal} {\bibinfo  {journal}
  {Astronomy \& Astrophysics}\ } (\bibinfo {year} {forthcoming})}\BibitemShut
  {NoStop}%
\bibitem [{\citenamefont {{Younsi, Z.}}\ \emph {et~al.}(2012)\citenamefont
  {{Younsi, Z.}}, \citenamefont {{Wu, K.}},\ and\ \citenamefont {{Fuerst, S.
  V.}}}]{Younsi2012}%
  \BibitemOpen
  \bibfield  {author} {\bibinfo {author} {\bibnamefont {{Younsi, Z.}}},
  \bibinfo {author} {\bibnamefont {{Wu, K.}}}, \ and\ \bibinfo {author}
  {\bibnamefont {{Fuerst, S. V.}}},\ }\href {\doibase
  10.1051/0004-6361/201219599} {\bibfield  {journal} {\bibinfo  {journal}
  {A\&A}\ }\textbf {\bibinfo {volume} {545}},\ \bibinfo {pages} {A13} (\bibinfo
  {year} {2012})}\BibitemShut {NoStop}%
\bibitem [{\citenamefont {Fuerst}\ and\ \citenamefont
  {Wu}(2004)}]{Fuerst_2004}%
  \BibitemOpen
  \bibfield  {author} {\bibinfo {author} {\bibfnamefont {S.~V.}\ \bibnamefont
  {Fuerst}}\ and\ \bibinfo {author} {\bibfnamefont {K.}~\bibnamefont {Wu}},\
  }\href {\doibase 10.1051/0004-6361:20035814} {\bibfield  {journal} {\bibinfo
  {journal} {Astronomy {\&} Astrophysics}\ }\textbf {\bibinfo {volume} {424}},\
  \bibinfo {pages} {733} (\bibinfo {year} {2004})}\BibitemShut {NoStop}%
\bibitem [{\citenamefont {Fuerst}\ and\ \citenamefont
  {Wu}(2007)}]{Fuerst_2007}%
  \BibitemOpen
  \bibfield  {author} {\bibinfo {author} {\bibfnamefont {S.~V.}\ \bibnamefont
  {Fuerst}}\ and\ \bibinfo {author} {\bibfnamefont {K.}~\bibnamefont {Wu}},\
  }\href {\doibase 10.1051/0004-6361:20066008} {\bibfield  {journal} {\bibinfo
  {journal} {Astronomy {\&} Astrophysics}\ }\textbf {\bibinfo {volume} {474}},\
  \bibinfo {pages} {55} (\bibinfo {year} {2007})}\BibitemShut {NoStop}%
\bibitem [{\citenamefont {{Thorne}}(1974)}]{1974ApJThorne}%
  \BibitemOpen
  \bibfield  {author} {\bibinfo {author} {\bibfnamefont {K.~S.}\ \bibnamefont
  {{Thorne}}},\ }\href {\doibase 10.1086/152991} {\bibfield  {journal}
  {\bibinfo  {journal} {\apj}\ }\textbf {\bibinfo {volume} {191}},\ \bibinfo
  {pages} {507} (\bibinfo {year} {1974})}\BibitemShut {NoStop}%
\bibitem [{\citenamefont {Shipley}\ and\ \citenamefont
  {Dolan}(2016)}]{Shipley_2016}%
  \BibitemOpen
  \bibfield  {author} {\bibinfo {author} {\bibfnamefont {J.~O.}\ \bibnamefont
  {Shipley}}\ and\ \bibinfo {author} {\bibfnamefont {S.~R.}\ \bibnamefont
  {Dolan}},\ }\href {\doibase 10.1088/0264-9381/33/17/175001} {\bibfield
  {journal} {\bibinfo  {journal} {Classical and Quantum Gravity}\ }\textbf
  {\bibinfo {volume} {33}},\ \bibinfo {pages} {175001} (\bibinfo {year}
  {2016})}\BibitemShut {NoStop}%
\bibitem [{\citenamefont {Ayzenberg}\ \emph {et~al.}(2023)\citenamefont
  {Ayzenberg} \emph {et~al.}}]{Ayzenberg:2023hfw}%
  \BibitemOpen
  \bibfield  {author} {\bibinfo {author} {\bibfnamefont {D.}~\bibnamefont
  {Ayzenberg}} \emph {et~al.},\ }\href@noop {} {\  (\bibinfo {year} {2023})},\
  \Eprint {http://arxiv.org/abs/2312.02130} {arXiv:2312.02130 [astro-ph.HE]}
  \BibitemShut {NoStop}%
\bibitem [{\citenamefont {Gurvits}\ \emph {et~al.}(2021)\citenamefont {Gurvits}
  \emph {et~al.}}]{Gurvits:2019ioq}%
  \BibitemOpen
  \bibfield  {author} {\bibinfo {author} {\bibfnamefont {L.~I.}\ \bibnamefont
  {Gurvits}} \emph {et~al.},\ }\href {\doibase 10.1007/s10686-021-09714-y}
  {\bibfield  {journal} {\bibinfo  {journal} {Exper. Astron.}\ }\textbf
  {\bibinfo {volume} {51}},\ \bibinfo {pages} {559} (\bibinfo {year} {2021})},\
  \Eprint {http://arxiv.org/abs/1908.10767} {arXiv:1908.10767 [astro-ph.IM]}
  \BibitemShut {NoStop}%
\bibitem [{\citenamefont {Fromm}\ \emph {et~al.}(2021)\citenamefont {Fromm},
  \citenamefont {Mizuno}, \citenamefont {Younsi}, \citenamefont {Olivares},
  \citenamefont {Porth}, \citenamefont {De~Laurentis}, \citenamefont {Falcke},
  \citenamefont {Kramer},\ and\ \citenamefont {Rezzolla}}]{Fromm:2021flr}%
  \BibitemOpen
  \bibfield  {author} {\bibinfo {author} {\bibfnamefont {C.~M.}\ \bibnamefont
  {Fromm}}, \bibinfo {author} {\bibfnamefont {Y.}~\bibnamefont {Mizuno}},
  \bibinfo {author} {\bibfnamefont {Z.}~\bibnamefont {Younsi}}, \bibinfo
  {author} {\bibfnamefont {H.}~\bibnamefont {Olivares}}, \bibinfo {author}
  {\bibfnamefont {O.}~\bibnamefont {Porth}}, \bibinfo {author} {\bibfnamefont
  {M.}~\bibnamefont {De~Laurentis}}, \bibinfo {author} {\bibfnamefont
  {H.}~\bibnamefont {Falcke}}, \bibinfo {author} {\bibfnamefont
  {M.}~\bibnamefont {Kramer}}, \ and\ \bibinfo {author} {\bibfnamefont
  {L.}~\bibnamefont {Rezzolla}},\ }\href {\doibase 10.1051/0004-6361/201937335}
  {\bibfield  {journal} {\bibinfo  {journal} {Astron. Astrophys.}\ }\textbf
  {\bibinfo {volume} {649}},\ \bibinfo {pages} {A116} (\bibinfo {year}
  {2021})},\ \Eprint {http://arxiv.org/abs/2101.08618} {arXiv:2101.08618
  [astro-ph.HE]} \BibitemShut {NoStop}%
\bibitem [{\citenamefont {{The Sage Developers}}(2021)}]{sagemath}%
  \BibitemOpen
  \bibfield  {author} {\bibinfo {author} {\bibnamefont {{The Sage
  Developers}}},\ }\href@noop {} {\emph {\bibinfo {title} {{S}ageMath, the
  {S}age {M}athematics {S}oftware {S}ystem ({V}ersion 9.3.2)}}} (\bibinfo
  {year} {2021}),\ \bibinfo {note} {{\tt https://www.sagemath.org}}\BibitemShut
  {NoStop}%
\end{thebibliography}%

\end{document}